\documentclass[sn-mathphys,Numbered]{sn-jnl}
\usepackage{graphicx}%
\usepackage{multirow}%
\usepackage{mathtools}
\usepackage{amsmath,amssymb,amsfonts}%
\usepackage{amsthm}%
\usepackage{mathrsfs}%
\usepackage[title]{appendix}%
\usepackage{xcolor}%
\usepackage{textcomp}%
\usepackage{manyfoot}%
\usepackage{booktabs}%
\usepackage{algpseudocode}%
\usepackage{listings}%
\usepackage{float}
\usepackage{booktabs} 
\usepackage[linesnumbered, ruled, vlined]{algorithm2e}
\usepackage{caption}
\usepackage{subcaption}
\usepackage{tcolorbox}
\usepackage{hyperref}
\hypersetup{
    colorlinks,
    citecolor=blue,
    filecolor=black,
    linkcolor=red,
    urlcolor=green
}

\theoremstyle{thmstyleone}%
\newtheorem{theorem}{Theorem}
\theoremstyle{thmstyletwo}%

\theoremstyle{thmstylethree}%
\newtheorem{definition}{Definition}%

\begin{document}

\title{Minimizing Regret in Billboard Advertisement under Zonal Influence Constraint   }

\author{\fnm{Dildar} \sur{Ali}}\email{2021rcs2009@iitjammu.ac.in}
\author*{\fnm{Suman} \sur{Banerjee}}\email{suman.banerjee@iitjammu.ac.in}

\author{\fnm{Yamuna} \sur{Prasad}}\email{yamuna.prasad@iitjammu.ac.in}

\affil{\orgdiv{Department of Computer Science and Engineering}, \orgname{Indian Institute of Technology Jammu}, \orgaddress{\postcode{181221}, \state{Jammu and Kashmir}, \country{India}}}

\abstract{In a typical \emph{billboard advertisement} technique, a number of digital billboards are owned by an \emph{influence provider}, and many advertisers approach the influence provider for a specific number of views of their advertisement content on a payment basis. If the influence provider provides the demanded or more influence, then he will receive the full payment or else a partial payment. In the context of an influence provider, if he provides more or less than an advertiser's demanded influence, it is a loss for him. This is formalized as ‘Regret’, and naturally, in the context of the influence provider, the goal will be to allocate the billboard slots among the advertisers such that the total regret is minimized. In this paper, we study this problem as a discrete optimization problem and propose four solution approaches. The first one selects the billboard slots from the available ones in an incremental greedy manner, and we call this method the Budget Effective Greedy approach. In the second one, we introduce randomness with the first one, where we perform the marginal gain computation for a sample of randomly chosen billboard slots. The remaining two approaches are further improvements over the second one. We analyze all the algorithms to understand their time and space complexity. We implement them with real-life trajectory and billboard datasets and conduct a number of experiments. It has been observed that the randomized budget effective greedy approach takes reasonable computational time while minimizing the regret.}


\maketitle

\section{Introduction}
\begin{figure}[t]
    \centering
    \includegraphics[width=90mm]{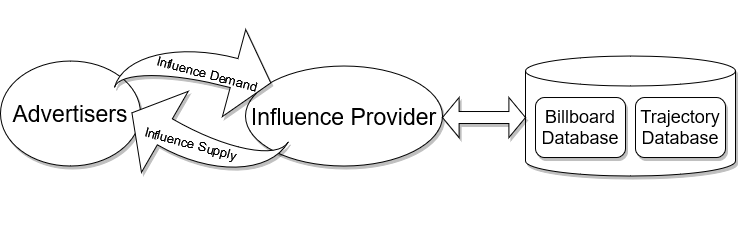}
     \caption{Schematic Diagram}
    \label{Fig:1}
\end{figure}

Among many other advertisement techniques, digital billboards have emerged as an effective approach for the out-of-home advertisement technique \footnote{\url{https://www.lamar.com/howtoadvertise/Research/}}. In a recent market survey \footnote{\url{https://topmediadvertising.co.uk/billboard-advertising-statistics/}}, it has been reported that this advertisement approach leads to more return on investment. In general, if an advertiser requests some influence from an influence provider depending upon their budget, the influence provider provides the required influence and gains profit. Figure \ref{Fig:1} shows a schematic diagram of the process. Most of the existing literature \cite{8295265,wang2022data,zhang2020towards,wang2019efficiently,10.14778/3352063.3352067,lotfi2017multi} on the influence maximization problem in billboard advertisement is solved from the advertiser's perspective. The key problem addressed in these studies is to choose $k$ many influential billboard slots to maximize the influence. However, from the influence provider's point of view, billboard slots must be allocated to minimize the total regret (formally defined in Section \ref{Sec:BPD}). Hence, deviating from the existing literature, in this paper, we study the problem of allocating billboard slots to minimize the overall regret. To the best of our knowledge, other than the study by \cite{zhang2021minimizing}, there does not exist any other study in this direction.

\paragraph{\textbf{Background}} Now-a-days, billboards are digital and allocated slot-wise to the advertisers for displaying their advertisement content. In practice, several advertisers approach an influence provider for some influence demand on a payment basis. The payment rule is as follows: if the influence provider provides the demanded or more influence, then the full payment will be made else a partial payment may be based on a prorata basis. In this setting, for the influence provider, it will be a loss whether he provides more or less influence (compared to the demanded one). This loss has been mathematically formalized as regret. Detailed discussion on this is deferred till Section \ref{Sec:BPD}. The allocation of the billboard slots needs to be done so that the whole regret is minimized. Next, we discuss the motivation of our study.  
 
\paragraph{\textbf{Motivation}} At present, a very limited amount of literature on regret analysis in billboard advertisements is available. In all the existing literature \cite{ali2022influential,ali2023efficient,zhang2019optimizing,zhang2020towards,zhang2021minimizing}, it has been considered that the influence at every zone is equally meaningful. However, it may not be the case always in practice. Consider the following scenario where we know the demographic information regarding a city's population. Then, it is quite natural that it is not meaningful to display the advertisement content of a costly product in a place where we know that the population is economically backward classes. Also, it is important to observe that this only makes sense for the billboards that are placed in residential zones. Hence, to incorporate this concept, we divide the billboard slots into two groups: those in crowded locations and those in residential zones. Now, from the advertiser's point of view, they may have the demand of zonal specific influence requirements. This means instead of having a total influence demand, advertisers post zone-specific influence demand. Consider the geographical region under consideration $\mathcal{Z}$ is divided into $k$ many non-overlapping zones ${z}_{1}, {z}_{2}, \ldots, {z}_{k}$ and advertiser $a_i$ expresses their zone specific influence requirement $\sigma_{i}({z}_{1}), \sigma_{i}({z}_{2}), \ldots, \sigma_{i}({z}_{k})$. So, studying the billboard slot allocation problem under the zone-specific influence demand constraint is important.

\paragraph{\textbf{Problem Definition}} Consider an influence provider with access to a billboard database containing information about several digital billboards and a trajectory database containing the location information over different time stamps of a group of people. A number of advertisers are approaching the influence provider to have a given amount of influence on a payment basis. As mentioned previously, the payment scheme is as follows: "If the influence provider can provide the required or more influence as asked by the advertiser, then the influence provider will get the full payment or else a partial payment will be made". Now, we analyze the situation from the influence provider's point of view. If the influence provider provides less influence, it will be a loss because the advertiser will make a partial payment. Similarly, if it provides more influence than required, it also causes loss to the influence provider. Because the influence provider does not get any incentive, on the other hand, this extra influence could have been provided to the other advertisers whose influence demand has not been fulfilled yet. Both these losses have been formalized in terms of ‘regret’. This study aims to allocate the billboard slots to minimize the total regret while maintaining the zone-specific influence demand. In this paper, we study this problem as a discrete optimization problem and propose four solution methodologies.

\section{Related Work}
This section will discuss the existing literature on regret minimization and influence maximization that are loosely related to our proposed work. We divide our literature survey into three major categories: Influential Zone Selection, Trajectory Driven Influence Maximization, and Regret Minimization.

\subsection{Influential Zone Selection}
The influential zone or site selection problems gain popularity due to its vast application area. In this direction, Cabello et al.\cite{cabello2006reverse,cabello2010facility} first introduce the MaxBRNN problem. They present an Euclidean distance-based strategy to solve the problem of locating all points in a dataset whose closest neighbor is a specific query point. Later, Du et al. \cite{10.1007/11535331_10} solve optimal location query in spatial databases. They use $R^{*}$- tree and $OL$-tree-based solution using Manhattan distance. In a similar direction, Xia et al. \cite{10.5555/1083592.1083701} address the problem of top-$t$ most influential sites selection and propose a new approach by extending the work of finding RNN proposed by Stanoi et al. \cite{10.5555/645927.672207} in which $R$-tree is called many times and that is the drawback of their work. To overcome this, Xia et al. proposed a novel algorithm called `$TopInfluentialSites$' in which the $R$-tree will be called only once. Next, Zhou et al. \cite{5767892} extended the MaxBRNN problem to the MaxBRkNN problem, which identifies an ideal region where establishing a service site ensures that the greatest number of consumers will regard the site to be one of their $k$ nearest service sites. Later, Huang et al. \citep{10.1145/2063576.2063971} studied the top-$k$ most influential location selection problem and proposed two branch and bound-based solution approaches using different geometric properties to prune the search space and perform better than a naive-based approach. Extensive experiments on real-world and synthetic datasets show the efficiency of their proposed methods.
\subsection{Trajectory Driven Influence Maximization}
The advancement of mobile and sensors increases the availability of trajectory datasets \cite{10.1145/3440207}. In recent years, Guo et al. \cite{7929916} have studied the problem of finding $k$ optimal trajectories rather than identifying $k$ influential sites due to the efficiency of influence maximization over the trajectory. The primary objective is to identify optimal trajectories associated with certain advertisements to maximize their impact on target audiences. Subsequently, Zhang et al. \cite{10.1145/3292500.3330829} conducted a study that addressed the issue of identifying a collection of highly influential billboards capable of attracting a greater number of trajectories while adhering to specific budgetary limitations. They present a user-defined parameter-based termination technique while introducing a tangent line-based algorithm to determine the top bound of billboard influence. It obtains an approximation guarantee of $\frac{\theta}{2}(1-\frac{1}{e})$. In a similar direction, a study by Wang et al. \cite{8604082} found that the outdoor advertising business faces challenges in effectively delivering its intended influence to the intended audience. A divide-and-conquer-based search technique was implemented to address this discrepancy. Instead of just considering billboards and trajectory, they considered some aspects, including advertisement content, traffic behavior, and mobility transition. Next, Zahradka et al.\cite{zahradka2021price} did a case study analysis of the cost of billboard advertising in the different regions of the Czech Republic. Zhang et al.\cite{zhang2018trajectory} studied the trajectory-driven influential billboard placement problem where a set of billboards, along with its location and cost, is given. The goal here is to choose a subset of the billboards within the budget that influence the largest number of trajectories. Later, Ali et al. \cite{ali2022influential,ali2023influential} studied the problem of selecting influential billboard slots and introduced a submodularity graph-based solution approach. Their main contribution is to prune less influential billboard slots and select the Top-$k$ most influential slots. Further, Wang et al. \cite{10.1145/3495159} studied the problem of the Targeted Outdoor Advertising Recommendation (TOAR) problem considering the facts of user profiles and advertisement topics. Their main contribution was a targeted influence model that characterizes the advertising influence spread along with user mobility. Based on the divide and conquer approach, they developed two solution strategies. Implementation with real-world datasets shows the efficiency and effectiveness of the proposed solution approaches. There are some loosely related studies with our work also. Liu et al.\cite{Liu2017SmartAdPVA} developed a visualization tool using large-scale trajectory data to place the billboards. 

\subsection{Regret Minimization}
Until now, very little literature exists in the context of regret minimization. In the direction of database query optimization, a $k$-regret query was examined \cite{10.14778/1920841.1920980, 6816699, 10.1145/3183713.3196903} which addresses decision-making based on multiple criteria. Their solution aims to overcome the constraints of both the top-$k$ query, which requires an exact utility function from the user, and the skyline query, which cannot control the size of the result. A $k$-regret query is designed to select $k$ tuples from a given database in such a manner that ensures the utility of the user's preferred tuple among these $k$ selections is only marginally lower than the usefulness of their preferred tuple in the entire database, regardless of the precise utility function utilized.  Further, Xie et al. \cite{10.1145/3299869.3300068} studied the problem of minimizing the regret ratio by enhancing user interaction in selecting the best tuples from the whole database. In social network advertisement, Aslay et al. \cite{10.14778/2752939.2752950} introduce a new problem domain that involves allocating users to advertisers to promote commercial posts. Their approach leverages the network effect and considers several practical considerations, including ad relevance, social proof’s impact, user attention span, and the constraints imposed by restricted advertiser expenditures. In a similar direction, few studies in the context of billboard advertisement consider the minimization of regret causes due to providing influence to the advertiser. Next, Zhang et al. \cite{zhang2021minimizing} studied the problem of regret minimization and proposed several solution methodologies. Experimentation with real-world datasets showed the efficiency of the approaches. Recently, Ali et al. \cite{ali2023efficient} introduced an efficient algorithm to minimize the total regret of a billboard provider in a multi-advertiser setting. To the best of our knowledge, no study exists related to regret minimization under zonal influence constraint for online or offline advertisement. Motivated by these observations, we studied the problem of regret minimization under zonal influence constraint by satisfying zone-specific influence demand in multi-advertiser settings.

\paragraph{\textbf{Our Contributions}} To the best of our knowledge, this is the first study on regret minimization for billboard advertisements considering zonal influence constraint. In particular, we make the following contributions:
\begin{itemize}
    \item We study the Regret Minimization in Billboard Advertisement Problem in multi-advertiser settings for which the literature is very limited.
    \item We propose four efficient heuristic solutions: Budget-Effective Greedy, Randomized Budget-Effective Greedy, Randomized Synchronous Greedy, and Randomized Advertiser Exchange algorithm for the regret minimization problem.
    \item We perform an extensive set of experiments with real-world datasets and compare the performance of the proposed solutions with the available strategies in the literature.
\end{itemize}

\paragraph{\textbf{Organization of the Paper}} The rest of the paper is organized as follows. Section \ref{Sec:BPD} describes required background concepts and defines the problem formally. Section \ref{Sec:PS} describes the proposed solution approach with illustration and analysis. Section \ref{Sec:EE} contains an experimental setup with different parameter settings and discussion. Section \ref{Sec:RD} contains the experimental evaluations and analysis of the proposed solution approaches. Section \ref{Sec:CFD} concludes this study and gives future research directions. Next, we discuss background and problem definition.

\section{Background and Problem Definition} \label{Sec:BPD}
In this section, we describe the background of the problem and formally define our problem. For any positive integer $k$, $[k]$ denotes the set $\{1,2, \ldots, k\}$ and for any two positive integer $x$ and $y$, with $x \leq y$, $[x,y]$ denotes the set $\{x, x+1, \ldots, y\}$. Initially, we start by describing the notion of billboard and billboard slots.

\subsection{Billboards and Billboard Slots} Consider $\ell$ billboards $\mathcal{B}=\{b_1, b_2, \ldots, b_{\ell}\}$ are places at different locations of a city. Each billboard $b_i \in \mathcal{B}$ runs for the duration $[T_1,T_2]$, and these billboards are allocated to different advertisers for a fixed duration denoted by $\Delta$. We call each fixed-duration slot a billboard slot. We denote a billboard slot as a tuple of the form $(b_i,[t,t+\Delta])$ where $b_i \in \mathcal{B}$ and $t \in \{T_1, T_1+\Delta+1, T_1+2 \Delta+1, \ldots, \frac{T_2-T_1}{\Delta} - \Delta+1 \}$. We denote the set of all billboard slots as $\mathbb{BS}$. Given a subset of billboard slots $\mathcal{S} \subseteq \mathbb{BS}$, we denote its influence as $I(\mathcal{S})$ and can be computed in different ways \cite{zhang2018trajectory,zhang2021minimizing}. We consider the influence is happening by the rule of the triggering model stated in Definition \ref{IBS}

\begin{definition}[Influence of Billboard Slots]\label{IBS}
Given a subset of billboard slots $\mathcal{S}$ and a trajectory database $\mathcal{D}$, the influence of $\mathcal{S}$ under the triggering model can be computed using Equation No. \ref{Eq:Influence}.
\begin{equation} \label{Eq:Influence}
    I(\mathcal{S})= \underset{t_j \in \mathcal{D}}{\sum} 1 \ - \ \underset{b_i \in \mathbb{BS}}{\prod} (1- Pr(b_i,t_j))
\end{equation}
where $Pr(b_i,t_j)$ denotes the influence probability of the trajectory $t_j$ by the billboard slot $b_i$.
\end{definition}
In this study, we assume that for every $b_i \in \mathbb{BS}$ and $t_j \in \mathcal{D}$, the value of $Pr(b_i,t_j)$ is known. In the literature \cite{zhang2020towards,zhang2021minimizing}, there are standard methods for computing this value. Next, we describe the notion of Zonal influence constraint.

\subsection{Zonal Influence Constraint} 
Consider that the geographical region under consideration $\mathcal{Z}$ is partitioned into $k$ many disjoint zones $z_1, z_2, \ldots, z_k$. Naturally, any billboard $b_i \in \mathcal{B}$ will belong to one of the $k$ zones. All the slots associated with the billboard will belong to the zone in which the billboard belongs to. Hence, the set of billboard slots $\mathbb{BS}$ can be partitioned into $k$ parts $\mathbb{BS}_{1}, \mathbb{BS}_{2}, \ldots, \mathbb{BS}_{k}$ such that for any $i \in [k]$, the partition $\mathbb{BS}_{i}$ contains the billboard slots that belongs to the zone $z_i$ only. For any advertiser $a_i \in \mathcal{A}$, can be represented by a tuple $(a_i, \sigma_i,u_i)$, which means that the advertiser $a_i$ is opting for the influence of amount $\sigma_i$ by the payment $u_i$. In this study, we consider that the influence demand of each advertiser is also specific to each zone. Hence, $\sigma_i$ can be replaced by the tuple $(\sigma^{1}_i, \sigma^{2}_i, \ldots, \sigma^{k}_i)$ where $\sigma^{j}_{i}$ denotes the influence demand of the advertiser $a_i$ for the zone $z_j$. Hence, for any $i \in [n]$, $\sigma_i =\sum_{j=1}^{k} \sigma^{j}_{i}$. Now, we formally define the zonal influence constraint.

\begin{definition} [Zonal Influence Constraint]\label{ZIC}
Let, $\mathbb{BS}_{a_i}$ denote the set of billboard slots allocated to the advertiser $a_i$. $\mathbb{BS}^{z_j}_{a_i}$ is the subset of $\mathbb{BS}_{a_i}$, contains the billboard slots that belongs to the zone $z_j$. Now, we say that the allocation satisfies the zonal influence constraint for the advertiser $a_i$ and zone $z_j$ if and only if $I(\mathbb{BS}^{z_j}_{a_i}) \geq \sigma^{j}_{i}$.  
\end{definition}
Next, we describe the regret model that we have considered in our study for billboard slot allocation.

\subsection{The Regret Model} As mentioned previously, there are $n$ advertisers, $\mathcal{A}=\{a_1, a_2, $ $ \ldots,a_n\}$ and one influence provider $\mathcal{X}$. Each advertiser $a_i$ submits a campaign proposal to the Influence Provider $\mathcal{X}$ as follows: ``The advertiser $a_i$ has the influence demand of $\sigma^{1}_{i}, \sigma^{2}_{i}, \ldots, \sigma^{k}_{i}$ over the zones $z_{1}, z_{2}, \ldots, z_{k}$ on the basis of payment $u_i$". The influence provider $\mathcal{X}$ has the advertiser database $\mathbb{A}$ whose content are the tuples of the form $ (a_i,<\sigma^{1}_{i}, \sigma^{2}_{i}, \ldots, \sigma^{k}_{i}>,u_i)$ for all $i \in [n]$. Let $\mathbb{BS}^{z_j}_{a_i}$ be the set of billboard slots allocated to the advertiser $a_i$ and belongs to zone $z_j$. Now, as per the payment rule for all $z_j \in \mathcal{Z}$, if $I(\mathbb{BS}^{z_j}_{a_i}) \geq \sigma^{j}_{i}$, then the payment of amount $u_i$ will be made else a partial payment (e.g., pro-rata basis which will be lesser than the $u_i$) will be made." Now, there are a few important points to highlight.

\begin{itemize}
    \item The first one is for any advertiser and for any zone if the influence provided by $\mathcal{X}$ is less than what is expected, then this is a loss for $\mathcal{X}$ because a partial payment will be received. We call this regret as the \emph{Unsatisfied Regret}.
    \item On the other hand, if $\mathcal{X}$ provides more influence than what is expected, that also leads to a loss for $\mathcal{X}$. The reason is as follows. Assume that for the advertiser $a_i$ and zone $z_j$, $I(\mathbb{BS}^{z_j}_{a_i}) > \sigma^{j}_{i}$. Now, for providing the excess influence $I(\mathbb{BS}^{z_j}_{a_i}) - \sigma^{j}_{i}$ to the advertiser $a_i$ for the zone $z_j$, the influence provider does not get any extra incentive. Also, it may so happen that for any other advertiser $a_x \in \mathcal{A} \setminus \{a_i\}$ at the zone $z_j$, $I(\mathbb{BS}^{z_j}_{a_x}) < \sigma^{j}_{x}$. So, some of the billboard slots from $\mathbb{BS}^{z_j}_{a_i}$ can be included to $\mathbb{BS}^{z_j}_{a_x}$ such that $I(\mathbb{BS}^{z_j}_{a_x}) \geq \sigma^{j}_{x}$. We call the regret of this kind the \emph{Excessive Regret}.
 \end{itemize}

Combining these two cases, we define the regret model considered in this paper in Definition \ref{Def:Reg_Model}.
\begin{definition}[The Regret Model] \label{Def:Reg_Model}
    Let, for the advertiser $a_i \in \mathcal{A}$, the set of allocated billboard slots is $\mathbb{BS}_{a_i}$ and the regret associated with this allocation for the zone $z_{j}$ is denoted by $\mathcal{R}^{z_j}_{a_i}$. This quantity can be defined by the following conditional equation:
    \[
    \mathcal{R}^{z_j}_{a_i} = 
\begin{cases}
    u_{i} \cdot (1- \gamma \cdot \frac{I(\mathbb{BS}^{z_j}_{a_i})}{\sigma^{j}_{i}}),& \text{if }  \sigma^{j}_{i} > I(\mathbb{BS}^{z_j}_{a_i}) \\
    u_{i} \cdot \frac{I(\mathbb{BS}^{z_j}_{a_i}) - \sigma^{j}_{i}}{\sigma^{j}_{i}},              & \text{otherwise}
\end{cases}
\]
Here, the fraction $\frac{I(\mathbb{BS}^{z_j}_{a_i})}{\sigma^{j}_{i}}$ is part of the satisfied influence for the advertiser $a_i$ from the zone $z_{j}$ and $\gamma$ is a parameter which is called as the penalty ratio due to the unsatisfied demand. 
\end{definition}
The developed solution methodologies in this paper are independent of the choice of $\gamma$, and it has been explained further in Section \ref{Sec:EE} of this paper.

\subsection{Allocation of Billboard Slots}\label{SubSec: Allocation_slots}
The task is to allocate the billboard slots among the advertisers. The goal is to minimize the total regret while the following two constraints are satisfied:
\begin{itemize}
\item \emph{Zonal Influence Constraint :} This constraint means that for any advertiser $a_i \in \mathcal{A}$, and zone $z_j \in \mathcal{Z}$, $I(\mathbb{BS}^{z_j}_{a_i}) \geq \sigma^{j}_{i}$.
\item\emph{Disjointness Constraint :} For any two advertisers $a_x$ and $a_y$, let $\mathbb{BS}_{a_x}$ and $\mathbb{BS}_{a_y}$ denotes the set of billboard slots allocated to $a_x$ and $a_y$, respectively. The disjointness constraint implies that any billboard slot $(b_i,[t,t+\Delta]) \in \mathbb{BS}$ can not be allocated to more than one advertiser. Hence,  $\mathbb{BS}_{a_x} \cap \mathbb{BS}_{a_y} = \emptyset$.
\end{itemize}

Let $\mathcal{L}$ denote the set of all possible allocations, and any arbitrary allocation of all $n$ advertisers is denoted as $(\mathbb{BS}_{a_1}, \mathbb{BS}_{a_2}, \ldots, \mathbb{BS}_{a_n})$. Now, we define the notion of feasible allocation in Definition \ref{Def:FA}.

\begin{definition}[Feasible Allocation] \label{Def:FA}
An allocation of the billboard slots is said to be a feasible allocation if it satisfies both the Zonal Influence Constraint and the disjointness constraint.
\end{definition}
As mentioned previously, the objective of the allocation is to minimize the total regret. So, we define the notion of total regret for an allocation in Definition \ref{Def:TR}.

\begin{definition} [Total Regret Associated with an Allocation] \label{Def:TR}
    Given an allocation of the billboard slots $\mathcal{Y}=\{\mathbb{BS}_{a_1}, \mathbb{BS}_{a_2}, \ldots, \mathbb{BS}_{a_n}\}$ the total regret associated with the allocation $\mathcal{Y}$ is denoted by $\mathcal{R}(\mathcal{Y})$ and defined as the sum of the regret associated with the individual advertisers. In turn, the regret due to the allocation of one advertiser can be obtained by summing up the corresponding zonal regrets. 
    \begin{equation}
\mathcal{R}(\mathcal{Y})= \underset{a_i \in \mathcal{A}}{\sum} \ \underset{z_j \in \mathcal{Z}}{\sum} \  \mathcal{R}^{z_j}_{a_i}
\end{equation} 
\end{definition}
It can be observed that the value of $ \mathcal{R}^{z_j}_{a_i}$ has already defined. We denote an optimal allocation of the billboard slots by $\mathcal{Y}^{OPT}$. As mentioned previously, the goal of the influence provider is to find the optimal allocation of the billboard slots such that total regret is minimized. We call this problem the \textsc{Regret Minimization with Zonal Influence Constraint} Problem. This problem has been stated in Definition \ref{Def:Problem}.

\begin{definition}[Regret Minimization with Zonal Influence Constraint] \label{Def:Problem}
Given the billboard slot information $\mathbb{BS}$, a trajectory database $\mathcal{D}$, and the advertiser database $\mathbb{A}$, the goal of this problem is to create an allocation $\mathcal{Y}=\{\mathbb{BS}_{a_1}, \mathbb{BS}_{a_2}, \ldots, \mathbb{BS}_{a_n}\}$ in which the set of billboard slots $\mathbb{BS}_{a_i}$ are allocated to the advertiser $a_{i}$ such that the total regret is minimized. Mathematically, this problem can be posed as follows:

\begin{equation}
\mathcal{Y}^{OPT} = \underset{\mathcal{Y}_{i} \in \mathcal{L}(\mathcal{Y})}{argmin} \ \mathcal{R}(\mathcal{Y}_{i})
\end{equation}
\end{definition}
Here the constraint is for all $\mathbb{BS}_{a_i}$ and $\mathbb{BS}_{a_j}$ they need to be disjoint; i.e., $\mathbb{BS}_{a_i} \cap \mathbb{BS}_{a_j} = \emptyset$. Now, from the computational point of view, this problem can be written as follows:

\begin{center}
\begin{tcolorbox}[title=\textsc{Regret Minimization in Billboard Advertisement} Problem, width=12.5cm]
\textbf{Input:} Billboard Slots set $\mathbb{BS}$, Influence Function $I()$, Trajectory Database $\mathcal{D}$, Advertiser Database $\mathbb{A}$.

\textbf{Problem:} Find out an optimal allocation $\mathcal{Y}^{OPT} = \{\mathbb{BS}_{a_1}, \mathbb{BS}_{a_2}, \ldots, \mathbb{BS}_{a_n}\}$ of the billboard slots that minimizes the overall regret.
\end{tcolorbox}
\end{center}
The regret minimization problem in the context of billboard advertisement has been studied by Zhang et al. \cite{zhang2021minimizing} without the zonal influence constraint. They had an inappoximability result stated in Theorem \ref{Th:1}. 

\begin{theorem} \label{Th:1}
\textsc{Regret Minimization with Zonal Influence  Constraint} Problem is NP-hard to solve optimally, and it is also hard to obtain any constant factor approximation algorithm.  
\end{theorem}

This means that the same inappoximability result will also hold for our problem as well. Next, we proceed to describe the proposed solution methodologies.

\section{Proposed Solution} \label{Sec:PS}
The hardness of the problem implies that any efficient algorithm with a theoretical guarantee for optimal regret does not exist unless P = NP. Now, we define the notion of budget-effective advertiser in Definition \ref{Def:1}.

\begin{definition} [Budget Effective Advertiser] \label{Def:1}
Given any two advertisers $a_i, a_j \in \mathcal{A}$ with their respective proposal $(a_i,\sigma_{i},u_{i})$ and  $(a_j,\sigma_{j},u_{j})$, $a_{i}$ is said to be more budget effective than $a_{j}$ if $(\frac{u_{i}}{\sigma_{i}} > \frac{u_{j}}{\sigma_{j}})$ holds.
\end{definition}

 Considering its hardness, we propose two efficient algorithms: (1) Budget Effective Greedy and (2) Randomized Budget Effective Greedy that will satisfy more budget-effective advertisers while minimizing the total regret for the influence provider's perspective. 

\subsection{Budget-Effective Greedy (BG)}
We introduce a heuristic Greedy in Algorithm \ref{Algo:1}.  First, initialize a set of empty billboard slots set in Line No. $1$. Next, sort each advertiser in descending order based on budget over demand. Now, in Line No. $3$ to $11$, for each advertiser and its corresponding zones, \texttt{While Loop} iterates for assigning billboard slots to the advertisers that best reduce the total regret and satisfy the zonal influence requirement. The final billboard slot set is returned after all the advertisers are satisfied or billboard slots run out.

\begin{algorithm}[h!]
\scriptsize
\SetAlgoLined
\KwData{Trajectory Database $\mathcal{D}$, Billboard Slot Information $\mathbb{BS}$, Advertiser Database $\mathbb{A}$, and the Influence Function $I()$.}
\KwResult{  An allocation $\mathcal{Y}$ of the billboard slots that minimizes the total regret}
$\text{Initialize} \ \mathcal{Y} \leftarrow \{\mathcal{S}_{1}, \mathcal{S}_{2},\ldots,\mathcal{S}_{|\mathcal{A}|}\} \ \text{of empty lists}$\; 
 $\mathcal{A} \leftarrow \text{Sort each advertiser $a_{i} \in \mathcal{A}$ based on descending order of } \frac{u_i}{\sigma_{i}} $\;
 \For {each $a_{i} \in \mathcal{A}$} {
 \For {each $z_{j} \in a_{i_z}$}{
 \While{$\sigma_{i} > I(S_{i}) ~and~ \mathcal{BS}_{z_j} \neq \phi$}{
 $s^{*} \leftarrow \underset{s \in \mathcal{BS}_{z_j}}{argmax} \ \frac{\mathcal{R}(\mathcal{S}_{i})- \mathcal{R}(\mathcal{S}_{i} \cup \{s\})}{\sigma(\{s\})}$\;
 $\mathcal{S}_{i} \leftarrow \mathcal{S}_{i} \cup \{s^{*}\}$\;
 $\mathcal{BS}_{z_j} \leftarrow \mathcal{BS}_{z_j} \setminus \{s^{*}\}$\;
 }}}
 return $ \mathcal{Y}$;
 \caption{Budget-Effective Greedy Algorithm for Regret Minimization Problem}
 \label{Algo:1}
\end{algorithm}

\textbf{Complexity Analysis.} Now, we analyze the time and space requirement of Algorithm \ref{Algo:1}. In-Line No. $1$, initializing $\mathcal{Y}$ as a set of empty billboard slots set will take $\mathcal{O}(n)$ time. Sorting advertisers in Line No. $2$ will take $\mathcal{O}(n \log n)$ time. It is clear that \texttt{For Loop} at Line No. $3$ will iterate for $n$ number of advertisers, and it will take $\mathcal{O}(n)$ time to execute. Line No. $4$ \texttt{For Loop} will execute at most $k$ times, and it takes $\mathcal{O}(k)$ time. Now, in the worst case, \texttt{While Loop} at Line No. $5$ will execute at most the number of billboard slots a zone has, and to compute the regret in Line No. $6$, we need to compute the influence for $l$ number of slots that takes $\mathcal{O}(n.k.\ell.t)$ time where $t$ is the number of tuple in the trajectory database. Line No. $7$ and $8$ will take $\mathcal{O}(n.k)$ time. Hence, Line No. $3$ to $11$ will take $\mathcal{O}(n.k.\ell.t + n.k)$ time to execute. So, the total time taken by Algorithm \ref{Algo:1} is $\mathcal{O}(n + n \log n + n.k.\ell.t + n.k)$ i.e., $\mathcal{O}(n \log n + n.k.\ell.t)$. The extra space is taken by Algorithm \ref{Algo:1} to store the billboard slots list $\mathcal{Y}$ and  Advertisers list $\mathcal{A}$ both will take $\mathcal{O}(n)$ space. As $k$ number of zones is there, to store them, it will take $\mathcal{O}(k)$ extra space. In the worst case, $\mathcal{S}_{i}$ will take $\mathcal{O}(\ell)$ extra space as at most $\ell$ many billboard slots can be allocated to satisfy the zonal influence requirement of an advertiser. Hence, the total extra space required by Algorithm \ref{Algo:1} is $\mathcal{O}(n+n+k+\ell)$ i.e., $\mathcal{O}(n+k+\ell)$.
\par Though Algorithm \ref{Algo:1} is simple to understand and easy to implement, this algorithm is extremely inefficient, as mentioned in Section \ref{Sec:EE}. This is particularly due to the huge number of marginal gain computations. To eliminate this problem, we propose the randomized budget effective greedy algorithm. 
\subsection{Randomized Budget-Effective Greedy (RG)}
Algorithm \ref{Algo:2} represents the randomized budget-effective greedy heuristic search in the form of pseudocode. At first, in Line No. $1$, we initialize an empty set of billboard slots set for each advertiser. Next, each advertiser is sorted in descending order based on the advertiser's budget over influence demand. Each advertiser has influence demand over different zones, and to satisfy this quantity, in Line No. $3$ \texttt{For Loop} iterates over all the advertisers, and in Line No. $4$, \texttt{For Loop} iterates over different zone for each advertiser. Now, in Line No. $5$, billboard slots for a particular zone are sorted in ascending order according to their individual influence value. Next, in Line No. $8$ to $13$ \texttt{While Loop} iterates over all the slots of a zone to satisfy the zonal influence demand of an advertiser. From this \texttt{While Loop}, a maximum number of billboard slots set that satisfy the zonal influence demand of an advertiser is generated. So, from Line No. $14$ to $19$, a greedy heuristic is applied to fulfill the advertiser's influence demand by assigning slots according to zonal influence requirement and best reduce the total regret. Now, in Line No. $14$, for each advertiser with corresponding zonal information, \texttt{While Loop} iterates till influence demand for a particular zone is not satisfied or slots run out. At each iteration, it samples out $\frac{|\mathcal{BS}_{z_j}|}{|\mathcal{S}_{i}^{'}|} \log \frac{1}{\epsilon}$ many elements form $\mathcal{BS}_{z_j}$, and from the sample out slots pick one which minimizes the regret. Here, $|\mathcal{BS}_{z_j}|$, and $|\mathcal{S}_{i}^{'}|$ is the number of slots in zone $j$, and the maximum number of billboard slots required from a zone to satisfy an advertiser $a_{i}$, respectively, and $\epsilon$ is a user-defined parameter. This procedure will continue till all the advertisers are satisfied or no more billboard slots remain, and finally, the billboard slot set $\mathcal{Y}$ will be returned.

\begin{algorithm}[h!]
\scriptsize
\SetAlgoLined
\KwData{Trajectory Database $\mathcal{D}$, Billboard Slot Information $\mathbb{BS}$, Advertiser Database $\mathbb{A}$, and the Influence Function $I()$.}
\KwResult{  An allocation $\mathcal{Y}$ of the billboard slots that minimizes the total regret}
$\text{Initialize} \ \mathcal{Y} \leftarrow \{\mathcal{S}_{1}, \mathcal{S}_{2},\ldots,\mathcal{S}_{|\mathcal{A}|}\} \ \text{of empty lists}$\; 
 $\mathcal{A} \leftarrow \text{Sort each advertiser $a_{i} \in \mathcal{A}$ based on descending order of } \frac{u_i}{\sigma_{i}} $\;
 \For {each $a_{i} \in \mathcal{A}$} {
 \For {each $z_{j} \in a_{i_z}$}{
  $\mathcal{BS}_{z_j}\leftarrow \text{Sort slots in ascending order based on individual influence; }$\\
 $c \longleftarrow 1$\;
 $\mathcal{S}_{i}^{'} \longleftarrow \emptyset$; $\mathcal{S}_{i} \longleftarrow \emptyset$; $I(\mathcal{S}_{i}^{'}) \longleftarrow 0 $; $I(\mathcal{S}_{i}) \longleftarrow 0 $\;
 \While{$I(\mathcal{S}_{i}^{'}) < \sigma_{i}$ and $\mathcal{BS}_{z_j} \neq \emptyset$}{
 $\mathcal{S}_{i}^{'} \longleftarrow \mathcal{S}_{i}^{'} \cup \mathcal{BS}_{z_j}[c]$\;
 $I(\mathcal{S}_{i}^{'}) \longleftarrow I(\mathcal{S}_{i}^{'}) + I(\mathcal{BS}_{z_j}[c])$\;
 $\mathcal{BS}_{z_j} \longleftarrow \mathcal{BS}_{z_j} \setminus (\mathcal{BS}_{z_j}[c])$\;
 $c \longleftarrow c+1$
 }
 \While{$\sigma_{i} > I(S_{i}) ~and~ \mathcal{BS}_{z_j} \neq \emptyset$}{
 $\mathcal{P} \leftarrow \text{Sample }\frac{|\mathcal{BS}_{z_j}|}{|\mathcal{S}_{i}^{'}|} \log \frac{1}{\epsilon}\text{ many elements from }\mathcal{BS}_{z_j} \setminus \mathcal{S}_{i}$\;
 $s^{*} \leftarrow \underset{s \in \mathcal{P}}{argmax} \ \frac{\mathcal{R}(\mathcal{S}_{i})- \mathcal{R}(\mathcal{S}_{i} \cup \{s\})}{\sigma(\{s\})}$\;
 $\mathcal{S}_{i} \leftarrow \mathcal{S}_{i} \cup \{s^{*}\}$\;
 $\mathcal{BS}_{z_j} \leftarrow \mathcal{BS}_{z_j} \setminus \{s^{*}\}$\;
 }}}
 return $ \mathcal{Y}$;
 \caption{Randomized Budget-Effective Greedy Algorithm for Regret Minimization Problem}
 \label{Algo:2}
\end{algorithm}

\textbf{Complexity Analysis.} Now, we analyze the time and space requirement of Algorithm \ref{Algo:2}. In-Line No. $1$ initializing $\mathcal{Y}$ will take $\mathcal{O}(n)$ time, and Line No. $2$ will take $\mathcal{O}(n \log n)$ time to sort the advertisers. Now, in Line No. $3$ \texttt{For Loop} executes for $\mathcal{O}(n)$ time, and in Line No. $4$ \texttt{For Loop} executes for $\mathcal{O}(k)$ time. To sort the advertiser in Line No. $5$ based on individual influence value will take $\mathcal{O}(n.k.\ell \log \ell)$ time, and Line No. $6$ and $7$ will take $\mathcal{O}(n.k)$ time for initialization. Now, in Line No. $8$ \texttt{While Loop} will execute for $\mathcal{O}(n.k.\ell)$ time as in the worst case, a zone has at most $\ell$ number of billboard slots. Line No. $9,11,12$ will take $\mathcal{O}(n.k)$ time execute, but Line No. $10$ will take $\mathcal{O}(n.k.\ell.t)$ as each time needs to compute the influence using Equation \ref{Eq:Influence}. So, Line No. $8$ to $13$ will take $\mathcal{O}(n.k + n.k. \ell. t)$ time in total. In the worst case, \texttt{While Loop} in Line No. $14$ will execute for $\mathcal{O}(n.k.\ell)$ time. In-Line No. $15$ to sample out $\frac{|\mathcal{BS}_{z_j}|}{|\mathcal{S}_{i}^{'}|} \log \frac{1}{\epsilon}$ many elements from $\mathcal{BS}_{z_j}$ will take $\mathcal{O}(n.k.\ell.{\frac{|\mathcal{BS}_{z_j}|}{\ell} \log \frac{1}{\epsilon}})$ time when $|\mathcal{S}_{i}^{'}| = \ell$ in the worst case. Calculating regret in Line No. $16$ will take $\mathcal{O}(n.k.\ell.t)$ time and Line No. $17,18$ will take $\mathcal{O}(n.k.\ell)$ computational time. So, from Line No. $14$ to $19$ will take $\mathcal{O}(n.k.\ell.{\frac{|\mathcal{BS}_{z_j}|}{\ell} \log \frac{1}{\epsilon}}+n.k.\ell.t + n.k.\ell)$ time in total. Hence, Algorithm \ref{Algo:2} will take $\mathcal{O}(n+ n \log n + n.k.\ell \log \ell + n.k + n.k + n.k. \ell. t + n.k.\ell.{\frac{|\mathcal{BS}_{z_j}|}{\ell} \log \frac{1}{\epsilon}}+n.k.\ell.t + n.k)$ i.e., $\mathcal{O}(n \log n + n.k.\ell \log \ell + n.k. \ell. t + n.k.\ell.{\frac{|\mathcal{BS}_{z_j}|}{\ell} \log \frac{1}{\epsilon}})$ time.

Now, the additional space requirement to store the lists $\mathcal{Y}$, $\mathcal{A}$, $\mathcal{BS}_{z_j}$, $\mathcal{S}_{i}^{'}$, and $\mathcal{S}_{i}$ will be of $\mathcal{O}(n)$, $\mathcal{O}(n)$, $\mathcal{O}(\ell)$, $\mathcal{O}(\ell)$, and $\mathcal{O}(\ell)$, respectively. To store zonal information of billboard slots will take $\mathcal{O}(k)$ space. Hence, the total extra space requirement for Algorithm \ref{Algo:2} will be of $\mathcal{O}(k+2n+3\ell)$, i.e., $\mathcal{O}(k+n+\ell)$.

\subsection{Randomized Synchronous Greedy Approach (RSG)}
Budget-effective greedy suffers from allocating billboard slots fairly, as most influential slots are allocated to some of the most budget-effective advertisers. The unsatisfied regret plays a significant role compared to excessive regret to produce total regret in our regret model described in Section \ref{Def:Reg_Model}. So, the advertisers who are unsatisfied due to billboard slots running out are not preferable to consider. To address this, we introduce a synchronous greedy approach in Algorithm \ref{Algo:3} in which slots are allocated synchronously to overcome the drawback of budget-effective greedy. First, in Line No. $1$, initialize a set of empty billboard slots set. In Line No. $2$, all advertisers are sorted according to their budget effectiveness. Next, in Line No. $3$ to $21$, synchronously allocate billboard slots to each advertiser individually, reducing the regret best. In Line No. $22$ to $34$, the number of unsatisfied advertisers is calculated after allocating billboard slots to them. Now, in Line No. $35$ to $43$ after allocation, if more than one advertiser is still unsatisfied, then release advertisers one by one according to less budget effectiveness and allocate the released slots to other unsatisfied ones. This procedure continues until all the remaining advertisers are satisfied or no further improvement occurs.

\begin{algorithm}[h!]
\scriptsize
\SetAlgoLined
\KwData{Trajectory Database $\mathcal{D}$, Billboard Slot Information $\mathbb{BS}$, Advertiser Database $\mathbb{A}$, and the Influence Function $I()$.}
\KwResult{  An allocation $\mathcal{Y}$ of the billboard slots that minimizes the total regret}
$\text{Initialize} \ \mathcal{Y} \leftarrow \{\mathcal{S}_{1}, \mathcal{S}_{2},\ldots,\mathcal{S}_{|\mathcal{A}|}\} \ \text{of empty lists}$\; 
 $\mathcal{A} \leftarrow \text{Sort each advertiser $a_{i} \in \mathcal{A}$ based on descending order of } \frac{u_i}{\sigma_{i}} $\;
\While{True}{  
 \For {each $a_{i} \in \mathcal{A}$} {
 \For {each $z_{j} \in a_{i_z}$}{
  $\mathcal{BS}_{z_j}\leftarrow \text{Sort slots in ascending order based on individual influence}$\;
 $c \longleftarrow 1;\mathcal{S}_{i}^{'} \longleftarrow \emptyset$; $\mathcal{S}_{i} \longleftarrow \emptyset$; $I(\mathcal{S}_{i}^{'}) \longleftarrow 0 $; $I(\mathcal{S}_{i}) \longleftarrow 0 $\;
 \While{$I(\mathcal{S}_{i}^{'}) < \sigma_{i}$ and $\mathcal{BS}_{z_j} \neq \emptyset$}{
 $\mathcal{S}_{i}^{'} \longleftarrow \mathcal{S}_{i}^{'} \cup \mathcal{BS}_{z_j}[c]$\;
 $I(\mathcal{S}_{i}^{'}) \longleftarrow I(\mathcal{S}_{i}^{'}) + I(\mathcal{BS}_{z_j}[c])$\;
 $\mathcal{BS}_{z_j} \longleftarrow \mathcal{BS}_{z_j} \setminus (\mathcal{BS}_{z_j}[c])$\;
 $c \longleftarrow c+1$
 }
 \While{$\sigma_{i} > I(S_{i}) ~and~ \mathcal{BS}_{z_j} \neq \emptyset$}{
 $\mathcal{P} \leftarrow \text{Sample }\frac{|\mathcal{BS}_{z_j}|}{|\mathcal{S}_{i}^{'}|} \log \frac{1}{\epsilon}\text{ many elements from }\mathcal{BS}_{z_j} \setminus \mathcal{S}_{i}$\;
 $s^{*} \leftarrow \underset{s \in \mathcal{P}}{argmax} \ \frac{\mathcal{R}(\mathcal{S}_{i})- \mathcal{R}(\mathcal{S}_{i} \cup \{s\})}{\sigma(\{s\})}$\;
 $\mathcal{S}_{i} \leftarrow \mathcal{S}_{i} \cup \{s^{*}\}$\;
 $\mathcal{BS}_{z_j} \leftarrow \mathcal{BS}_{z_j} \setminus \{s^{*}\}$\;
 }}}
$|\mathcal{A}_{u}| \leftarrow 0 $\;
\For {each $a_{i} \in \mathcal{A}$} {
$N_{z} \leftarrow 0; count \leftarrow 0$\;
 \For {each $z_{j} \in a_{i_z}$}{
 $N_{z} \leftarrow N_{z} + 1$\;
 \If{$\sigma_{i} < I(S_{i})$}{
 $count \leftarrow count +1$\;
 }}
 \If{$count \neq N_{z}$}{
 $| \mathcal{A}_{u}| \leftarrow |\mathcal{A}_{u}| + 1$
 }}
 \If{$| \mathcal{A}_{u}|~ \geq 2$}{
 $\mathcal{A}_{u} \leftarrow \text{Sort each advertiser $a_{i} \in \mathcal{A}_{u}$ based on ascending order of } \frac{u_i}{\sigma_{i}} $\;
$\mathcal{BS}_{z} \leftarrow S_{i} \in \mathcal{A}_{u}[1]$\;
$\mathcal{A} \longleftarrow \mathcal{A} \setminus \mathcal{A}_{u}[1]$\;
}
\Else
{
return $\mathcal{Y}$;
}}
 \caption{Randomized Synchronous Greedy Approach for Regret Minimization Problem}
 \label{Algo:3}
\end{algorithm}

\textbf{Complexity Analysis.} Now, we analyze the time and space requirements of Algorithm \ref{Algo:3}. In-Line No. $1$ initializing $\mathcal{Y}$ will take $\mathcal{O}(n)$ time and Line No. $2$, to sort the advertisers will take $\mathcal{O}(n\log n)$ time. Now, in Line No. $4$ \texttt{For Loop} executes for $\mathcal{O}(n)$ time, and in Line No. $5$ \texttt{For Loop} executes for $\mathcal{O}(k)$ time as there are $n$ number of advertisers and $k$ number of influential zones respectively. To sort the advertisers in Line No. $6$ based on their individual influence value will take $\mathcal{O}(n.k.\ell \log \ell)$ time, and Line No. $7$ will take $\mathcal{O}(n.k)$ time for initialization. Now, in Line No. $8$ \texttt{While Loop} will execute for $\mathcal{O}(n.k.\ell)$ time as in the worst case, a zone has at most $\ell$ number of billboard slots. Line No. $9,11,12$ will take $\mathcal{O}(n.k)$ time execute, but Line No. $10$ will take $\mathcal{O}(n.k.\ell.t)$ as each time needs to compute the influence using Equation \ref{Eq:Influence}. So, Line No. $8$ to $13$ will take $\mathcal{O}(n.k.\ell + n.k. \ell. t)$ time in total. In the worst case, \texttt{While Loop} in Line No. $14$ will execute for $\mathcal{O}(n.k.\ell)$ time. In-Line No. $15$ to sample out $\frac{|\mathcal{BS}_{z_j}|}{|\mathcal{S}_{i}^{'}|} \log \frac{1}{\epsilon}$ many elements from $\mathcal{BS}_{z_j}$ will take $\mathcal{O}(n.k.\ell.{\frac{|\mathcal{BS}_{z_j}|}{\ell} \log \frac{1}{\epsilon}})$ time when $|\mathcal{S}_{i}^{'}| = \ell$ in the worst case. Calculating regret in Line No. $16$ will take $\mathcal{O}(2.n.k.\ell.t)$ time and Line No. $17,18$ will take $\mathcal{O}(n.k.\ell)$ computational time. So, from Line No. $14$ to $19$ will take $\mathcal{O}(n.k.\ell.{\frac{|\mathcal{BS}_{z_j}|}{\ell} \log \frac{1}{\epsilon}}+n.k.\ell.t + n.k.\ell)$ time in total. So. from Line No. $4$ to $21$ will take $\mathcal{O}(n+ n.k.\ell \log \ell + n.k + n.k.\ell + n.k. \ell. t + n.k.\ell.{\frac{|\mathcal{BS}_{z_j}|}{\ell} \log \frac{1}{\epsilon}}+n.k.\ell.t + n.k)$ i.e., $\mathcal{O}(n.k.\ell \log \ell + n.k. \ell. t + n.k.\ell.{\frac{|\mathcal{BS}_{z_j}|}{\ell} \log \frac{1}{\epsilon}})$ time. Now, in Line No. $22$ to $34$, calculating the number of unsatisfied  advertisers will take $\mathcal{O}(n.k)$ time to execute as in the worst case, all the $n$ advertisers can be unsatisfied and in Line No. $36$ to sort all the advertisers will take $\mathcal{O}(n \log n)$ in the worst case. Next, in Line No. $37$ copying slots from released advertiser to billboard slots zone $\mathcal{BS}_{z}$ will take $\mathcal{O}(\ell)$ in the worst case and Line No. $38$ will take $\mathcal{O}(1)$ time. Hence, Line No. $35$ to $39$ will take $\mathcal{O}(n \log n + \ell)$ time. Hence, Algorithm \ref{Algo:3} will take $\mathcal{O}(n.k.\ell \log \ell + n.k. \ell. t + n.k.\ell.{\frac{|\mathcal{BS}_{z_j}|}{\ell} \log \frac{1}{\epsilon} + n \log n})$ time in total for allocating slots in one iteration of the \texttt{While Loop}.

Now, the additional space requirement to store the lists $\mathcal{Y}$, $\mathcal{A}$, $\mathcal{BS}_{z_j}$, $\mathcal{S}_{i}^{'}$, and $\mathcal{S}_{i}$ will be of $\mathcal{O}(n)$, $\mathcal{O}(n)$, $\mathcal{O}(\ell)$, $\mathcal{O}(\ell)$, and $\mathcal{O}(\ell)$, respectively. To store zonal information of billboard slots will take $\mathcal{O}(k)$ space. Hence, the total extra space requirement for Algorithm \ref{Algo:3} will be of $\mathcal{O}(k+2n+3\ell)$, i.e., $\mathcal{O}(k+n+\ell)$.

\subsection{Randomized Advertiser Exchange Approach (RAE)}
In the `RSG' approach, after allocating billboard slots, if more than two advertisers are still unsatisfied, we release advertisers one by one according to their budget effectiveness and allocate released billboard slots to the remaining unsatisfied ones. Now, after the allocation of billboard slots to the advertisers in the `RSG' approach, the excessive regret is huge compared to the unsatisfied regret. Now, another allocation of billboard slots to the advertiser may exist, as described in Definition \ref{Def:FA}, which may be better than the current allocation as we have used randomization during allocation, as shown in Algorithm \ref{Algo:3}. We introduce the Randomized Advertiser Exchange approach to address this issue after billboard slot allocation in the `RSG' approach. Here, our main objective is to achieve an allocation such that total regret is minimized. First, instead of an empty set of billboard slot set, the final non-empty advertisers set returned by Algorithm \ref{Algo:3} is considered the input of the ‘RAE’ approach. In Line No. $3$, make a copy of $\mathcal{Y}^{final}$ as $\mathcal{Y}^{temp}$, and Line No. $4$ \texttt{For Loop} will iterates for all the advertisers, $a_{i} \in \mathcal{A}$ whereas \texttt{For Loop} at Line No. $5$ iterates for all the advertisers $a_{j} \in \mathcal{A} \setminus \{a_{i}\}$. Now, in Line No. $6$ if there exist  $S_{i} \in a_{i}, S_{j} \in a_{j}$ such that exchange $[S_{i}, S_{j}]$ will  minimize the total regret of $\mathcal{Y}^{temp}$ then exchanging of slots between advertiser $a_{i}$ and $a_{j}$ is performed in Line No. $7$ to $9$. Therefore, after completion of \texttt{For Loop} in Line   No. $4$ to $12$ a new allocation of billboard slots to the advertisers is achieved. If total regret in new allocation $\mathcal{Y}^{temp}$ is less than   the total regret of previous allocation $\mathcal{Y}^{final}$ then $\mathcal{Y}^{temp}$ becomes $\mathcal{Y}^{final}$ and again iterates over Line No. $2$ to $13$ and this will continue till \texttt{While Loop} gets true but when the condition at Line No. $14$ gets false then Line No. $17$ to $19$ executes and \texttt{While Loop} breaks and return $\mathcal{Y}^{final}$.

\begin{algorithm}[h!]
\scriptsize
\SetAlgoLined
\KwData{Trajectory Database $\mathcal{D}$, Billboard Slot Information $\mathbb{BS}$, Advertiser Database $\mathbb{A}$, and the Influence Function $I()$.}
\KwResult{An allocation $\mathcal{Y}$ of the billboard slots that minimizes the total regret}
$\mathcal{Y}^{final} \leftarrow \text{RandomizedSynchronousGreedy}(\mathcal{D}, \mathbb{BS},\mathbb{A}, \mathcal{Y})$\;
\While {True}{
$\mathcal{Y}^{temp} \leftarrow \mathcal{Y}^{final}$

 \For {each $a_{i} \in \mathcal{A}$} {
 \For {each $a_{j} \in \mathcal{A} \setminus \{a_{i}\}$} {
 \If{$ \exists ~ S_{i} \in a_{i},~ S_{j} \in a_{j}~ \text{such that exchange}~ [S_{i}, S_{j}]~ will ~ minimize ~\mathcal{R}(\mathcal{Y}^{temp})$}{
 $a_{i} \leftarrow a_{i} \cup S_{j}$\;
 $a_{j} \leftarrow a_{j} \cup S_{i}$\;
 $a_{i} \leftarrow a_{i} \setminus S_{i}$\;
 $a_{j} \leftarrow a_{j} \setminus S_{j}$\;
 }}}
\If{ $\mathcal{R}(\mathcal{Y}^{temp}) < \mathcal{R}(\mathcal{Y}^{final}) $}{
$\mathcal{Y}^{final} \leftarrow \mathcal{Y}^{temp}$
}
\Else
{
return $ \mathcal{Y}^{final}$;
}}
\caption{Randomized Advertiser Exchange Approach for Regret Minimization Problem}
 \label{Algo:4}
\end{algorithm}

\textbf{Complexity Analysis.} Now, we analyze the time and space requirement of Algorithm \ref{Algo:4}. First, in Line No. $1$ `RSG' approach will take $\mathcal{O}(n.k.\ell \log \ell + n.k. \ell. t + n.k.\ell.{\frac{|\mathcal{BS}_{z_j}|}{\ell} \log \frac{1}{\epsilon} + n \log n})$ time in total as shown in Algorithm \ref{Algo:3}. Next, in Line No $3$ copying $\mathcal{Y}_{temp}$ to $\mathcal{Y}_{final}$ will take $\mathcal{O}(\ell)$ time in the worst case. In-Line No. $4$ \texttt{For Loop} will execute for $\mathcal{O}(n)$ time and \texttt{For Loop} at Line No. $5$ will execute for $n-1$ times as total $n$ number of advertisers are there. At Line No. $6$ in the worst case, exchanging slots between $a_{i}$ with $a_{j}$ will take $\mathcal{O}(\ell)$ time and to calculate regret of $\mathcal{Y}^{temp}$ we need to compute the influence of $\ell$ number of slots that will take $\mathcal{O}(n^{2}.\ell.t)$ time where $t$ is the number of tuple in the trajectory database. In Line No. $7$ to $10$ will take $\mathcal{O}(n^{2})$ time to execute. Hence, from Line No. $4$ to $13$ will take $\mathcal{O}(n^{2}.\ell.t + n^{2}.\ell +n^{2})$ time. In Line No. $14$ regret will be calculated twice and it will take $\mathcal{O}(2.n^{2}.\ell.t)$ time. Finally, in Line No. $15$ copying from $\mathcal{Y}^{temp}$ to $\mathcal{Y}^{final}$ will take $\mathcal{O}(n)$ time. Therefore, Algorithm \ref{Algo:4} will take total $\mathcal{O}(n^{2}.\ell.t + n^{2}.\ell +n^{2} + 2.n^{2}.\ell.t + n^{2})$ i.e., $\mathcal{O}(n^{2}.\ell.t)$ time to execute one iteration of the \texttt{While Loop} in Line no $2$.
Now, the additional space requirement for Algorithm \ref{Algo:4} is for $\mathcal{Y}^{final}$ and $\mathcal{Y}^{temp}$ will be $\mathcal{O}(n)$ and $\mathcal{O}(n)$ respectively. Hence, total space requirement of Algorithm \ref{Algo:4} will be $\mathcal{O}(n + n)$ i.e., $\mathcal{O}(n)$.

\subsection{An Illustrative Example}
Consider, an influence provider owns thirteen billboard slots, $\mathcal{BS}=\{ bs_1, bs_2,\ldots, bs_{13}\}$ with corresponding individual influence as reported in Table No. \ref{ETable:1}. These billboard slots are divided into three demographic zones i.e., $\mathcal{Z}_{1}, \mathcal{Z}_{2}, \mathcal{Z}_{3}$. Let, zone $\mathcal{Z}_{1}$, $\mathcal{Z}_{2}$ and $\mathcal{Z}_{3}$ contains billboard slots $ \{ bs_{1}, bs_{6}, bs_{11}, bs_{10}\}$, $ \{ bs_{2}, bs_{5}, bs_{8}, bs_{7}, bs_{13}\}$ and $ \{ bs_{3}, bs_{4}, bs_{9}, bs_{12}\}$ respectively. Now, five advertisers, $\mathcal{A}= \{a_1, a_2, \ldots a_5\}$ approach to the influence provider with their required zone-specific influence demand with budget constraints as shown in Table No. \ref{ETable:2}. To address this at first, the influence provider sorts the advertisers based on budget effectiveness in descending order i.e., $\{ a_{1}, a_{4}, a_{2}, a_{3}, a_{5}\}$ and allocate billboard slots to each advertiser one by one using `RG' approach to satisfy their zone specific influence requirements. After allocation, we found that advertisers $a_{3}$, and  $a_{5}$ are still unsatisfied due to billboard slots running out while the remaining advertisers are satisfied as presented in Table No. \ref{ETable:3}. After this allotment, the `RSG' approach applied to release unsatisfied advertisers one by one and allot released advertisers allocated slots to remaining unsatisfied advertisers. So, less budget effective advertiser $\{a_{5}\}$ removed and released slot, $\{bs_{13}\}$ is belongs to zone $\mathcal{Z}_{2}$ of $\{a_{5}\}$. But, advertiser  $\{a_{3}\}$ have influence requirements from $\mathcal{Z}_{3}$. So, advertisers $a_{3}$ still remain unsatisfied, and unsatisfied regret(UR) of advertisers $a_{3}$ increases as shown in Table No. \ref{ETable:4}. As our allocation scheme is based on randomization there may exist another better allocation on top of the current allocation. Next, the `RAE' approach is applied to the current allocation. We observe that exchange of billboard slots between advertiser $\{a_{3}\}$ and $\{a_{4}\}$ make the advertisers $\{a_{3}\}$ satisfied and reduce unsatisfied regret(UR) of advertiser $\{a_{3}\}$ and excessive regret (ER) of advertiser $\{a_{4}\}$. Finally, all the remaining advertisers i.e., $ \{a_1, a_2, \ldots a_4\}$ are satisfied and overall regret is minimized as shown in Table No. \ref{ETable:5}.

\begin{table}[ht]
\begin{subtable}[c]{0.2\linewidth}
\centering 
    \begin{tabular}{| c | c | c | c | c | c | c | c | c | c | c | c | c | c |}
 
    \hline
    $\mathcal{BS}_{i}$ & $bs_{1}$ & $bs_{2}$ & $bs_{3}$ & $bs_{4}$ & $bs_{5}$ & $bs_{6}$ & $bs_{7}$ & $bs_{8}$ & $bs_{9}$ & $bs_{10}$ & $bs_{11}$ & $bs_{12}$ & $bs_{13}$\\ \hline
    $\sigma(bs_{i})$ & 4 & 6 & 5 & 3 & 3 & 2 & 3 & 2 & 3 & 3 &2 &5 &3\\ \hline
    \end{tabular}
    \subcaption{\label{ETable:1} Billboard Slots information.}
\end{subtable}

\begin{subtable}[c]{0.5\textwidth}
\centering 
\begin{tabular}{ | c | p{1.8cm}| p{1.8cm} | p{1.8cm} | p{1.8cm}| p{1.8cm} | p{2cm} |}
\hline
    $\mathcal{A}$ & $a_{1}$ & $a_{2}$ & $a_{3}$ & $a_{4}$ & $a_{5}$\\ \hline
    $\sigma_{i}$ & $\sigma_{1}(\mathcal{Z}_{1})$ = 3;
    $\sigma_{1}(\mathcal{Z}_{2})$ = 2; 
    $\sigma_{1}(\mathcal{Z}_{3})$ = 2 
    & $\sigma_{2}(\mathcal{Z}_{1})$ = 3; 
    $\sigma_{2}(\mathcal{Z}_{2})$ = 3; 
    $\sigma_{2}(\mathcal{Z}_{3})$ = 3 
    & $\sigma_{3}(\mathcal{Z}_{1})$ = 1; 
    $\sigma_{3}(\mathcal{Z}_{2})$ = 5; 
    $\sigma_{3}(\mathcal{Z}_{3})$ = 4 
    & $\sigma_{4}(\mathcal{Z}_{1})$ = 1; 
    $\sigma_{4}(\mathcal{Z}_{2})$ = 1; 
    $\sigma_{4}(\mathcal{Z}_{3})$ = 2
    & $\sigma_{5}(\mathcal{Z}_{1})$ = 3; 
    $\sigma_{5}(\mathcal{Z}_{2})$ = 2; 
    $\sigma_{5}(\mathcal{Z}_{3})$ = 3  \\ \hline
    $u_{i}$ & \$15 & \$16 & \$15 & \$8 & \$7 \\ \hline
    \end{tabular}
    \subcaption{\label{ETable:2} Advertisers information.}
\end{subtable}

\begin{subtable}[c]{0.5\textwidth}
\centering 
    \begin{tabular}{ | c | p{1.8cm} | p{1.8cm} | p{1.8cm} | p{1.8cm} | p{1.8cm} |}
    \hline
    $\mathcal{A}$ & $a_{1}$ & $a_{2}$ & $a_{3}$ & $a_{4}$ & $a_{5}$ \\ \hline
    $\mathcal{BS}_{i}$ 
    & $\mathcal{Z}_{1} = \{ bs_{10} \}$; $\mathcal{Z}_{2} = \{ bs_{5} \}$; $\mathcal{Z}_{3} = \{ bs_{9} \}$ 
    & $\mathcal{Z}_{1} = \{ bs_{1} \}$; $\mathcal{Z}_{2} = \{ bs_{7} \}$; $\mathcal{Z}_{3} = \{ bs_{3} \}$  
    & $\mathcal{Z}_{1} = \{ bs_{11} \}$; $\mathcal{Z}_{2} = \{ bs_{2} \}$; $\mathcal{Z}_{3} = \{ bs_{4} \}$ 
    & $\mathcal{Z}_{1} = \{ bs_{6} \}$; $\mathcal{Z}_{2} = \{ bs_{8} \}$; $\mathcal{Z}_{3} = \{ bs_{12} \}$ 
    & $\mathcal{Z}_{1} = \{  \}$; $\mathcal{Z}_{2} = \{ bs_{13} \}$; $\mathcal{Z}_{3} = \{ \}$  \\ \hline
    Satisfied & Yes & Yes & No & Yes & No \\ \hline
    \end{tabular}
    \subcaption{\label{ETable:3}Initial allotment using the `RG' approach.}
\end{subtable}

\begin{subtable}[c]{0.5\textwidth}
\centering 
    \begin{tabular}{ | c | p{2.2cm} | p{2.2cm} | p{2.2cm} | p{2.2cm} |}
    \hline
    $\mathcal{A}$ & $a_{1}$ & $a_{2}$ & $a_{3}$ & $a_{4}$ \\ \hline
    $\mathcal{BS}_{i}$ 
    & $\mathcal{Z}_{1} = \{ bs_{10} \}$; $\mathcal{Z}_{2} = \{ bs_{5} \}$; $\mathcal{Z}_{3} = \{ bs_{9} \}$
    & $\mathcal{Z}_{1} = \{ bs_{1} \}$; $\mathcal{Z}_{2} = \{ bs_{7} \}$; $\mathcal{Z}_{3} = \{ bs_{3} \}$  
    & $\mathcal{Z}_{1} = \{ bs_{11} \}$; $\mathcal{Z}_{2} = \{ bs_{2} \}$; $\mathcal{Z}_{3} = \{ bs_{4} \}$ 
    & $\mathcal{Z}_{1} = \{ bs_{6} \}$; $\mathcal{Z}_{2} = \{ bs_{8} \}$; $\mathcal{Z}_{3} = \{ bs_{12} \}$  \\ \hline
Satisfied & Yes & Yes & No & Yes \\ \hline
Regret & ER: Yes, UR: No & ER: Yes, UR: No & ER: Yes, UR:Yes & ER:Yes, UR: No \\ \hline
    \end{tabular}
    \subcaption{\label{ETable:4}Allotment after the `RSG' approach.}
\end{subtable}

\begin{subtable}[c]{0.5\textwidth}
\centering 
    \begin{tabular}{ | c | p{2.2cm} | p{2.2cm} | p{2.2cm} | p{2.2cm} |}
    \hline
    $\mathcal{A}$ & $a_{1}$ & $a_{2}$ & $a_{3}$ & $a_{4}$ \\ \hline
    $\mathcal{BS}_{i}$ 
    & $\mathcal{Z}_{1} = \{ bs_{10} \}$; $\mathcal{Z}_{2} = \{ bs_{5} \}$; $\mathcal{Z}_{3} = \{ bs_{9} \}$
    & $\mathcal{Z}_{1} = \{ bs_{1} \}$; $\mathcal{Z}_{2} = \{ bs_{7} \}$; $\mathcal{Z}_{3} = \{ bs_{3} \}$ 
    & $\mathcal{Z}_{1} = \{ bs_{6} \}$; $\mathcal{Z}_{2} = \{ bs_{8} \}$; $\mathcal{Z}_{3} = \{ bs_{12} \}$ 
    & $\mathcal{Z}_{1} = \{ bs_{11} \}$; $\mathcal{Z}_{2} = \{ bs_{2} \}$; $\mathcal{Z}_{3} = \{ bs_{4} \}$   \\ \hline
Satisfied & Yes & Yes & Yes & Yes \\ \hline
Regret & ER: Yes, UR: No & ER: Yes, UR: No & ER: Yes, UR: No & ER: Yes, UR: No \\ \hline
    \end{tabular}
    \subcaption{\label{ETable:5} Final allotment after the `RAE' approach.}
\end{subtable}
\end{table}

\section{Experimental Setup} \label{Sec:EE}
In this section, we describe the experimental setup to determine the effectiveness and efficiency of the proposed solution methodologies. Initially, we start by describing the datasets.

\subsection{Dataset Descriptions}
We conduct experiments on two real-world trajectory datasets, check-ins for New York City (NYC) and Los Angeles (LA). A total of $227,428$ check-ins were recorded in NYC\footnote{\url{https://www.nyc.gov/site/tlc/about/tlc-trip-record-data.page}} between April 12, 2012, and February 16, 2013, while $74,170$ entries of user data from 15 streets in LA\footnote{\url{https://github.com/Ibtihal-Alablani}} included street names, GPS locations, timestamps, and other information. We have crawled billboard datasets from LAMAR\footnote{\url{http://www..lamar.com/InventoryBrowser}}, one of the largest billboard providers worldwide. All the essential characteristics of datasets are summarized in Table \ref{Table:Dataset_Description}. In trajectory information, $\mathcal{|T|}$, $\mathcal{|U|}$, and $Avg_{dist}$ represent trajectory size, number of unique users, and average distance each user covers. On the other hand, in billboard information $\mathcal{B}$, $\mathcal{BS'}$, $\mathcal{BS''}$, and $Avg_{dist}$ denote the number of billboards, the number of billboard slots, the number of non-zero influential billboard slots, and the average distance between each billboard place in the respective city. Additionally, we divide the billboard datasets for New York City into five geographic regions based on latitude and longitude: The Bronx, Brooklyn, Manhattan, Queens, and Staten Island, and Los Angeles billboard dataset is divided into $3$ different zones based on $15$ different streets.

\begin{table}[h]
\centering
\resizebox{\textwidth}{!}{%
\begin{tabular}{|c|ccc|cccc|}
\hline
\multirow{2}{*}{$\text{Dataset}$} & \multicolumn{3}{c|}{$\text{Trajectory Information}$} & \multicolumn{4}{c|}{$\text{Billboard Information}$} \\ \hline 
 & \multicolumn{1}{c|}{$|\mathcal{T}|$} & \multicolumn{1}{c|}{$|\mathcal{U}|$} & $Avg_{dist}$ & \multicolumn{1}{c|}{$|\mathcal{B}|$} & \multicolumn{1}{c|}{$|\mathcal{BS'}|$} & \multicolumn{1}{c|}{$|\mathcal{BS''}|$} & $Avg_{dist}$ \\ \hline
NYC & \multicolumn{1}{c|}{$227428$} & \multicolumn{1}{c|}{$1083$} & $3.12~ km$ & \multicolumn{1}{c|}{$716$} & \multicolumn{1}{c|}{$1031040$} & \multicolumn{1}{c|}{$11048$} & $15.07 ~km$ \\ \hline
LA & \multicolumn{1}{c|}{$74170$} & \multicolumn{1}{c|}{$2000$} & $0.61~km$ & \multicolumn{1}{c|}{1483} & \multicolumn{1}{c|}{$2135520$} & \multicolumn{1}{c|}{$4712$} & $10.26~ km$ \\ \hline
\end{tabular}%
}
\caption{Dataset Description}
\label{Table:Dataset_Description}
\end{table}

\subsection{Parameter Setting}
 In this study, the following parameter values need to be fixed, and we describe them briefly. All key parameters and their corresponding values are summarized in Table \ref{Table-2}.

\begin{table}[h]
\centering
    \begin{tabular}{ | p{2cm}| p{5.5cm}|}
    \hline
    Parameter & Values  \\ \hline
    $\delta$ & $40\%, 60\%, 80\%, \textbf{100\%}, 120\%$   \\ \hline
    $\lambda$ & $1\%, 2\%, \textbf{5\%}, 10\%, 20\%$  \\ \hline
    $\epsilon$ & $\textbf{0.01}, 0.05, 0.1, 0.15, 0.2$ \\ \hline
    $\gamma$ & $0, 0.25, \textbf{0.5}, 0.75, 1$  \\ \hline
    $\eta$ & $25m,50m,\textbf{100m},125m,150m$  \\ \hline
    \end{tabular}
    \caption{\label{Table-2} Key Parameters}
\end{table}

\paragraph{Demand-Supply Ratio ($\delta$)} It denotes the ratio of the global influence demand over the influence provider's supply, i.e., $\delta = \sigma^{\mathcal{A}} / \sigma^{h}$, where $\sigma^{\mathcal{A}}$ = $\sum_{a=1}^{\mathcal{|A|}} \sigma^{a} $ refers the global demand, and $\sigma^{h}$ refers the total influence supply, i.e., $\sigma^{h} = \sum_{b \in \mathcal{BS}} \sigma(b)$. 

\paragraph{Average-Individual Demand Ratio ($\lambda$)} It refers to the ratio of individual demand of advertisers and the influence provider's supply, i.e., $\lambda = \frac{\sigma^{\mathcal{A}} / |\mathcal{A}|}{\sigma^{h}}$. By adjusting the $\lambda$ value, we can control the individual demand of an advertiser. 
    
\paragraph{Penalty Ratio ($\gamma$)} It decides how much penalty should be imposed on an influence provider due to the advertiser's dissatisfaction. As described in Equation \ref{Def:Reg_Model}, two extreme case arises: (1) when $\gamma =0$, the influence provider does not receive any payment if the advertiser is not satisfied. (2) when $\gamma = 1$, the influence provider will receive the same fraction of payment as the fraction of influence it provides to the advertiser. 

\paragraph{Accuracy Speed-up parameter ($\epsilon$)} It is a user-defined parameter that decides the sample set size from which a slot is chosen to allocate in Algorithm \ref{Algo:2}. The $\epsilon$ value is set to $0.01, 0.05, 0.1, 0.15$, and $0.2$ to simulate different sample set sizes in our experiment.

\paragraph{Advertiser's Demand ($\sigma$)} Each advertisers demand in the Advertiser dataset is generated using $\lambda$, and $\sigma^{h}$. It can be represented as $\sigma^{a} = \lfloor \alpha. \sigma^{a}. \lambda \rfloor$, where $\alpha$ is a parameter randomly chosen between $0.8$ to $1.2$ to generate different demand values of the advertiser. 

\paragraph{Advertiser's Payment ($u$)} Previous studies \cite{10.14778/2752939.2752950,zhang2018trajectory} considered cost is proportional to its corresponding influence. So, we generate the advertiser's payment as $u = \lfloor \beta. \sigma^{a} \rfloor$, where $\beta$ is a factor randomly chosen between $0.9$ to $1.1$.

\paragraph{Environment Setup} All proposed and baseline methods have been implemented in Python using the Jupyter Notebook platform. All experiments are conducted in a Ubuntu-operated desktop system with 64 GB RAM and an Xeon(R) 3.50 GHz processor.

\paragraph{Performance Measurement} We have conducted each experiment three times, and average results are reported. In our experiment, we calculate two types of regret: Unsatisfied and Excessive, which leads to total regret.

\subsection{Methods in the Experimentation}

\subsubsection{Baseline Methodologies} \label{Sec:Baseline}
We have made a comparison of our proposed approaches with existing baseline solution methodologies.

\paragraph{Random Allocation} In this method, first, sort each advertiser in descending order based on budget effectiveness, i.e., total budget over total influence demand. Next, billboard slots are allocated uniformly at random to the advertisers one by one to fulfill their zone-specific influence demand. This allocation scheme will continue till all advertisers are satisfied or billboard slots run out.

\paragraph{Top-$k$ Allocation} At first, the individual influence of all billboard slots is computed, and based on influence value, slots are sorted in descending order. Similarly, advertisers are also sorted based on budget effectiveness in descending order. Next, from this sorted list, billboard slots are allocated to each advertiser one by one till the influence demand of all the advertisers is satisfied or no more slots remain to allocate.

\subsubsection{\textbf{Proposed Methodologies}} Here, we describe our proposed methods briefly.

\paragraph{Budget-Effective Greedy (RG)}
Budget Effective Greedy is the heuristic approach where, in each step, we compute the marginal gain for all billboard slots and pick one slot among them, which reduces the total regret best. This process will continue till all the advertisers are satisfied or billboard slots run out.

\paragraph{Randomized Budget-Effective Greedy (RG)} It is a randomized heuristic greedy approach where in each step, instead of computing marginal gain for all billboard slots, we compute marginal gain from randomly sampling out a billboard slot set $\mathcal{P}$ of size $\frac{|\mathcal{BS}_{z_j}|}{|\mathcal{S}_{i}^{'}|} \log \frac{1}{\epsilon}$, which in turn overlaps with $\mathcal{BS}_{z_j}$ with the probability of $(1 - \epsilon)$, and pick out billboard slot which reduces the regret best. These slots are allocated to the advertisers one by one until all advertisers are satisfied or no billboard slots remain to allocate. 


\paragraph{Randomized Synchronous Greedy (RSG)}
We introduce the Randomized Synchronous Greedy Approach to overcome the drawbacks of both the `BG' and `RG' approaches. The `BG' satisfied many advertisers with minimal regret but took substantial computational time. On the other hand, `RG' compromises with regret but reduces computational time. Both approaches suffer from higher total regret in which, in most cases, an unsatisfied penalty plays an important role. To resolve this issue, `RSG' first allocates billboard slots to the advertisers using the `RG' approach. Next, if more than two advertisers are still unsatisfied, remove them one by one according to less budget effectiveness. In that way, we can reduce unsatisfied regret. However, excessive regret increases, and the overall regret decreases compared to the `RG'.

\paragraph{Randomized Advertiser Exchange Approach (RAE)} The `RAE' approach is a randomized local search approach. After allocating billboard slots to the advertisers using the `RSG' approach, there may always be a chance that a better allocation exists compared to the current allocation. To address this, in the `RAE' approach, we exchange billboard slots among advertisers; if swapping slots will minimize total regret, then exchange occurs. In that way, we finally get a better allocation on top of the `RSG' approach. 

\subsection{Goals of Our Experimentation}
We fix the following research questions:
\begin{itemize}
\item{\textbf{RQ1:}} What will happen if the influence provider's maximum supply, $(\sigma^{h})$ is far below, near, or exceeded by the total global demand, $(\sigma^{\mathcal{A}})$ of all advertisers?

\item{\textbf{RQ2:}} Which kind of advertiser is more beneficial to the influence provider in terms of reducing regret?

\item{\textbf{RQ3:}} Varying $\delta$ with a fixed value of $\lambda$, and $|\mathcal{A}|$, how the regret value change?

\item{\textbf{RQ4:}} Varying $\delta$ with a fixed value of $\lambda$, and $|\mathcal{A}|$, how the computational cost change?

\item{\textbf{RQ5:}} How the penalty ratio, $\gamma$ and the speed-up parameter, $\epsilon$ impact on minimizing regret?
\end{itemize}

\section{Results and Discussions} \label{Sec:RD} 
 In this section, we discuss the experimental results and the impacts of different parameters on the NYC and LA datasets. We formulate total regret by combining two components: unsatisfied regret and excessive regret, as described in definition \ref{Def:Reg_Model}. To address RQ1, we vary the demand-supply ratio $(\delta)$, from $40\%$ to $120\%$. When $\delta$ value is very low, i.e., maximum influence supply $(\sigma^{h})$ is high and global demand $(\sigma^{\mathcal{A}})$ is low, almost all the advertisers are satisfied. However, when the $\delta$ value is higher, then $\sigma^{\mathcal{A}}$ is also higher or even exceeds the total influence supply. So, the number of satisfied advertisers becomes less. Next, to answer RQ2, we vary the average individual demand-supply ratio from $1\%$ to $20\%$, which represents the individual demand of advertisers from low to high. In the following, to address the research questions RQ3, RQ4, and RQ5, we present the experimental results of both the NYC and LA datasets.

\subsection{Effectiveness Study} \label{ES}
We represent our results under four significant cases of the NYC and LA datasets. To describe total regret in the results, we use stacked bars, and two components in each bar denote the percentage of unsatisfied regret and excessive regret, respectively. It is possible that in the results, only one component is in the stacked bar (e.g., when all the advertisers are fully satisfied, then unsatisfied regret is zero). Next, we discuss experimental observations over the NYC dataset.

\subsubsection{Observations over NYC dataset} \label{OBS_NYC}

\begin{figure*}[h]
    \centering
    \begin{tabular}{lclc}
       Unsatisfied Regret & \includegraphics[width=0.11\linewidth]{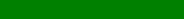} & Excessive Regret & \includegraphics[width=0.11\linewidth]{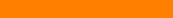} \\
    \end{tabular}
    \begin{tabular}{ccccc}     
        \includegraphics[width=0.17\linewidth]{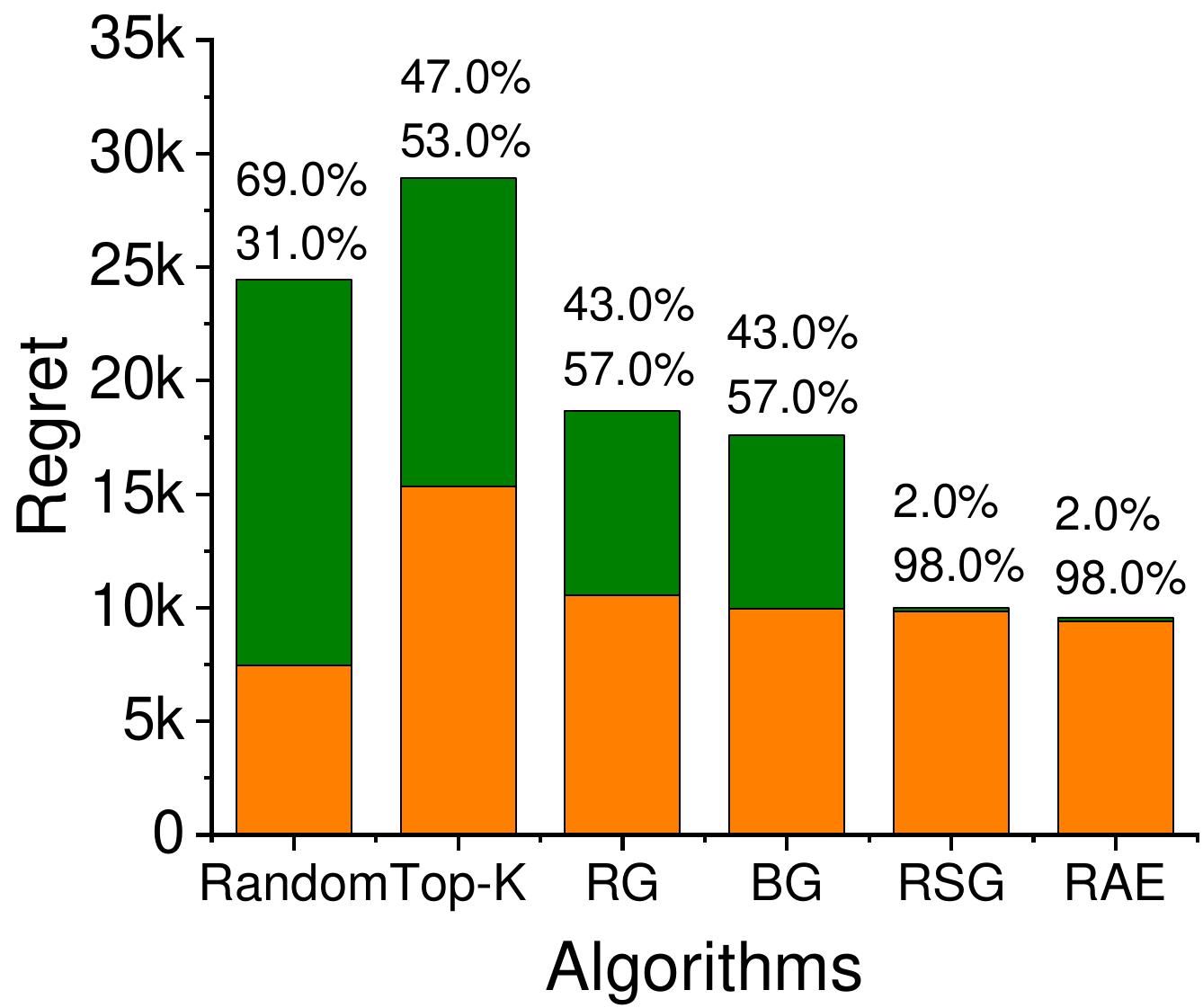} &
        \includegraphics[width=0.17\linewidth]{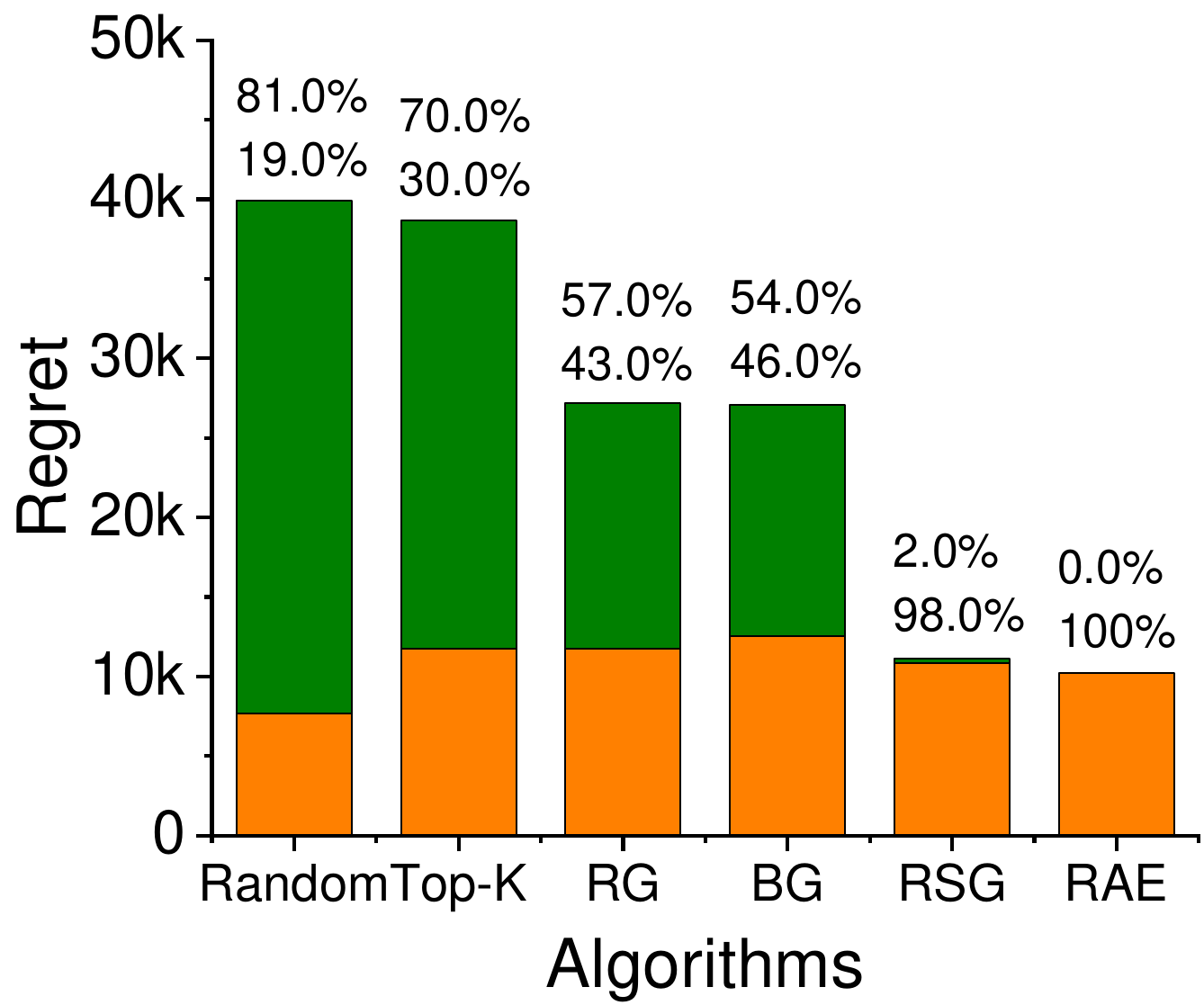} &
        \includegraphics[width=0.17\linewidth]{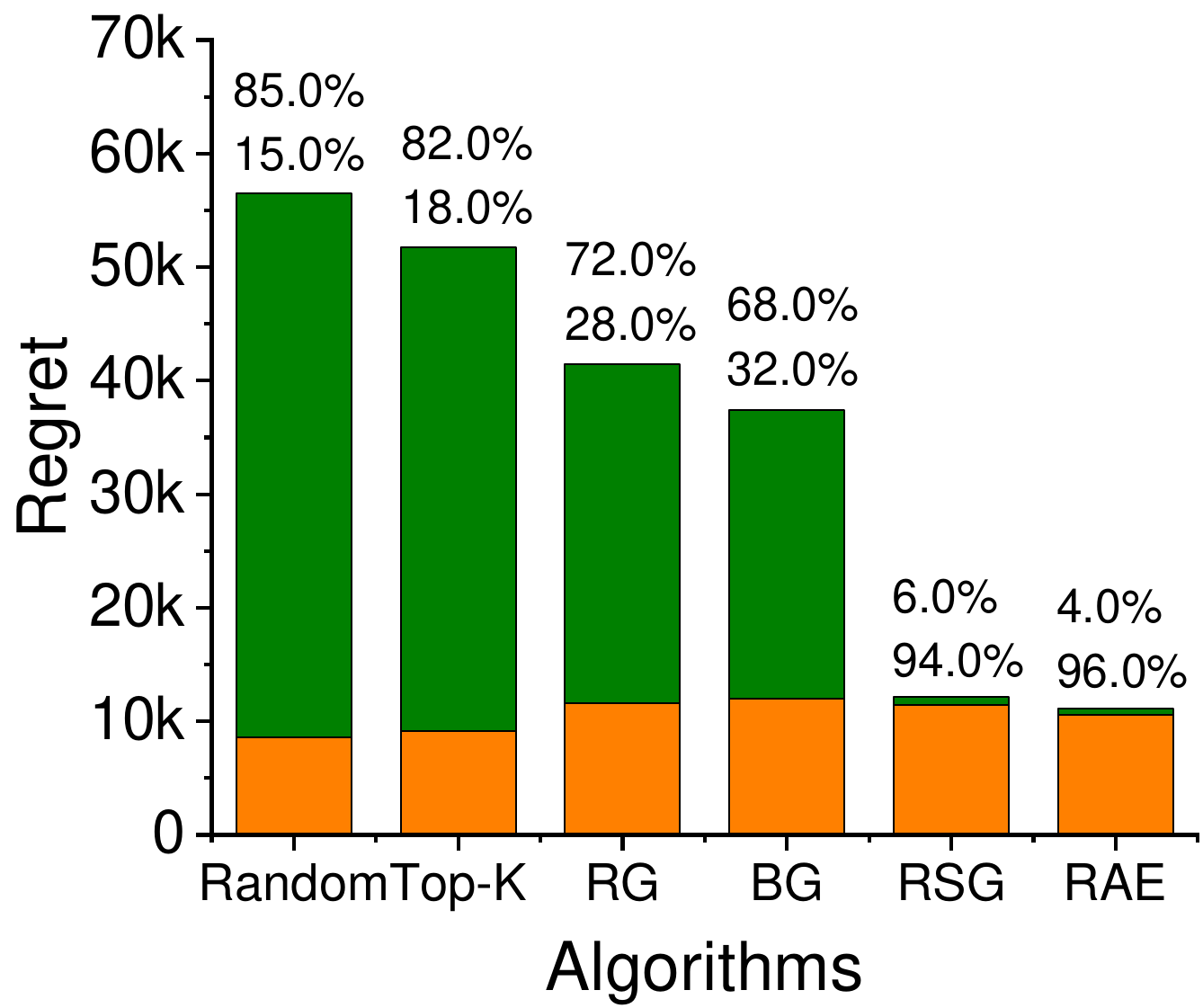} &
        \includegraphics[width=0.17\linewidth]{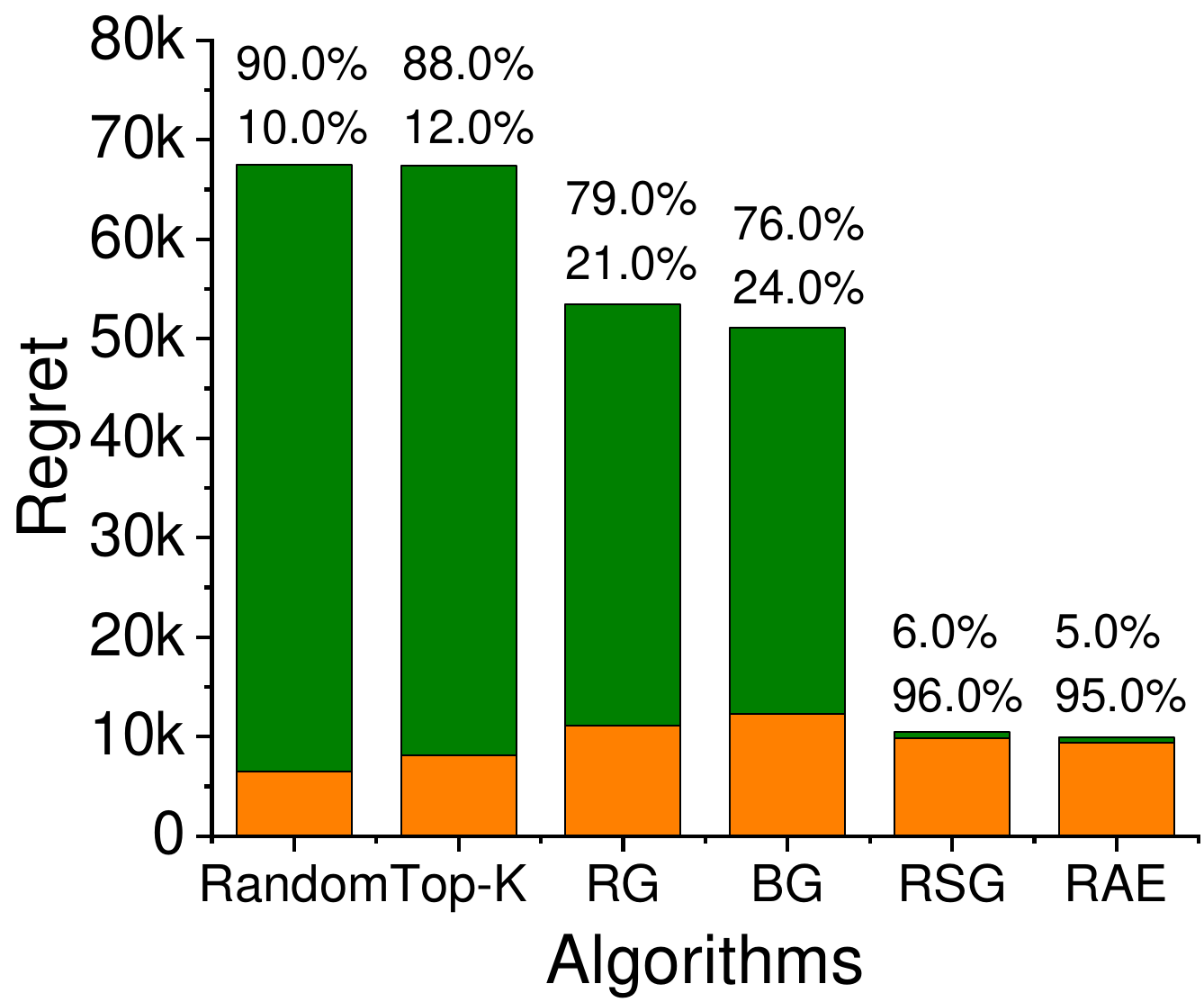} &
        \includegraphics[width=0.17\linewidth]{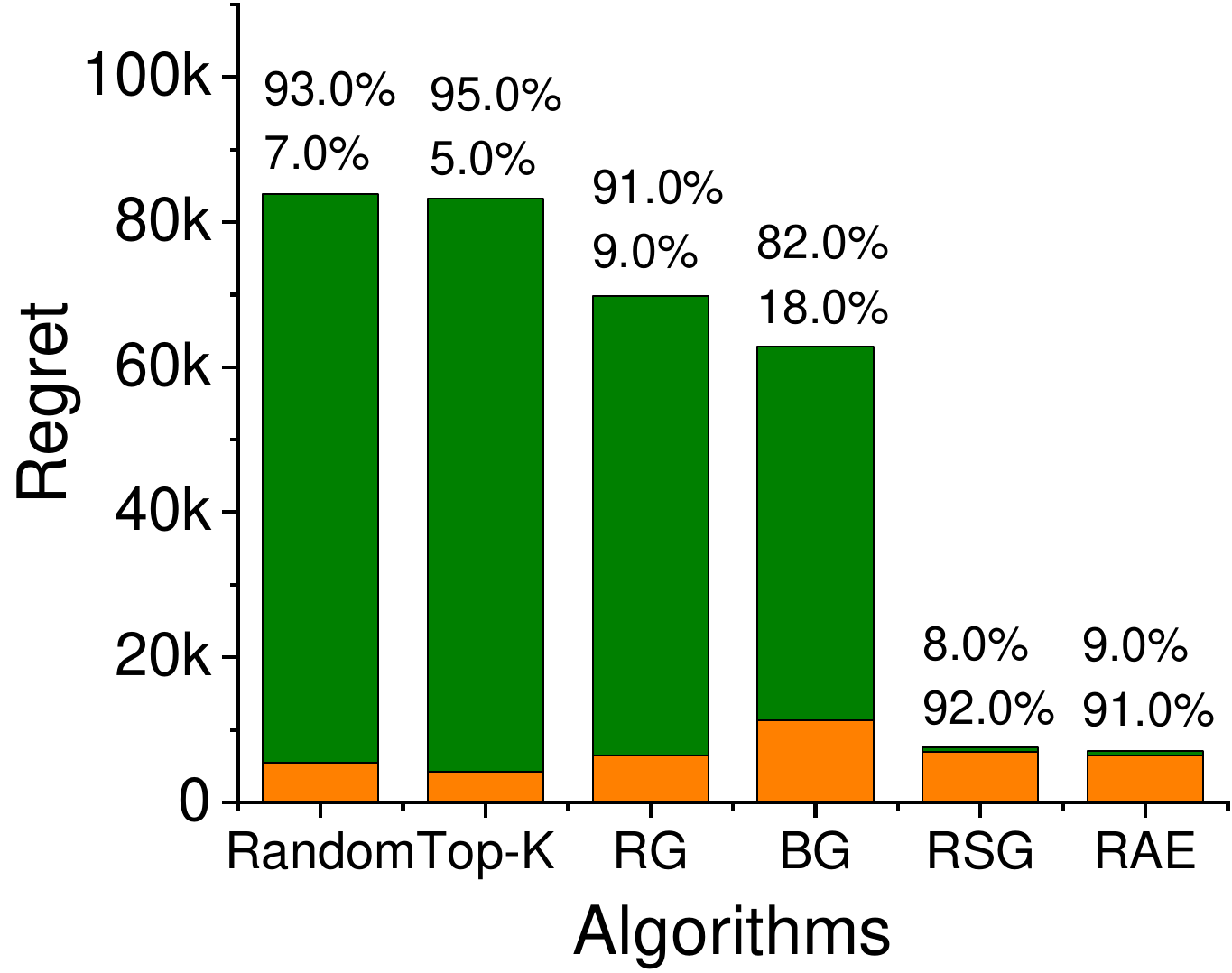} \\
        {\tiny (a) $\delta = 40 \%$} &
        {\tiny (b) $\delta= 60 \%$} &
        {\tiny (c) $\delta = 80 \%$} &
        {\tiny (d) $\delta = 100 \%$} &
        {\tiny (e) $\delta= 120 \%$} \\[5pt]

         \includegraphics[width=0.17\linewidth]{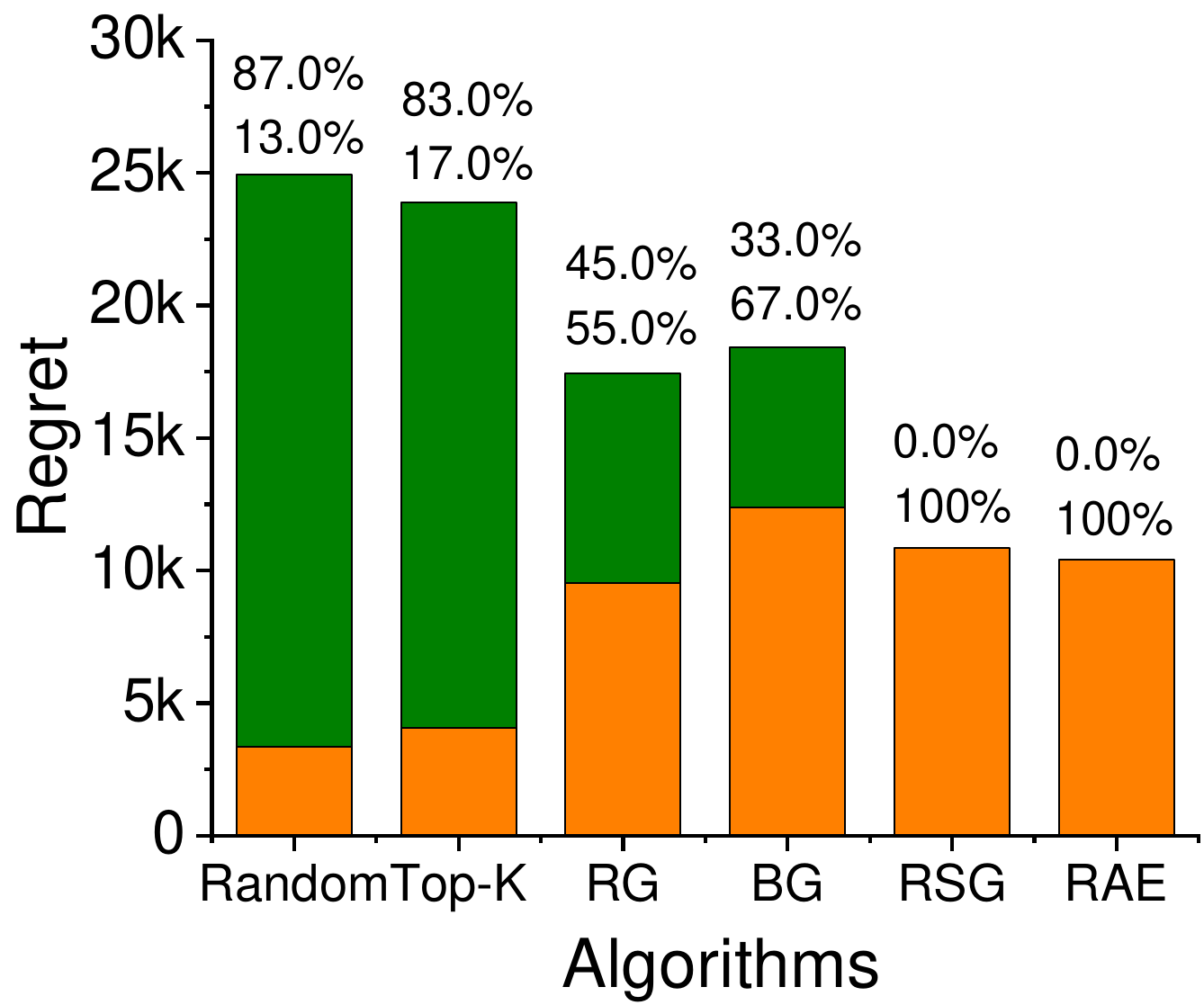} &
        \includegraphics[width=0.17\linewidth]{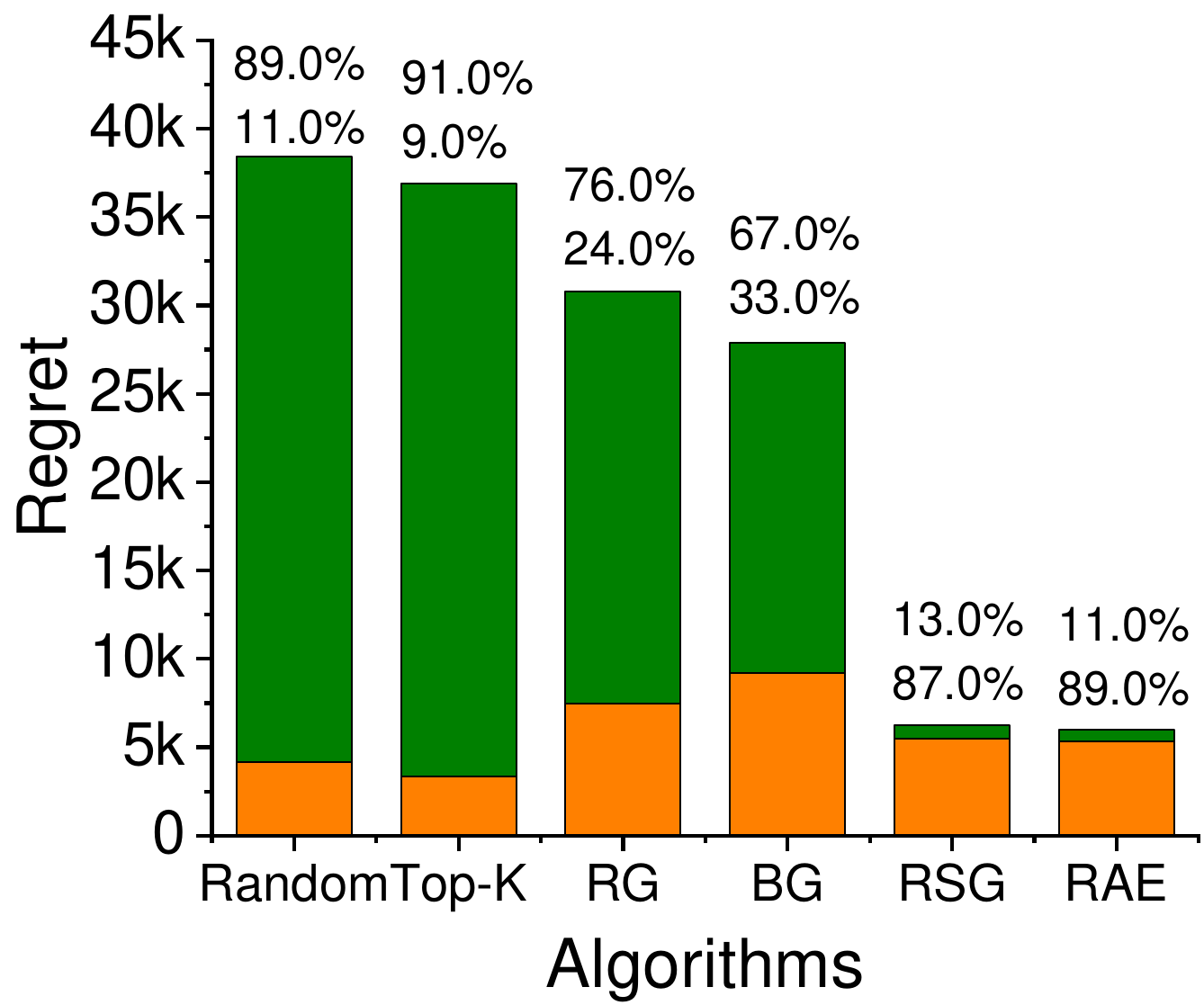} &
        \includegraphics[width=0.17\linewidth]{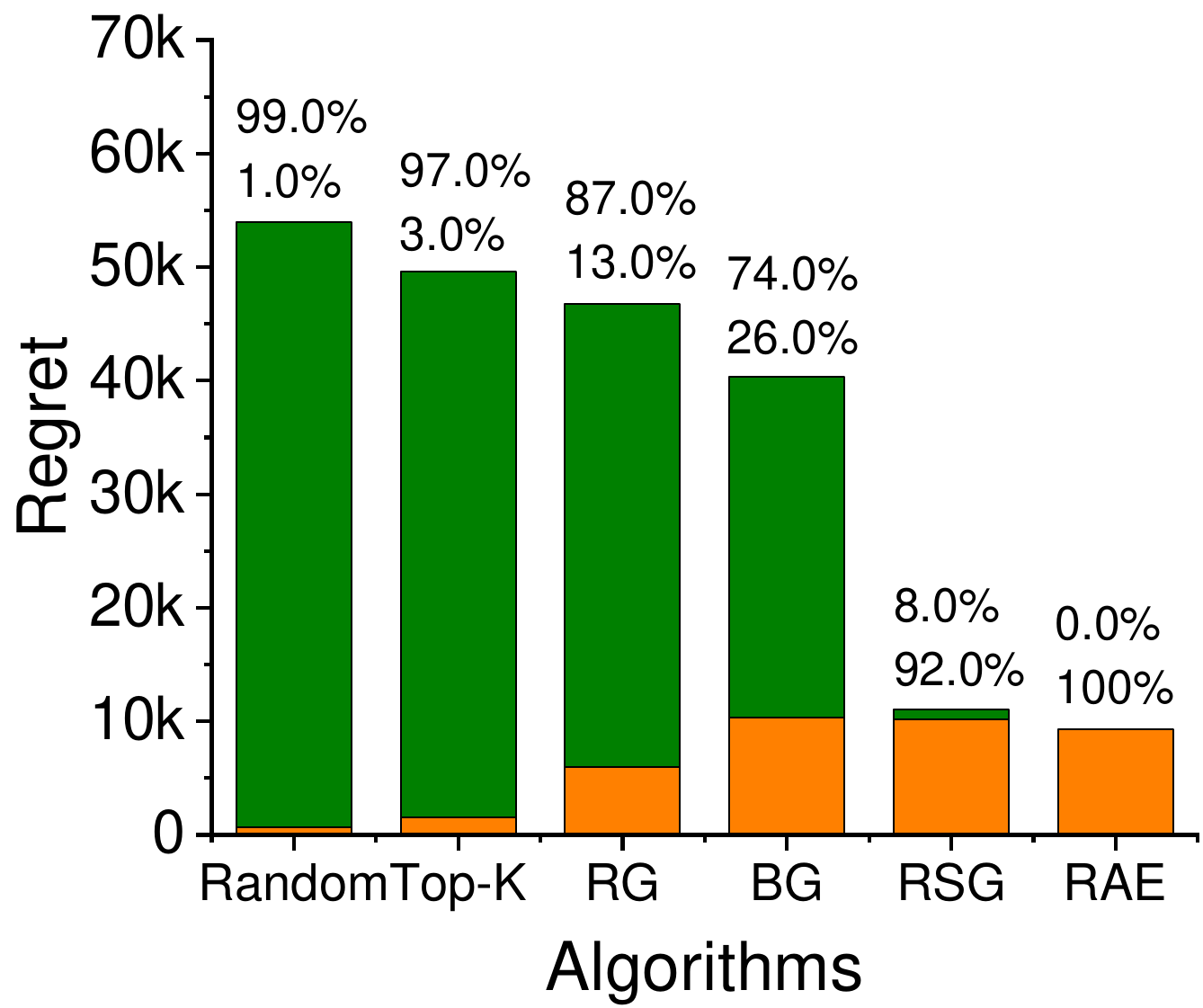} &
        \includegraphics[width=0.17\linewidth]{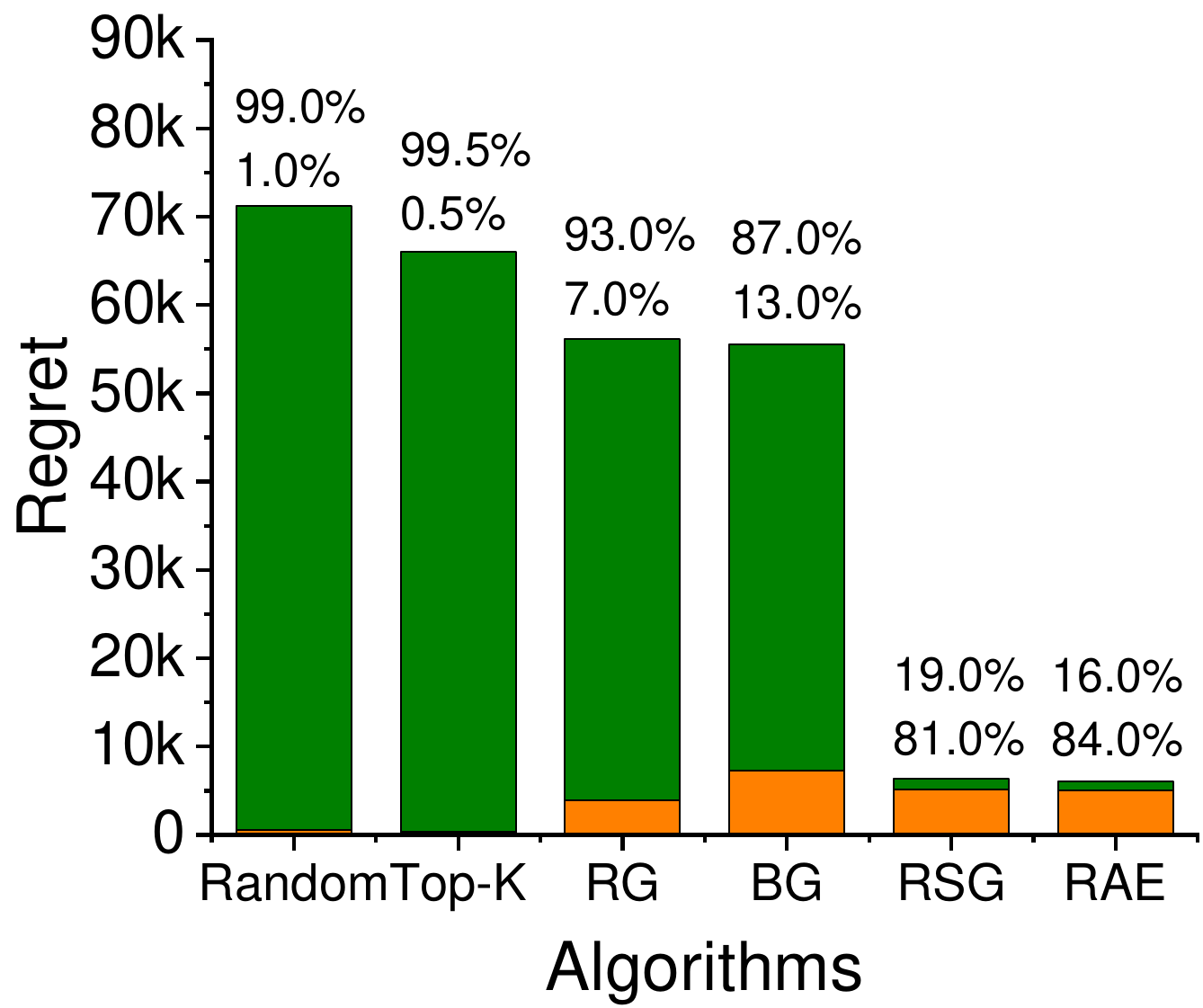} &
        \includegraphics[width=0.17\linewidth]{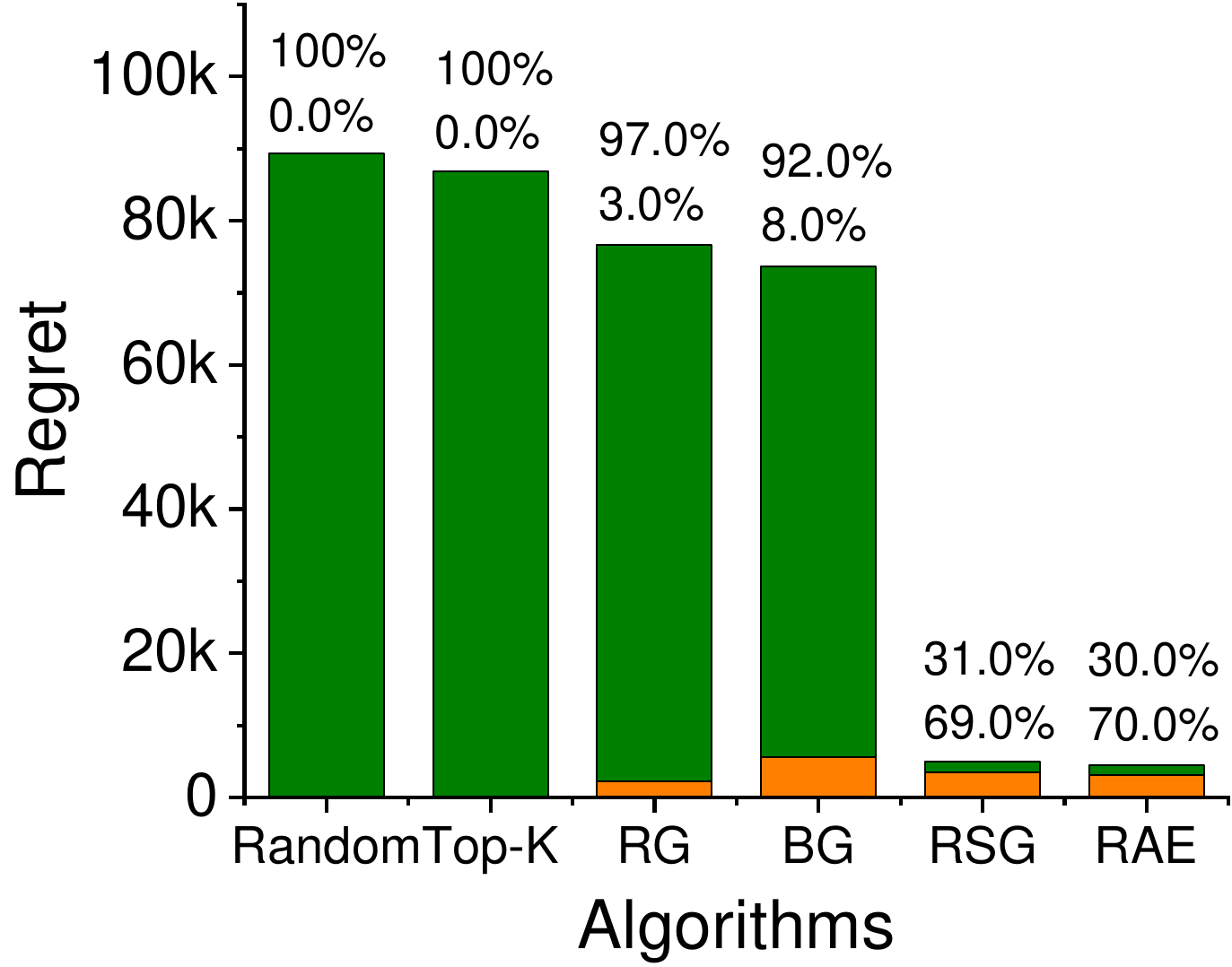} \\
        {\tiny (f) $\delta = 40 \%$} &
        {\tiny (g) $\delta= 60 \%$} &
        {\tiny (h) $\delta = 80 \%$} &
        {\tiny (i) $\delta = 100 \%$} &
        {\tiny (j) $\delta= 120 \%$} \\[5pt]
        
        \includegraphics[width=0.17\linewidth]{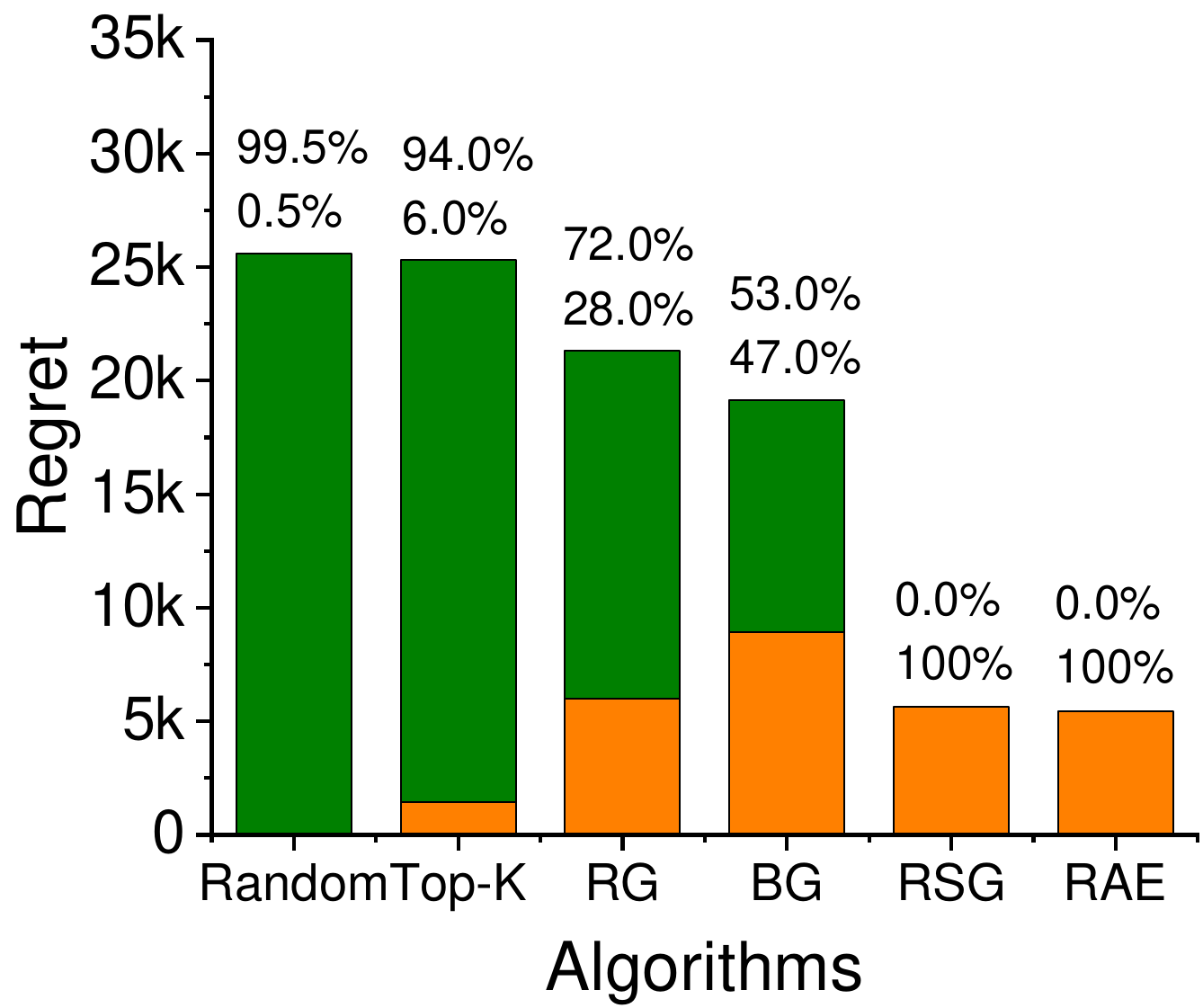} &
        \includegraphics[width=0.17\linewidth]{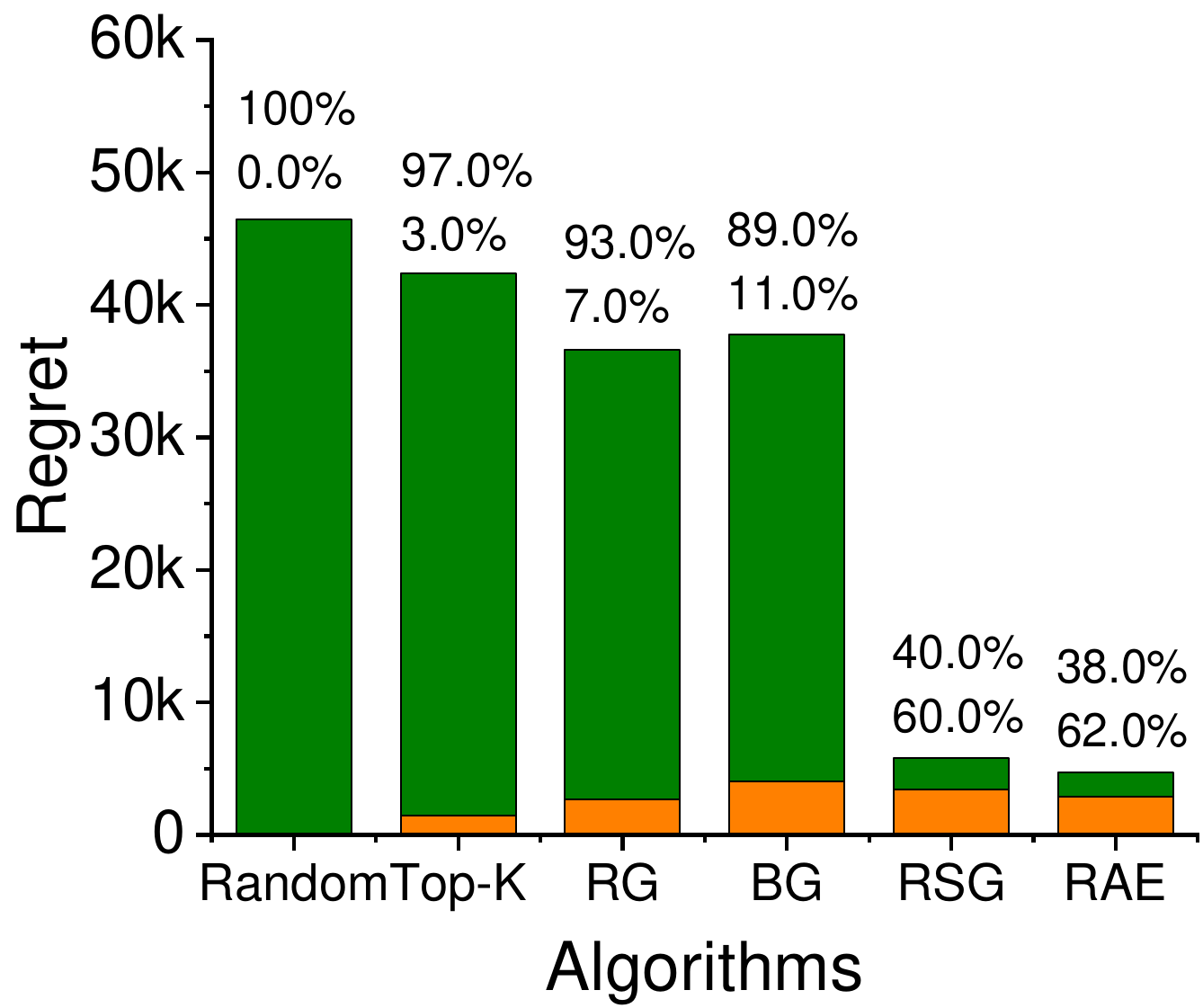} &
        \includegraphics[width=0.17\linewidth]{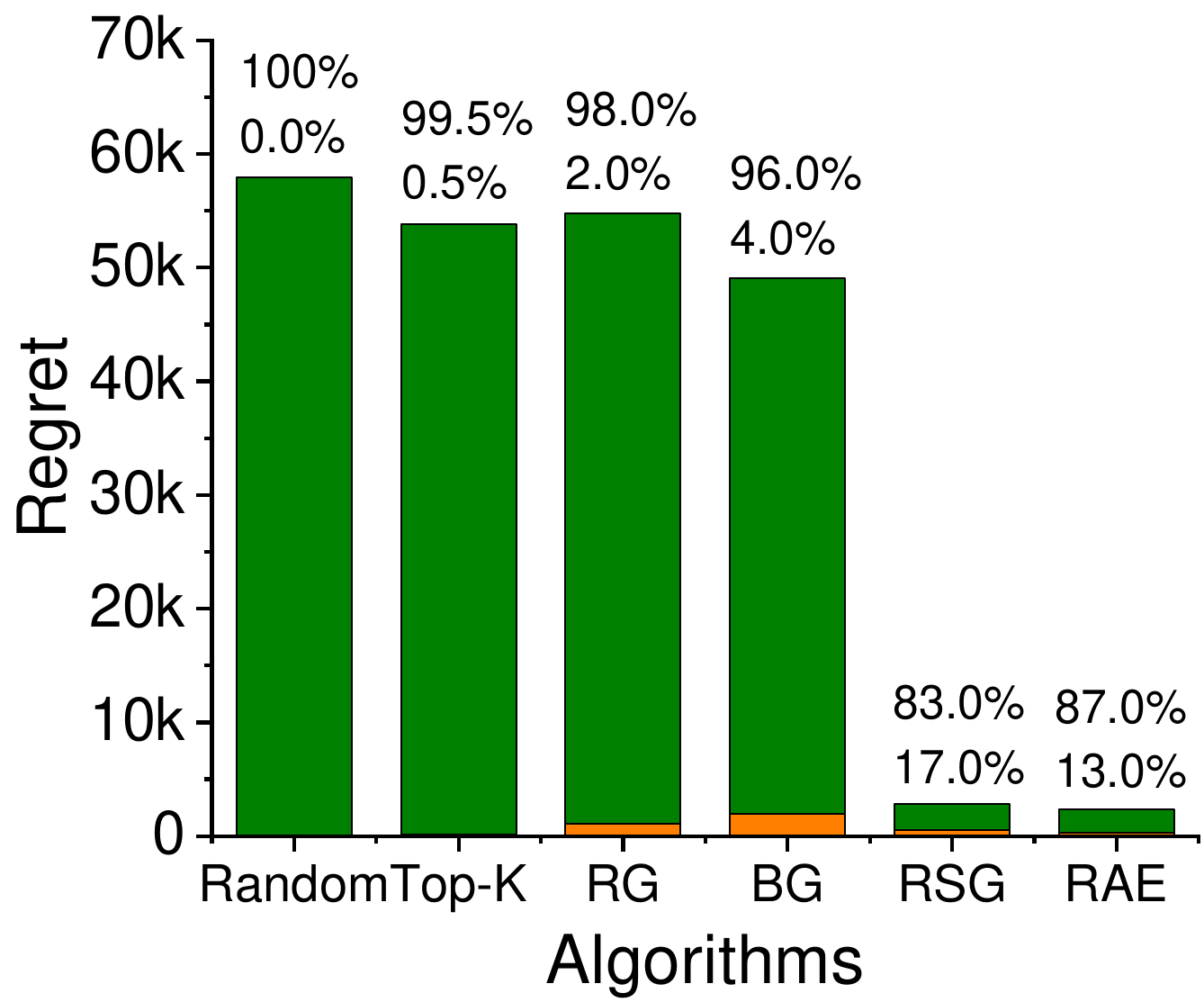} &
        \includegraphics[width=0.17\linewidth]{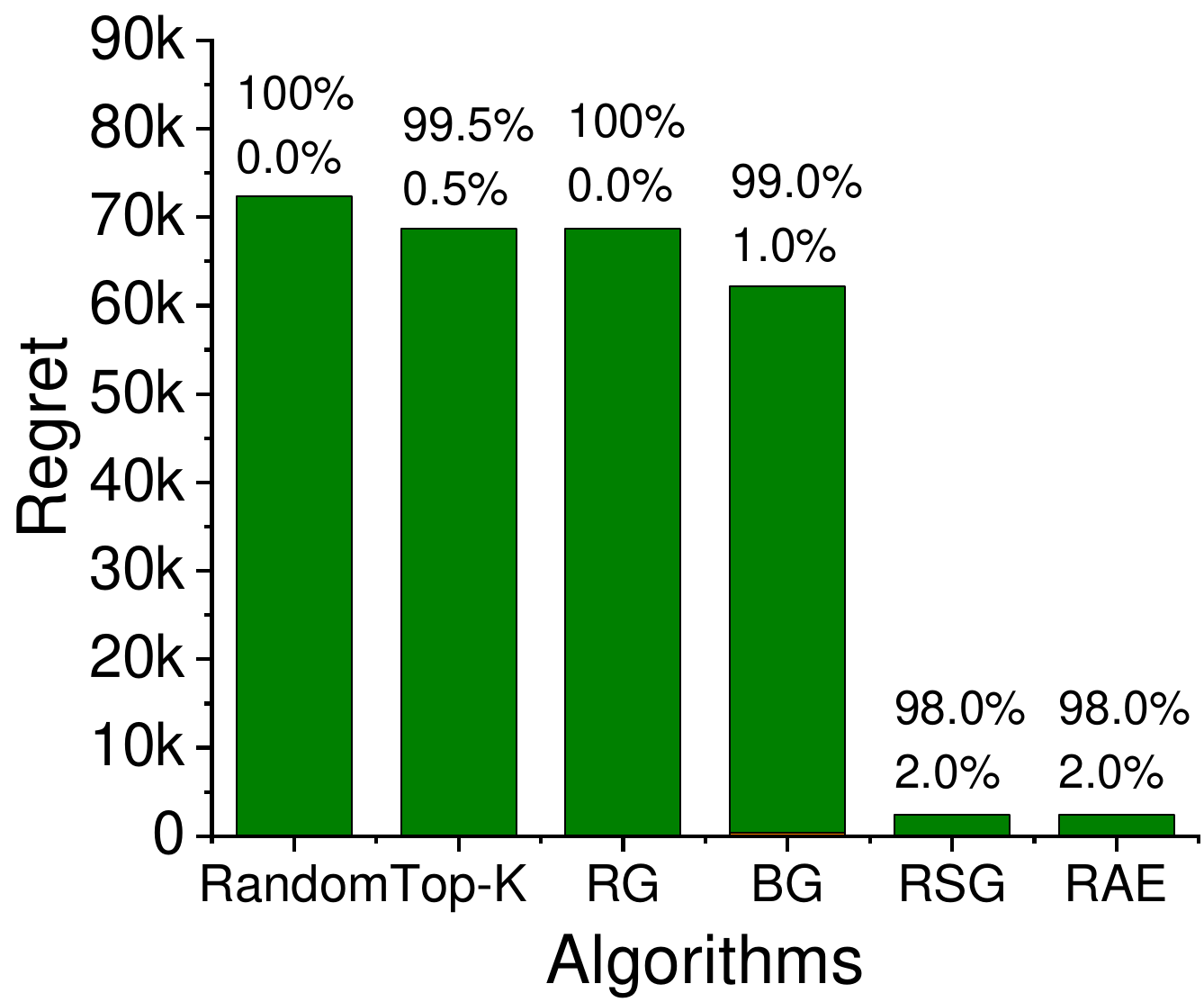} &
        \includegraphics[width=0.17\linewidth]{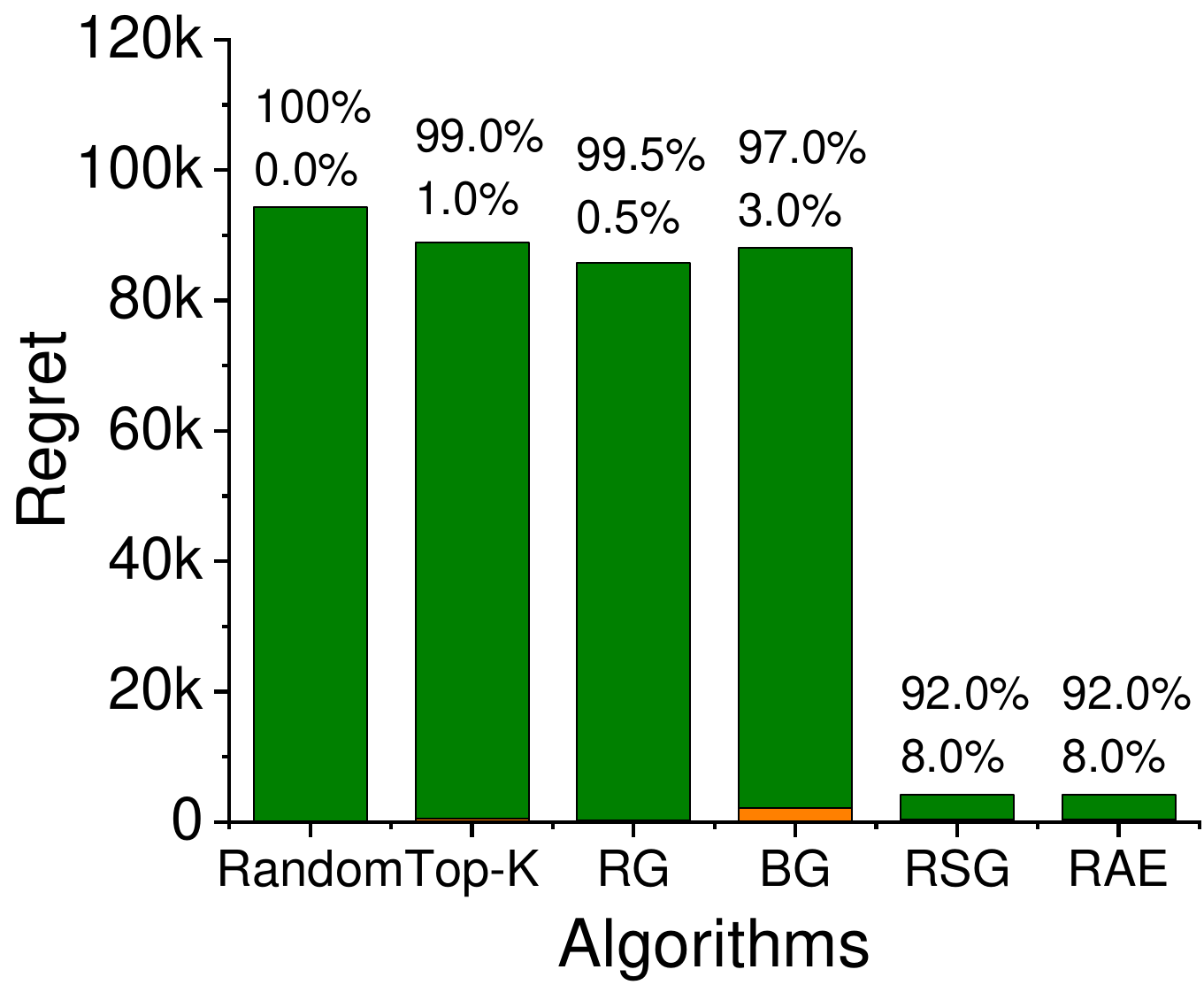} \\
        {\tiny (k) $\delta = 40 \%$} &
        {\tiny (l) $\delta= 60 \%$} &
        {\tiny (m) $\delta = 80 \%$} &
        {\tiny (n) $\delta = 100 \%$} &
        {\tiny (o) $\delta= 120 \%$} \\[5pt]
        
         \includegraphics[width=0.17\linewidth]{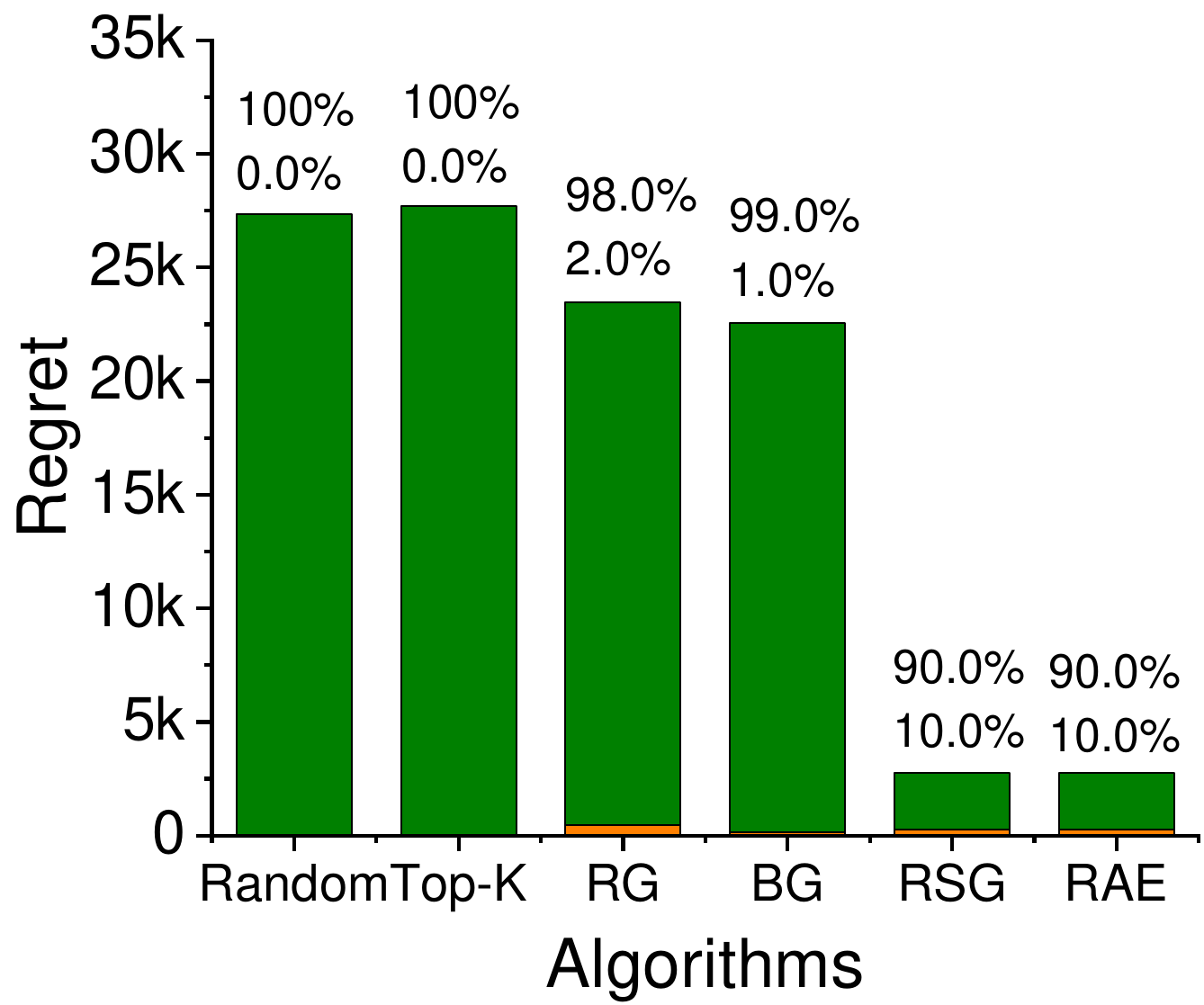} &
        \includegraphics[width=0.17\linewidth]{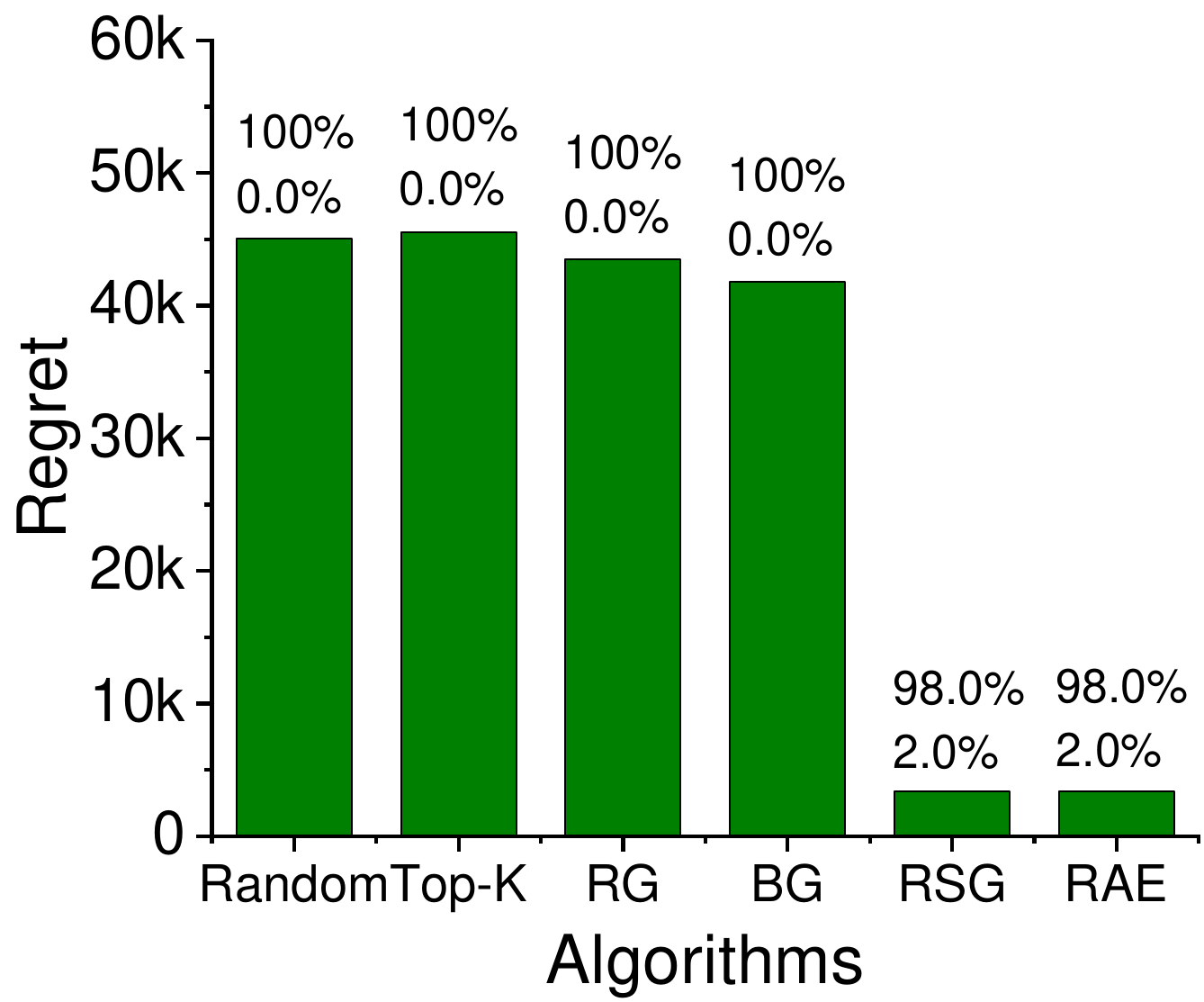} &
        \includegraphics[width=0.17\linewidth]{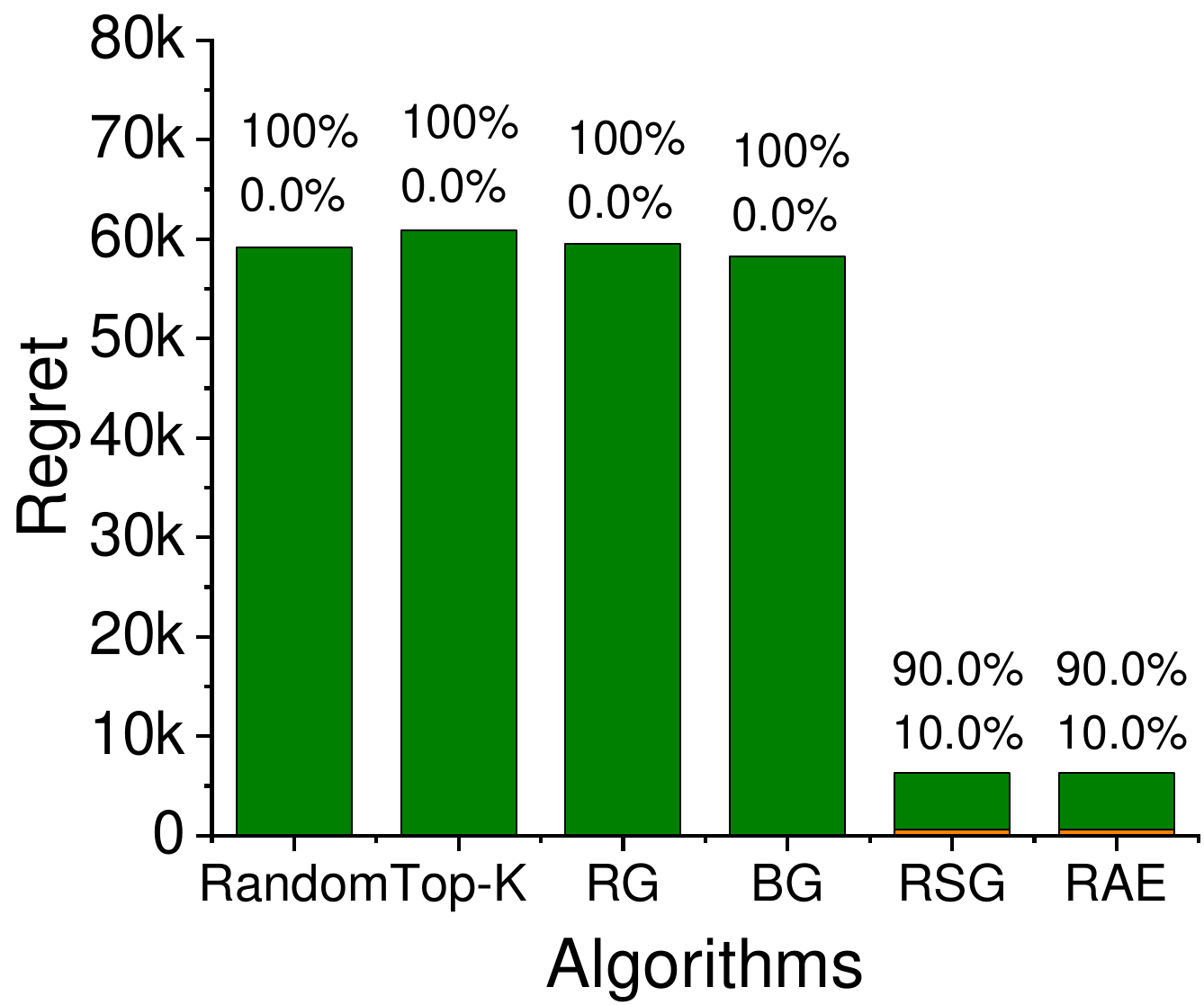} &
        \includegraphics[width=0.17\linewidth]{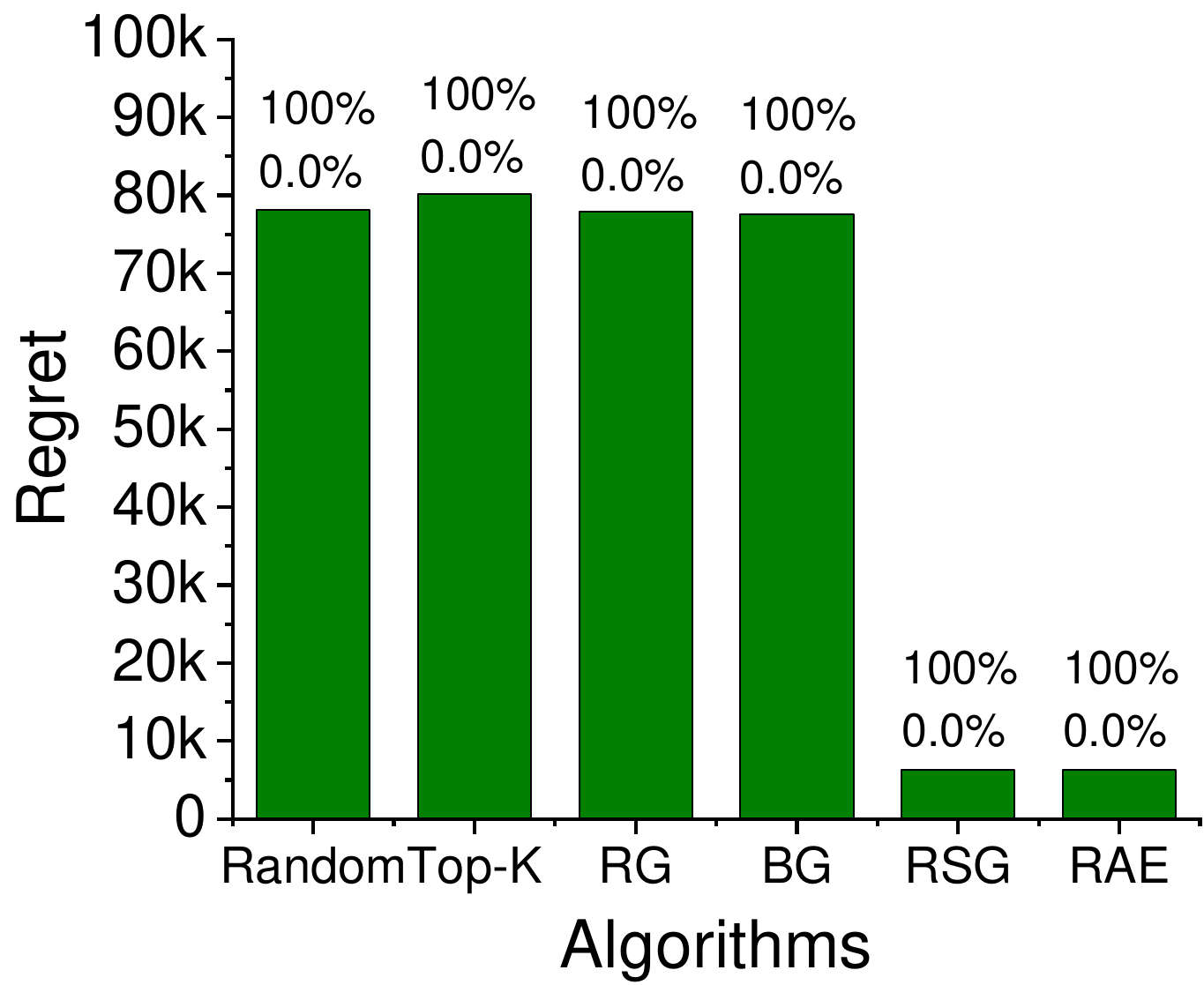} &
        \includegraphics[width=0.17\linewidth]{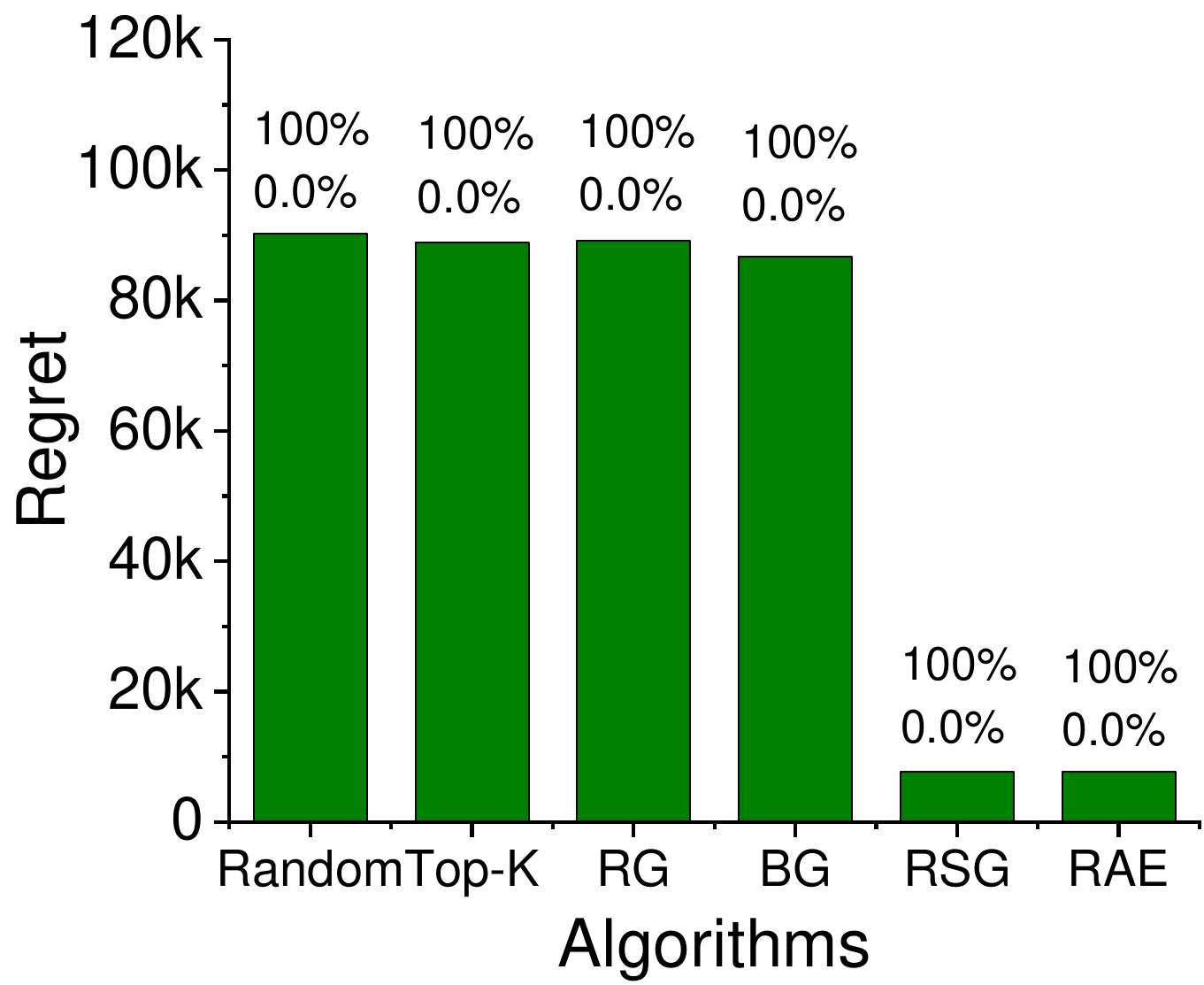} \\
        {\tiny (p) $\delta = 40 \%$} &
        {\tiny (q) $\delta= 60 \%$} &
        {\tiny (r) $\delta = 80 \%$} &
        {\tiny (s) $\delta = 100 \%$} &
        {\tiny (t) $\delta= 120 \%$} \\[5pt]
        
         \includegraphics[width=0.17\linewidth]{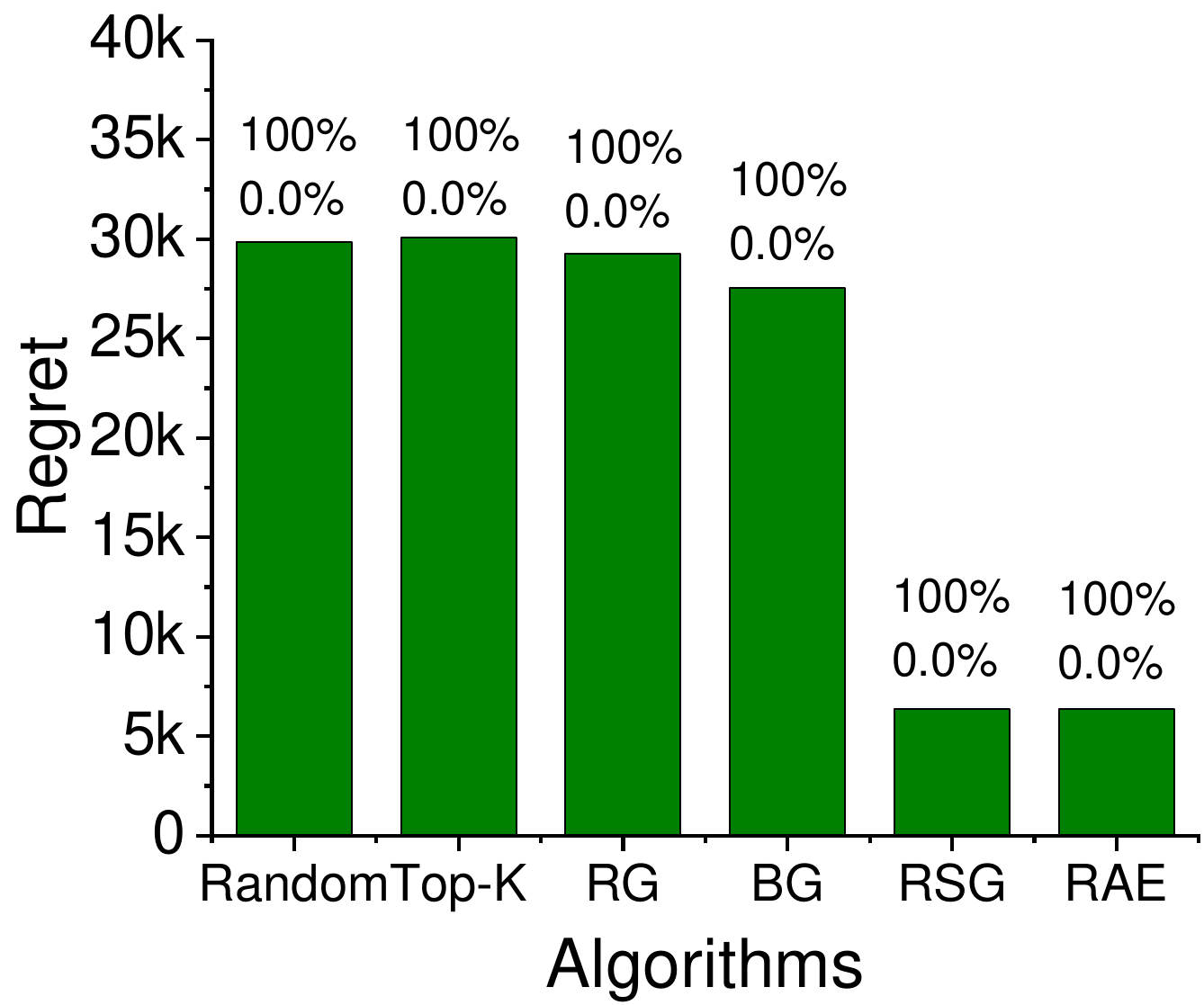} &
        \includegraphics[width=0.17\linewidth]{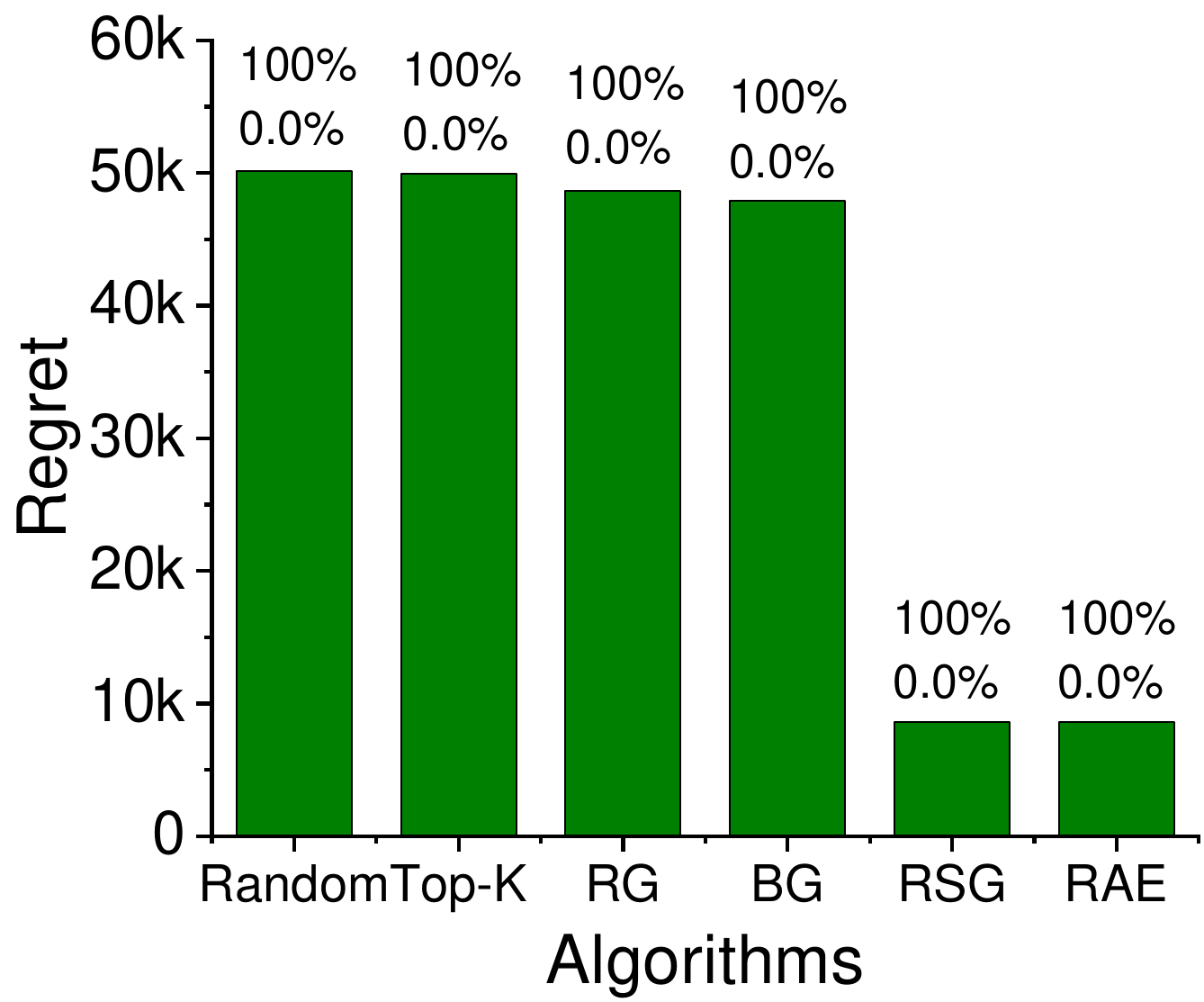} &
        \includegraphics[width=0.17\linewidth]{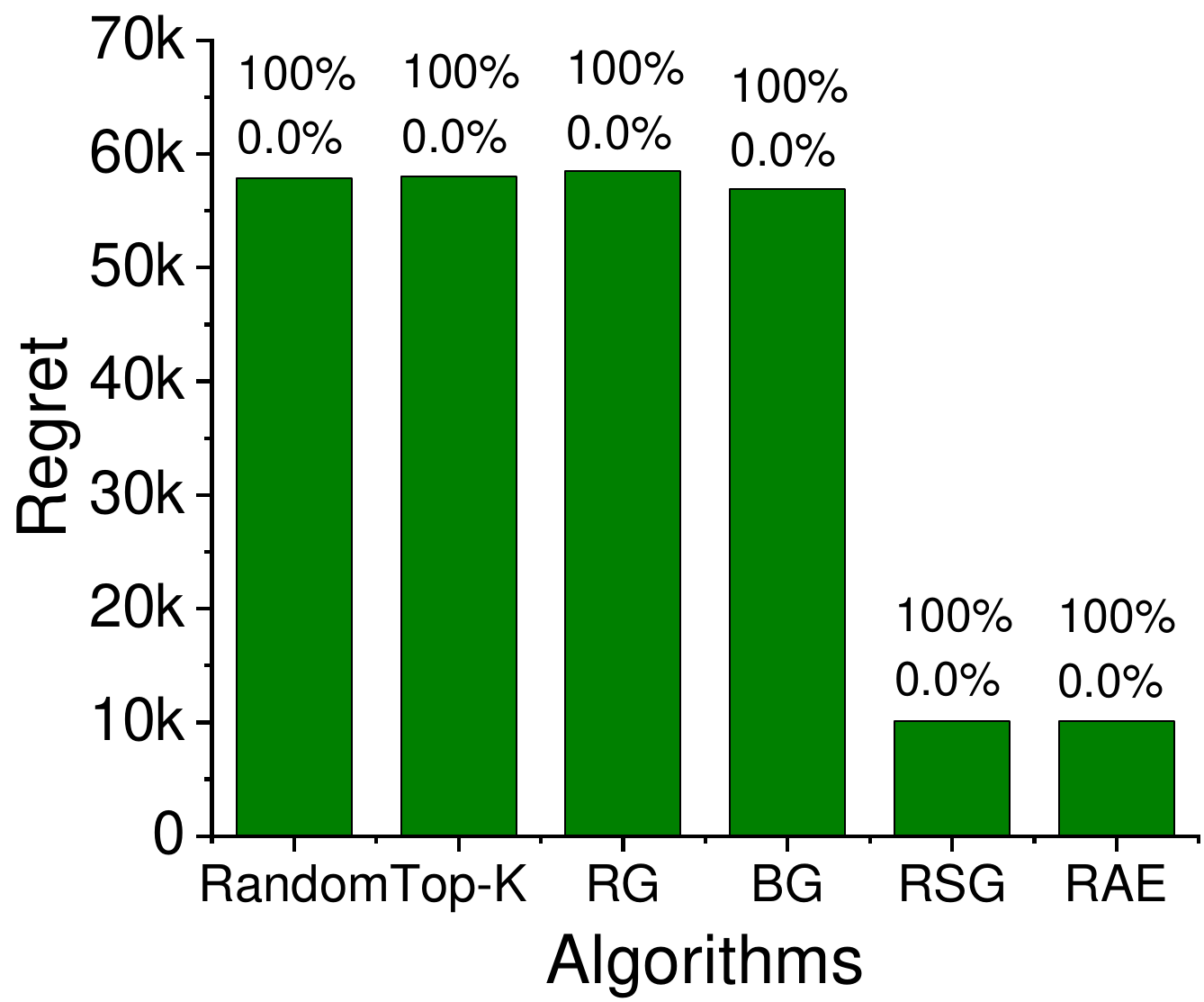} &
        \includegraphics[width=0.17\linewidth]{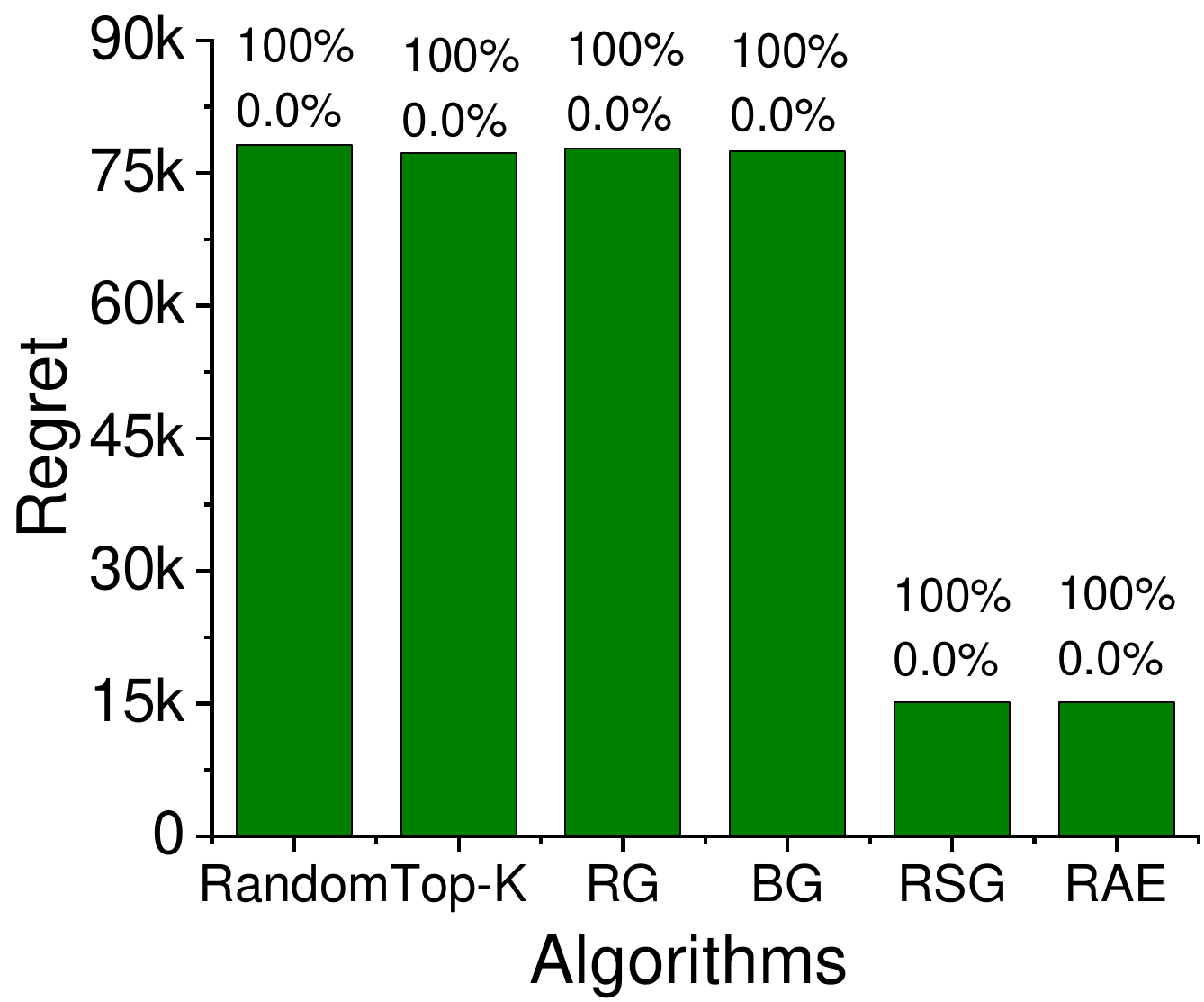} &
        \includegraphics[width=0.17\linewidth]{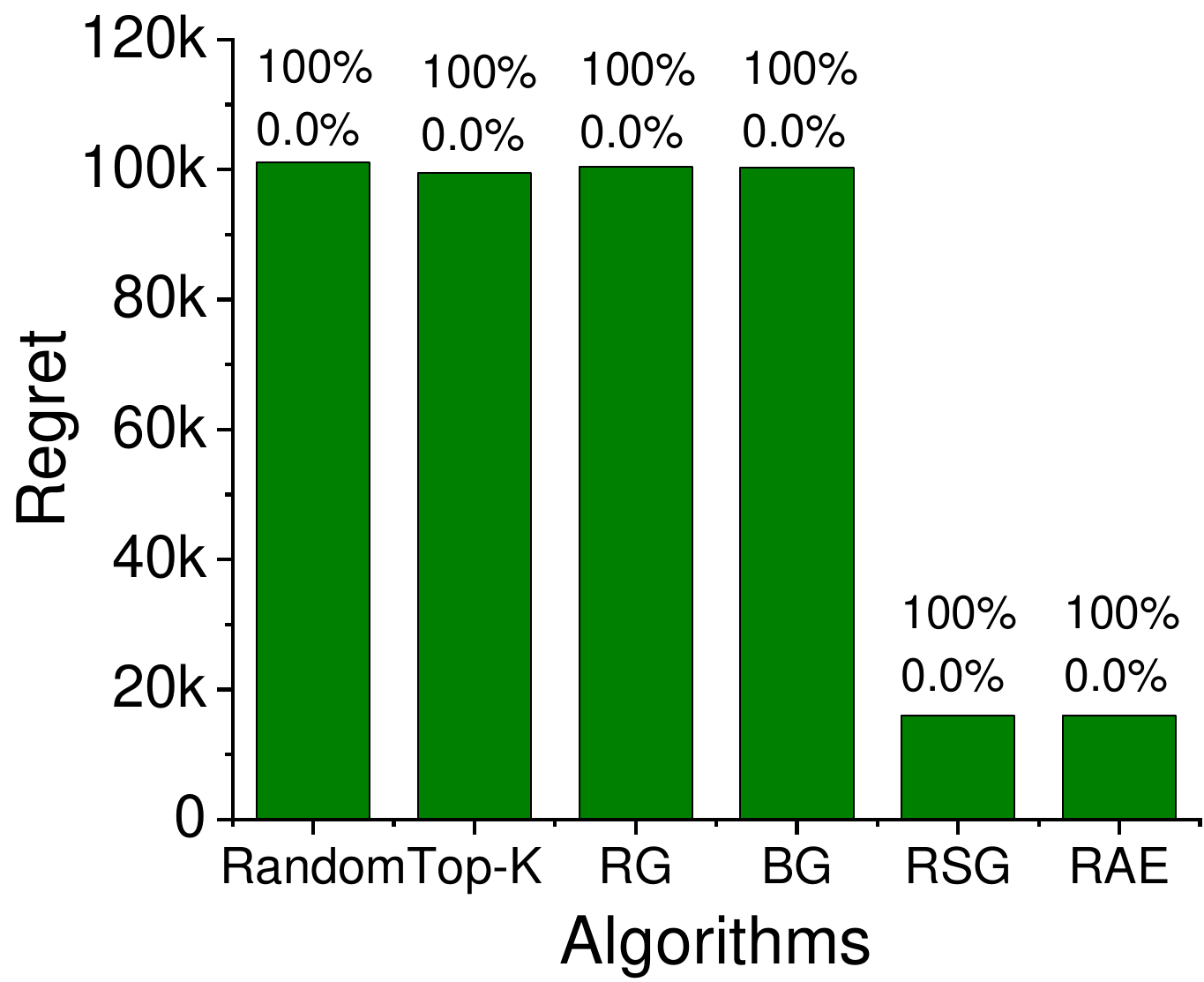} \\
        {\tiny (u) $\delta = 40 \%$} &
        {\tiny (v) $\delta= 60 \%$} &
        {\tiny (w) $\delta = 80 \%$} &
        {\tiny (x) $\delta = 100 \%$} &
        {\tiny (y) $\delta= 120 \%$} \\[5pt]
    \end{tabular}
   \caption{Regret varying $\delta$, when $\lambda = 1\%, \mathcal{|A|} = 100$ $(a, b, c, d, e)$, when $\lambda = 2\%, \mathcal{|A|} = 50$ $(f, g, h, i, j),$ when $\lambda = 5\%, \mathcal{|A|} = 20$ $(k,\ell,m, n,o)$, when $\lambda = 10\%, \mathcal{|A|} = 10$ $(p, q, r, s, t)$, when $\lambda = 20\%, \mathcal{|A|} = 5$ $(u, v, w, x, y),$ for NYC dataset }
    \label{Fig:NYC_Result}
\end{figure*}

\textbf{ Case 1: $\delta \leq 80\%,$ $\lambda \leq 5\%$ (parts $(a,b,c,f,g,h,k,\ell,m)$ of Figure \ref{Fig:NYC_Result}.} 
This refers to the situation where both global and individual demand are low. So, the influence provider has advertisers whose influence demand is minimal. As the $\delta$ value is low, the influence supply is higher than the global influence demand of the advertisers. However, in most cases, zone-specific influence demand is not satisfied for a few advertisers, and they remain unsatisfied. Total regret consists of excessive and unsatisfied regret in most cases except the `RSG' and `RAE'. Now, we have two main observations. First, when $\delta$ increases, excessive regret decreases, and unsatisfied regret increases. This is because when $\delta$ increases, the zone-specific global influence demand also increases. As every zone has only a limited number of billboard slots, the zone-specific influence demand of each advertiser is not satisfied. Second, in most experiments, `RSG' and `RAE' control the unsatisfied regret much better than `BG' and `RG'. The `RSG' can satisfy more advertisers because it releases unsatisfied advertisers individually and allocates released slots to the remaining unsatisfied ones. The `RAE' exchanges billboard slots between advertisers according to the solution of 'RSG'. So, `RAE' can satisfy more or an equal number of advertisers than 'RSG'.  Third, in the base case where $\delta = 40\%$ and $\lambda = 1\%$, i.e., the global influence demand is deficient compared to the influence supply. The influence provider has advertisers with small influence demand. However, a few advertisers are unsatisfied due to not fulfilling zone-specific influence demand. For example, `RG', `BG', `RSG' and `RAE' numbers of satisfied advertisers are $64$, $66$, $69$, and $69$, respectively. The `RSG' and `RAE' outperform `RG' by about $47\%$ and $49\%$, respectively.

\par

\textbf{ Case 2: $\delta \leq 80\%,$, $\lambda \geq 10\%$ (parts $(p,q,r,u,v,w)$ of Figure \ref{Fig:NYC_Result}.} 
This refers to a situation where global influence demand is low. However, individual influence demand is high. The influence provider has a small number of advertisers, and each advertiser's influence demand is high. We have three observations. First, as individual influence demand is high, excessive regret for all the proposed and baseline methods drops in most results. The number of satisfied advertisers is much less; as a result, unsatisfied penalty plays a significant role in total regret. Second, due to higher individual influence demand, influence providers allocate more billboard slots to the advertisers. When $\delta$ increases from $40\%$ to $80\%$, the unsatisfied regret rises for all the proposed and baseline approaches. Third, the ‘RSG’ and ‘RAE’ return minimum total regret by releasing advertisers individually. The ‘RAE’ has a more extensive search space to exchange billboard slots among advertisers to reduce total regret. For example, consider a scenario when $\delta = 40\%, \lambda = 10\%,$ and  $\mathcal{|A|} = 10$ then `RAE' outperforms `RG', and `BG' by about $89\%$, and $88\%$, respectively.

\textbf{ Case 3: $\delta \geq 100\%,$, $\lambda \leq 5\%$ (parts $(d,e,i,j,n,o)$ of Figure \ref{Fig:NYC_Result}.} This represents a scenario where global demand is very high, but individual influence demand is low. The influence provider has a large number of advertisers with small individual influence demand. We have two main observations. First, in the context of a significantly higher global demand, it is evident that none of the algorithms can satisfy the requirements of all advertisers. Hence, the unsatisfied penalty becomes higher and substantially contributes to total regret. Here, two cases arise: (1) when $\delta = 100\%$, the global demand and supply are equal. However, there is unsatisfied regret as the excessive regret can not be minimized. (2) when $\delta = 120\%$, the global influence demand exceeds the total influence supply. So, unsatisfied regret increases. Second, the ‘RSG’ and ‘RAE’ reduce total regret compared to the ‘RG’ and ‘BG’ but scarifies the majority of unsatisfied advertisers. For example, consider a scenario when $\delta = 100\%, \lambda = 1\%,$ and  $\mathcal{|A|} = 100$ then `RAE' outperforms `RG' and `BG' by about $82\%$, and $80\%$, respectively. However, the ‘RAE’ satisfies equal or more advertisers than ‘BG’ and ‘RG.’

\textbf{ Case 4: $\delta \geq 100\%,$, $\lambda \geq 10\%$ (parts $(s,t,x,y)$ of Figure \ref{Fig:NYC_Result}.} According to case $4$, we have a situation where both global and individual influence demand is very high. The influence provider has a small number of advertisers whose influence demand is very high. We have two main observations. First, the unsatisfied regret is very high with higher $\delta$ and $\lambda$ values. However, no excessive regret exists as the advertiser's zone-specific influence demand is unsatisfied. Hence, all the proposed and baseline approach suffers from higher unsatisfied regret. Second, when $\lambda$ value increases from $10\%$ to $20\%$, the individual demand also increases. So, the unsatisfied penalty becomes much higher. All the proposed and baseline approaches produce higher regret except `RSG' and `RAE'. The `RSG' reduces total regret and releases many unsatisfied advertisers.

\subsubsection{Observations over LA dataset} \label{OBS_LA}
In our experiments, we have different observations on the LA dataset compared to the NYC dataset. Here, we discuss experimental results over the LA dataset in four different cases. 

\begin{figure}[h]
    \centering
    \begin{tabular}{lclc}
       Unsatisfied Regret & \includegraphics[width=0.11\linewidth]{Result/Unsatisfied.png} & Excessive Regret & \includegraphics[width=0.11\linewidth]{Result/Excessive.png} \\
    \end{tabular}
    \begin{tabular}{ccccc}     
        \includegraphics[width=0.17\linewidth]{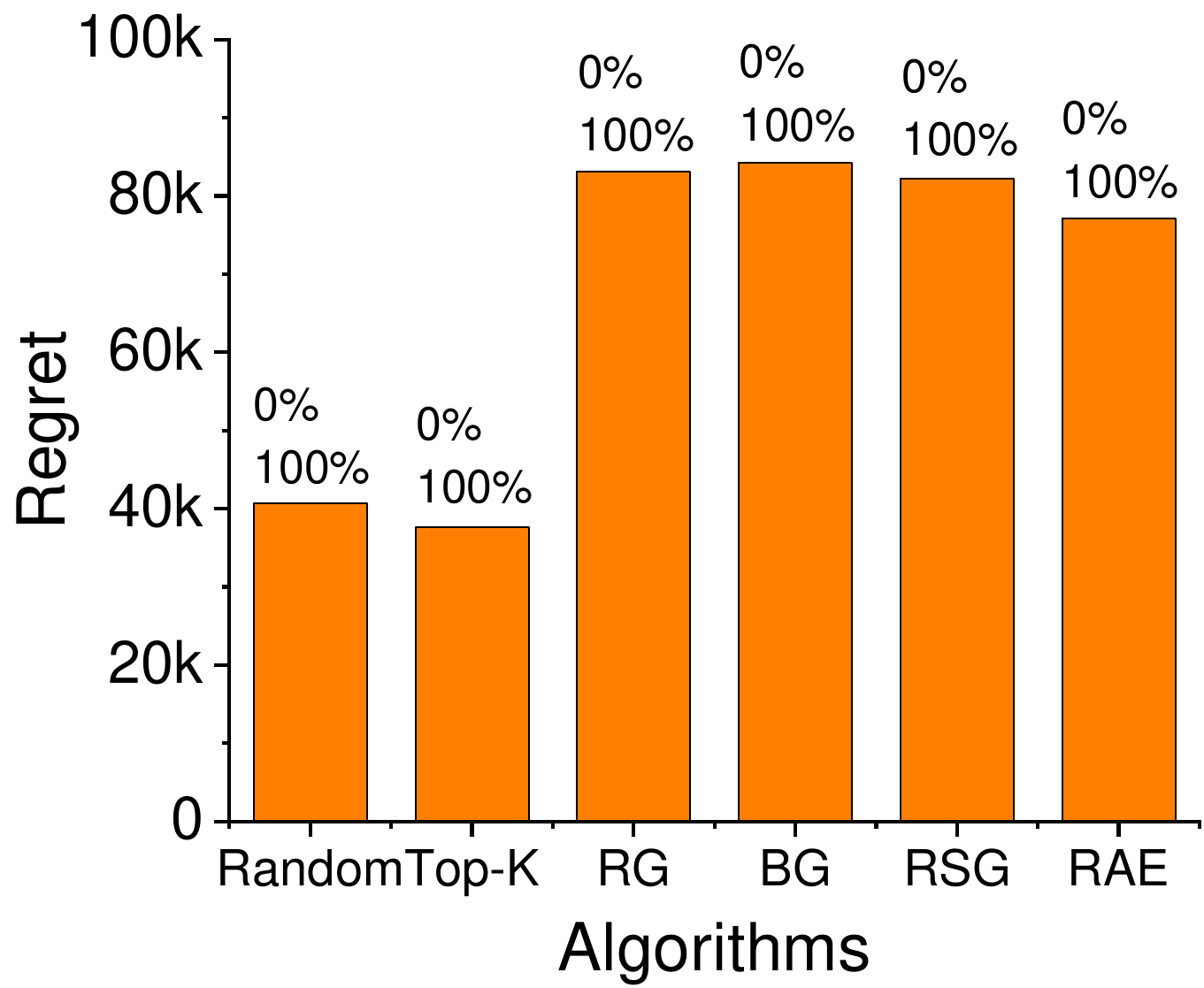} &
        \includegraphics[width=0.17\linewidth]{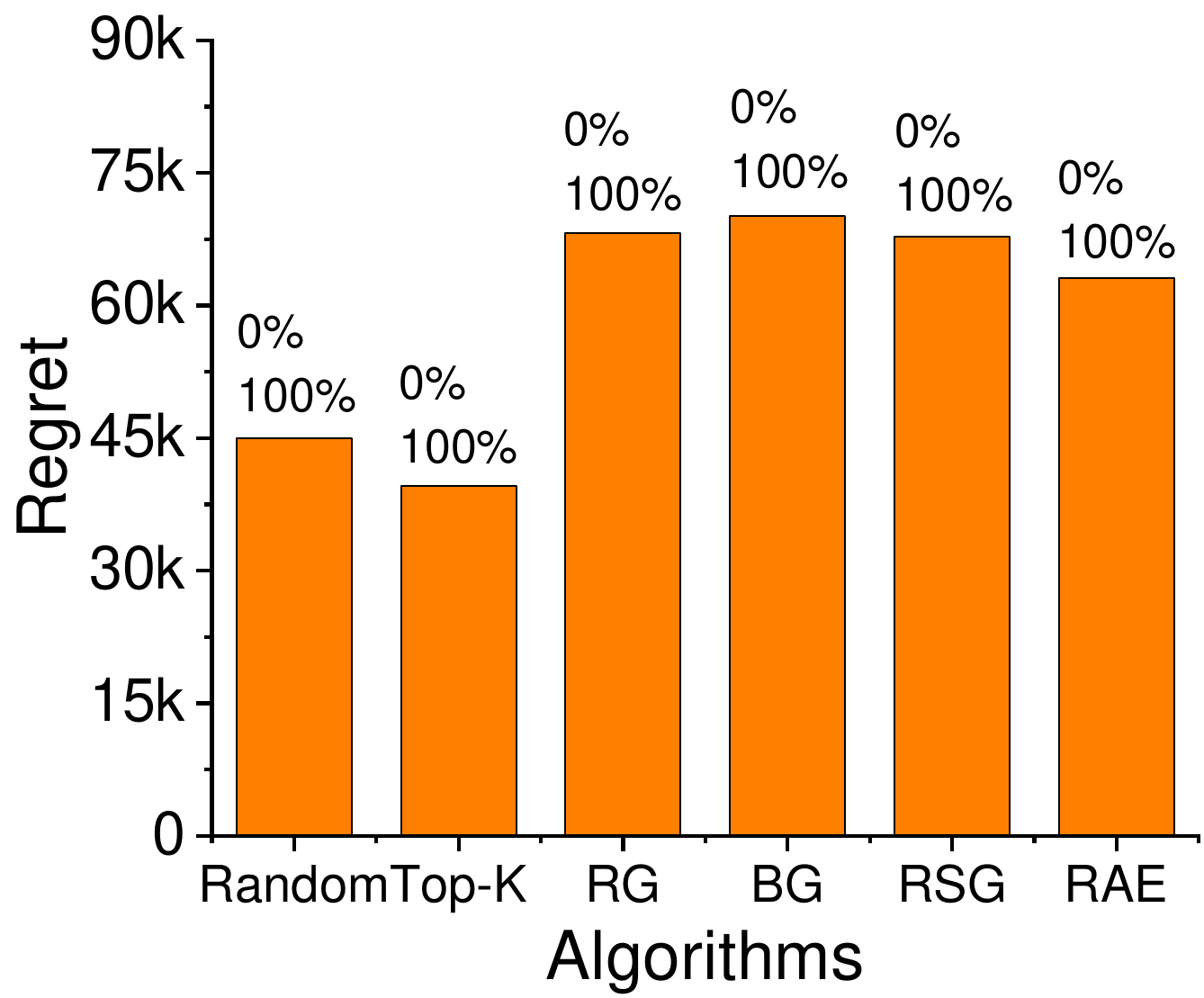} &
        \includegraphics[width=0.17\linewidth]{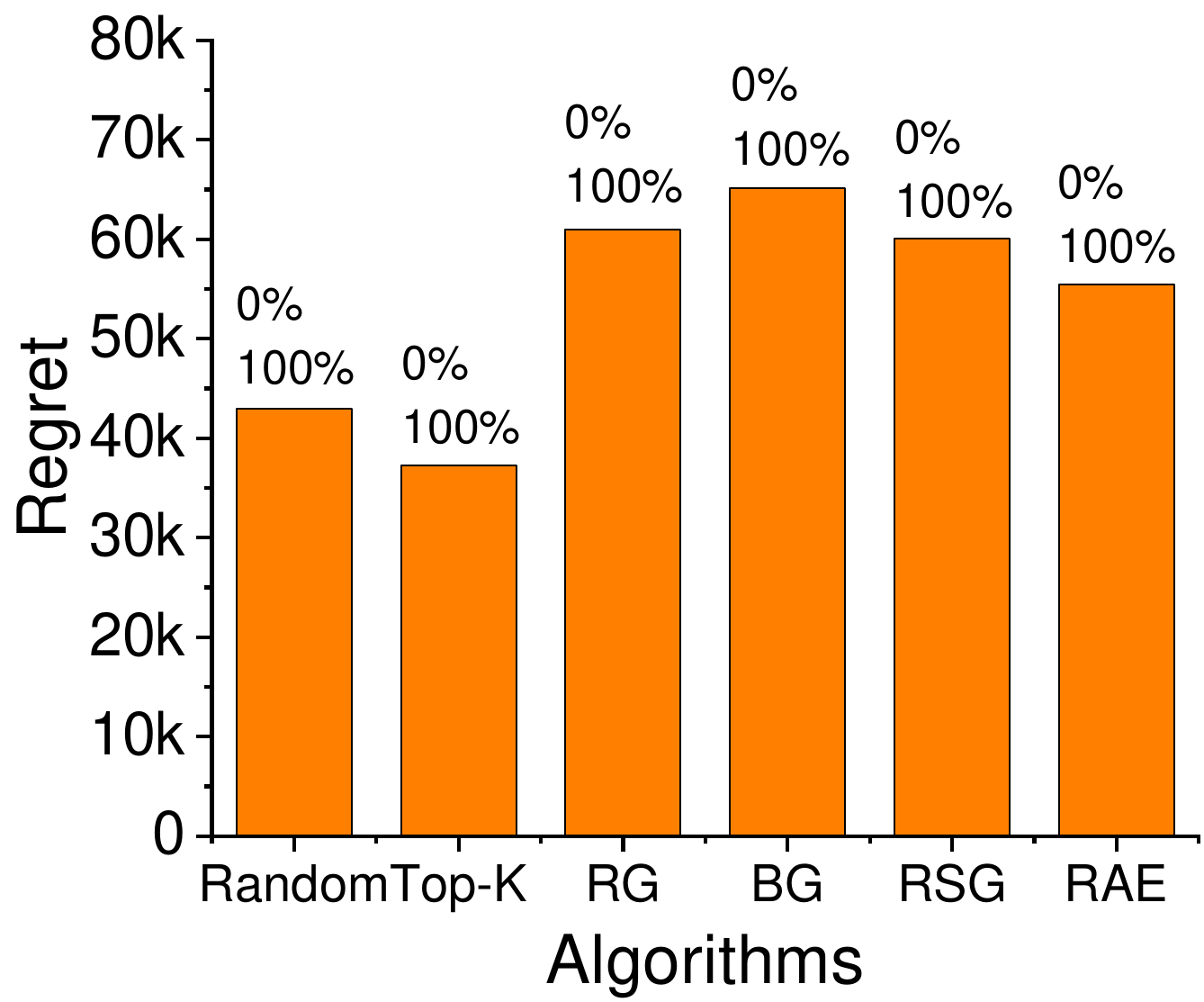} &
        \includegraphics[width=0.17\linewidth]{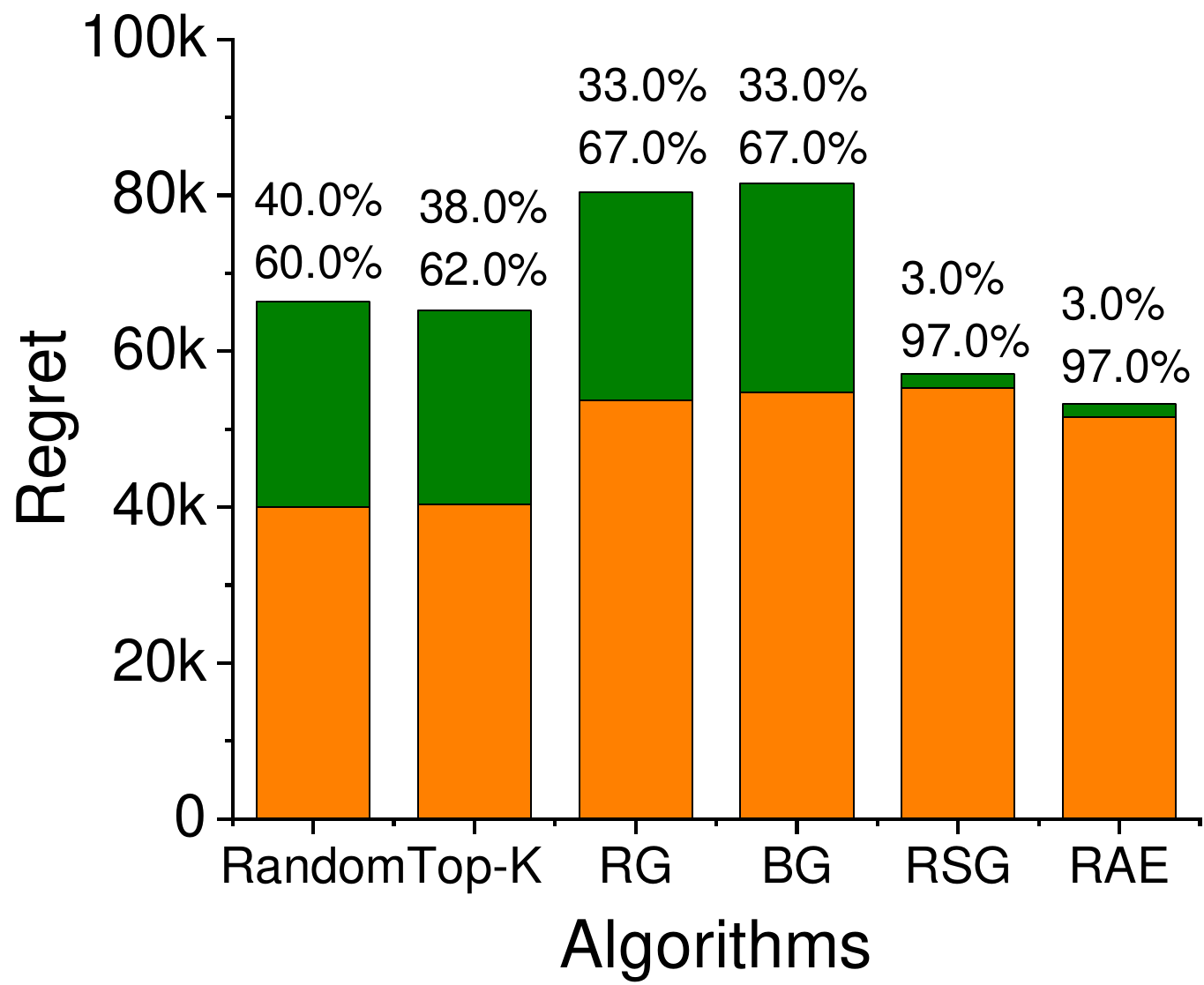} &
        \includegraphics[width=0.17\linewidth]{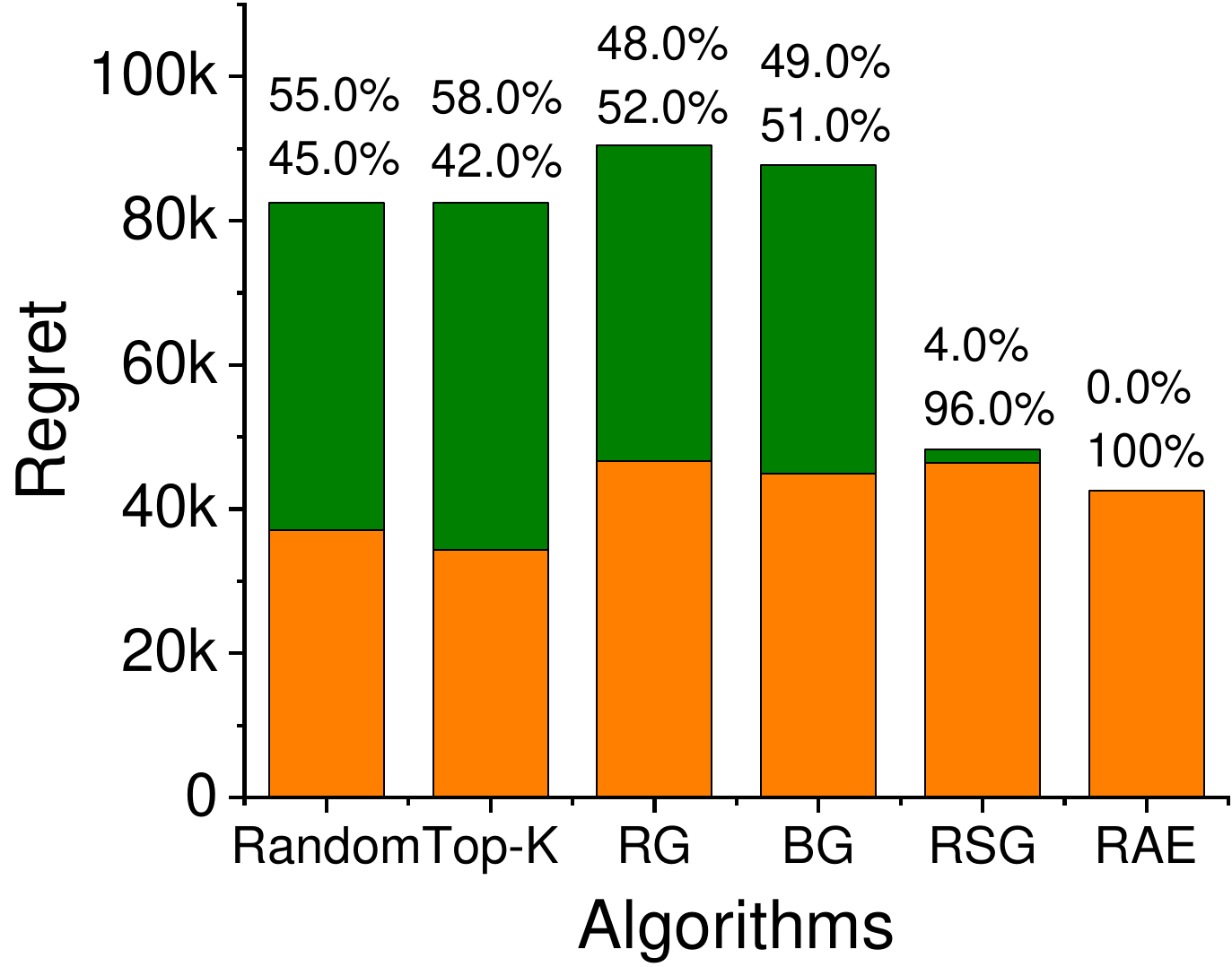} \\
        {\tiny (a) $\delta = 40 \%$} &
        {\tiny (b) $\delta= 60 \%$} &
        {\tiny (c) $\delta = 80 \%$} &
        {\tiny (d) $\delta = 100 \%$} &
        {\tiny (e) $\delta= 120 \%$} \\[5pt]

         \includegraphics[width=0.17\linewidth]{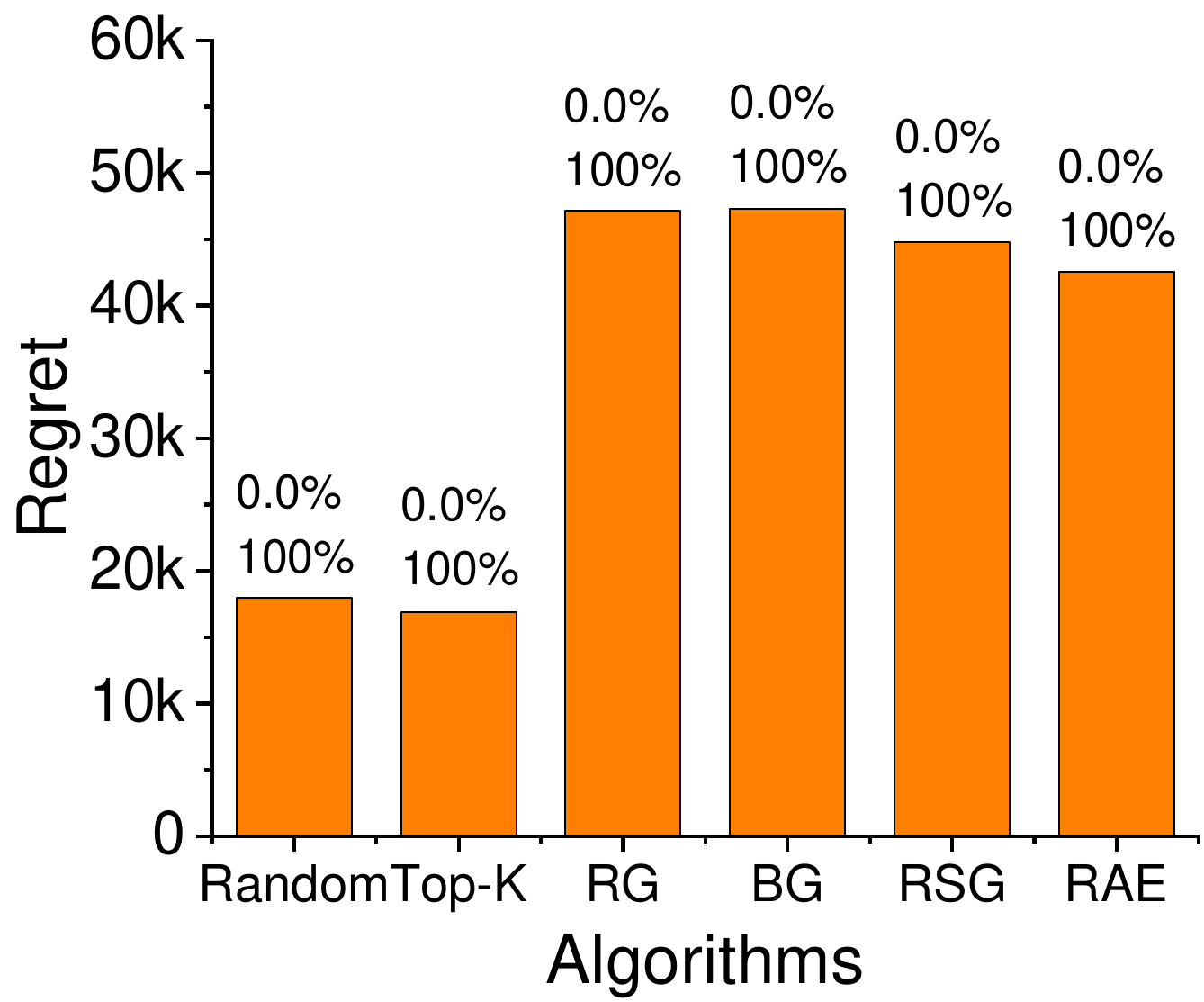} &
        \includegraphics[width=0.17\linewidth]{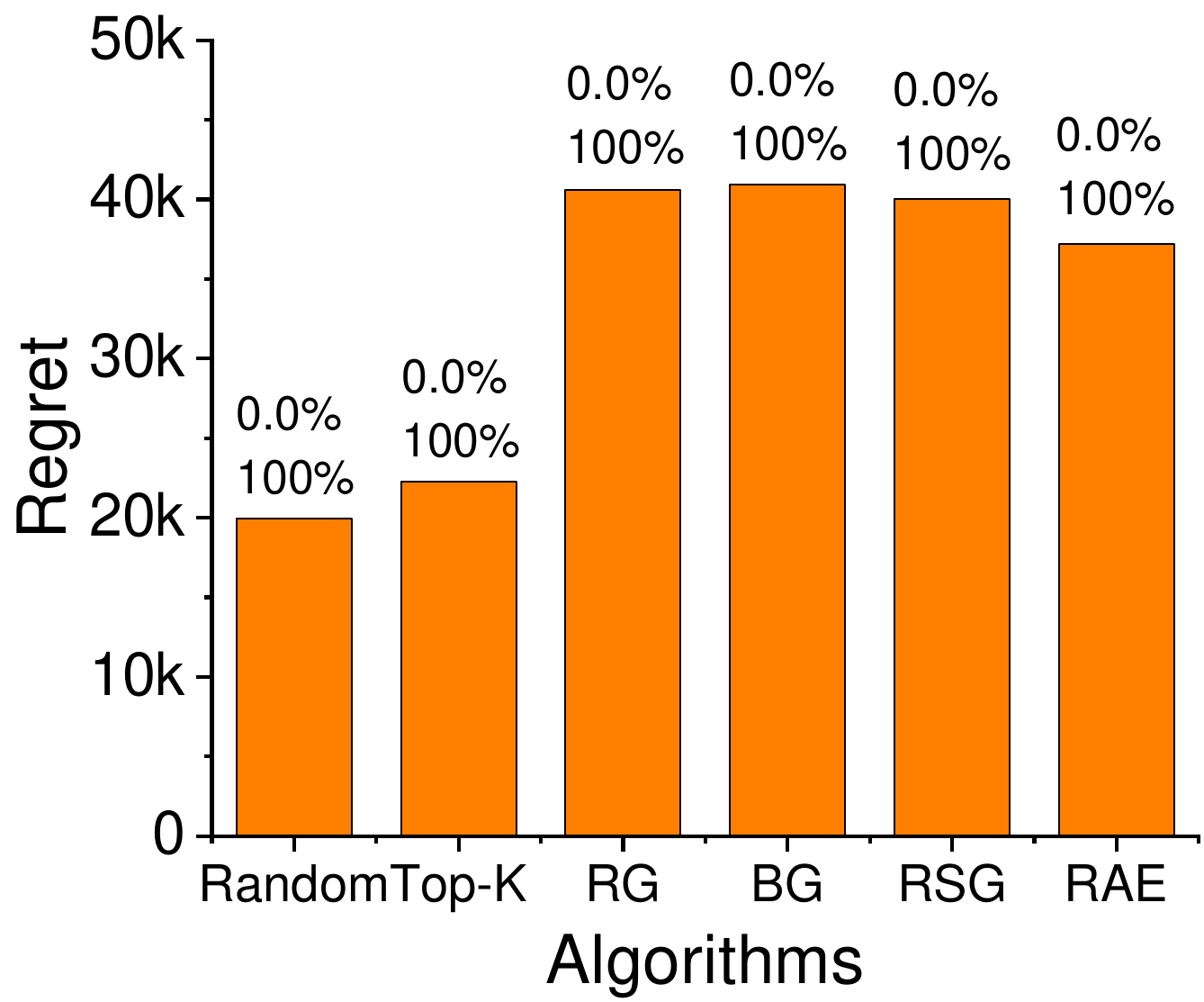} &
        \includegraphics[width=0.17\linewidth]{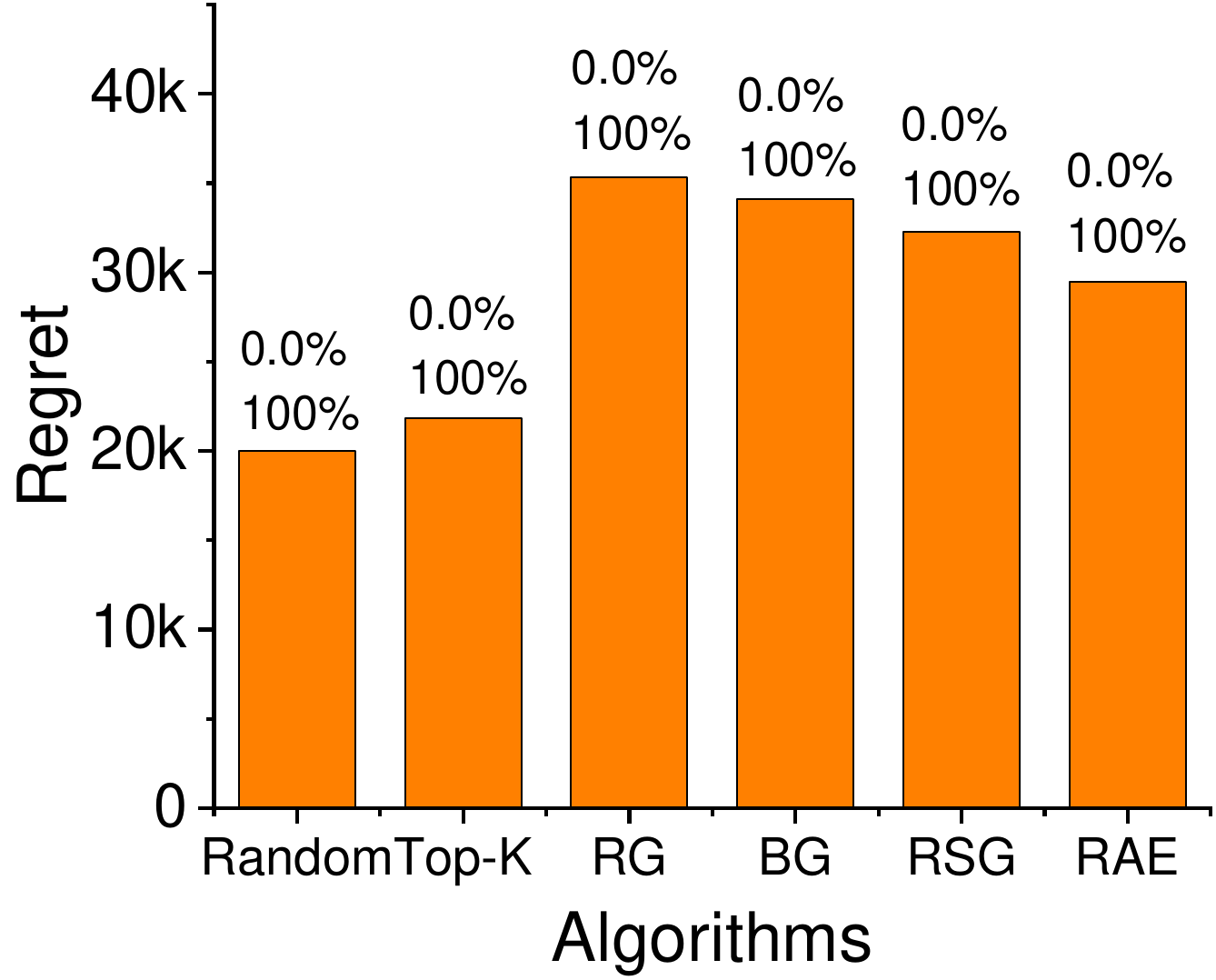} &
        \includegraphics[width=0.17\linewidth]{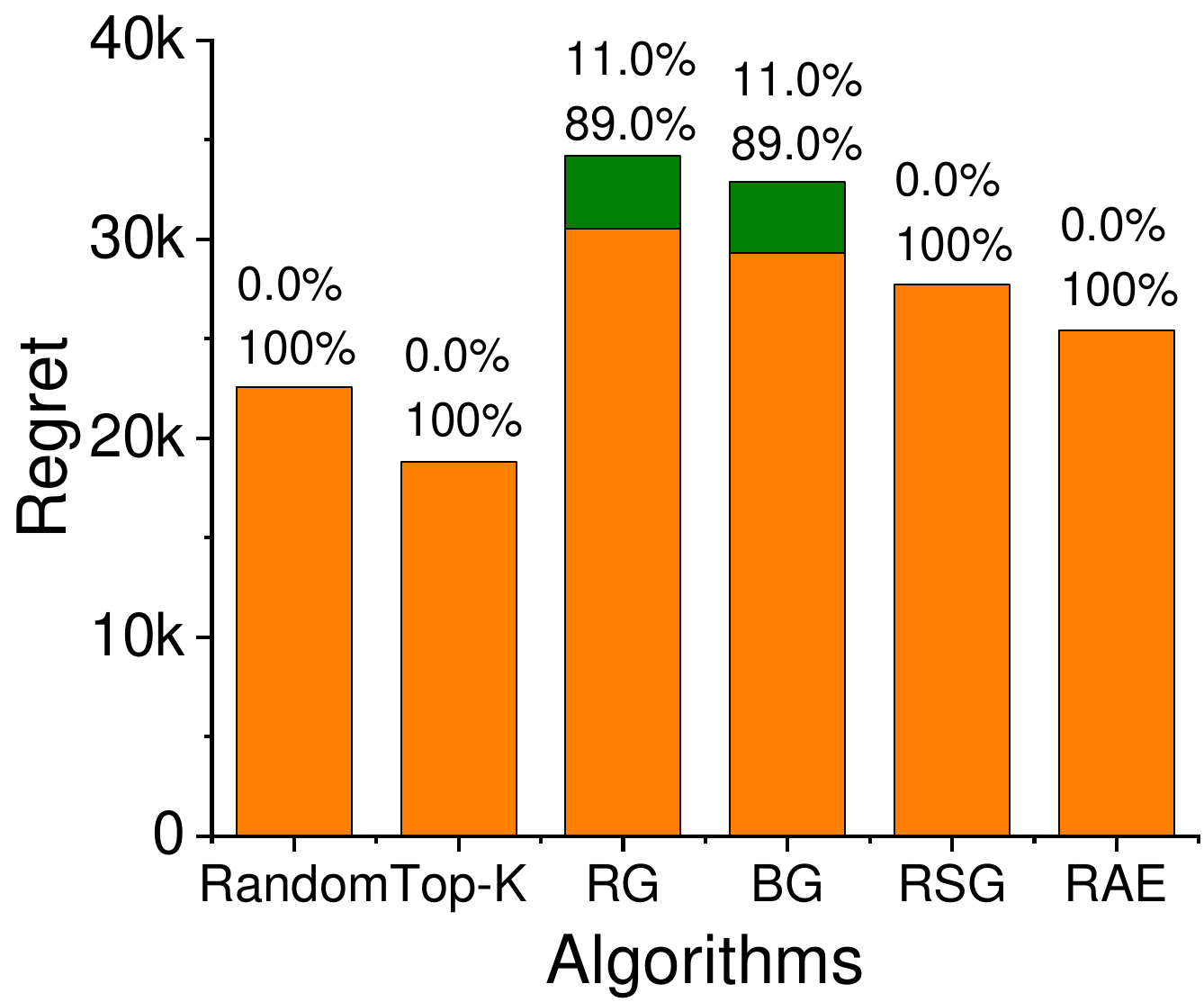} &
        \includegraphics[width=0.17\linewidth]{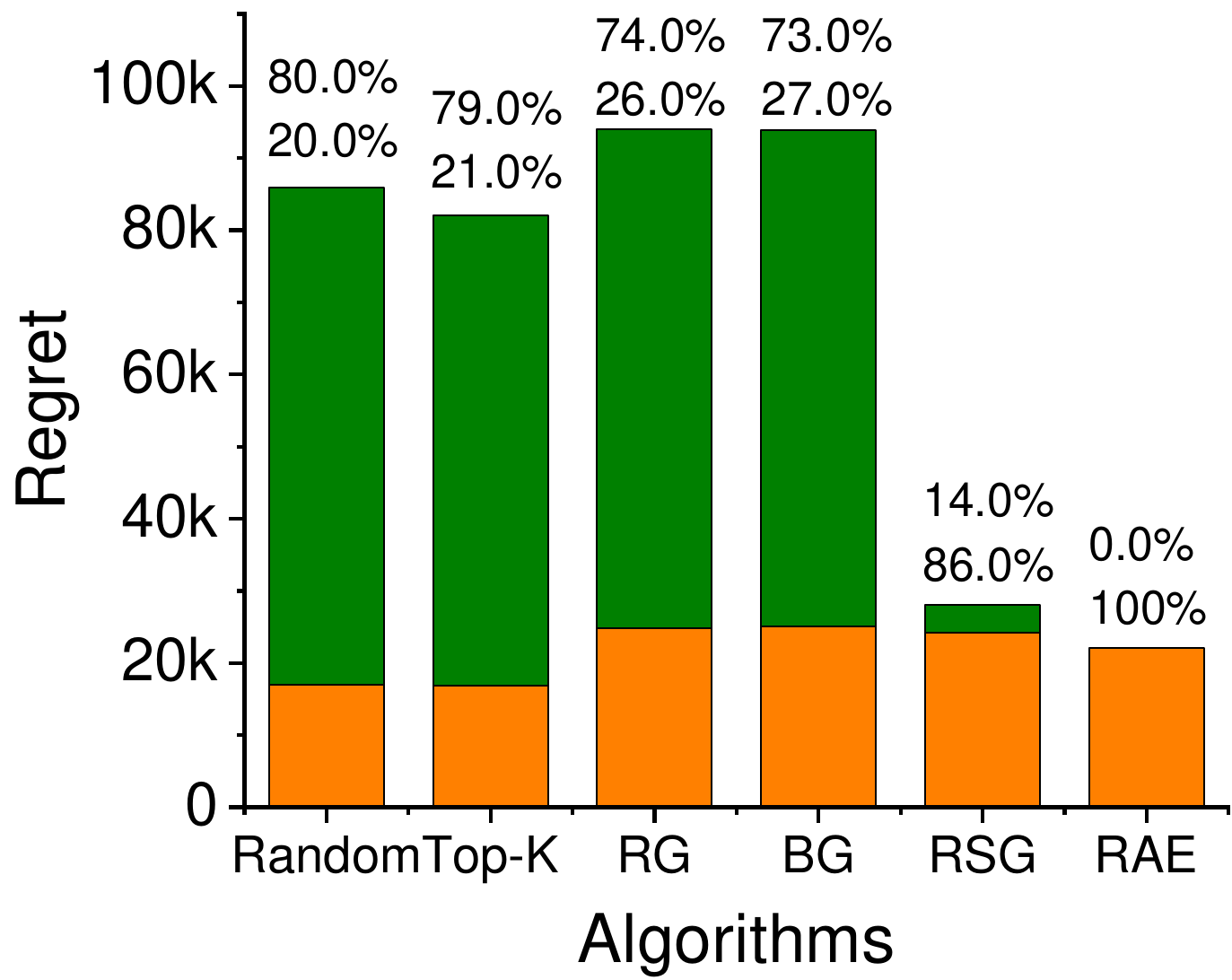} \\
        {\tiny (f) $\delta = 40 \%$} &
        {\tiny (g) $\delta= 60 \%$} &
        {\tiny (h) $\delta = 80 \%$} &
        {\tiny (i) $\delta = 100 \%$} &
        {\tiny (j) $\delta= 120 \%$} \\[5pt]
        
        \includegraphics[width=0.17\linewidth]{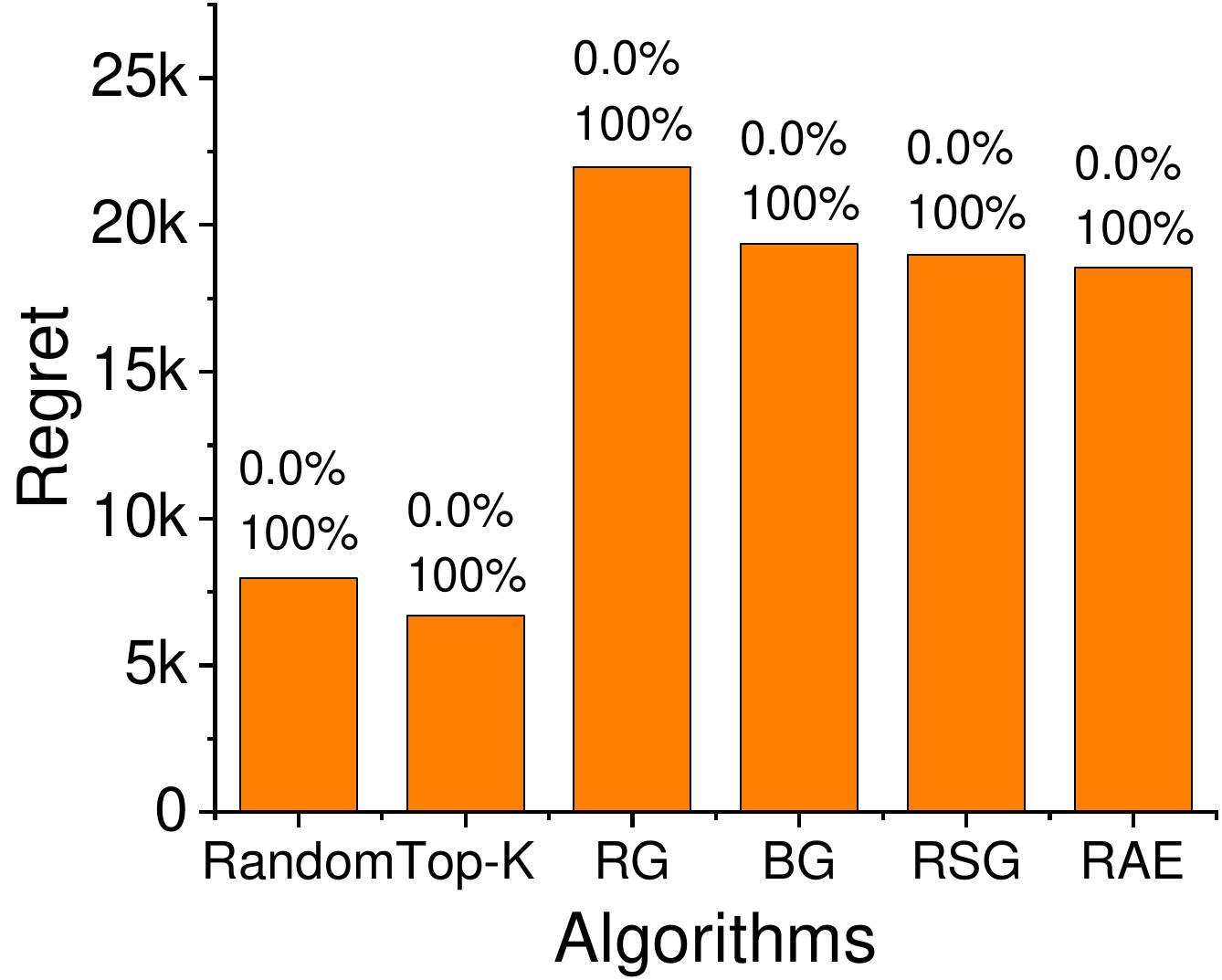} &
        \includegraphics[width=0.17\linewidth]{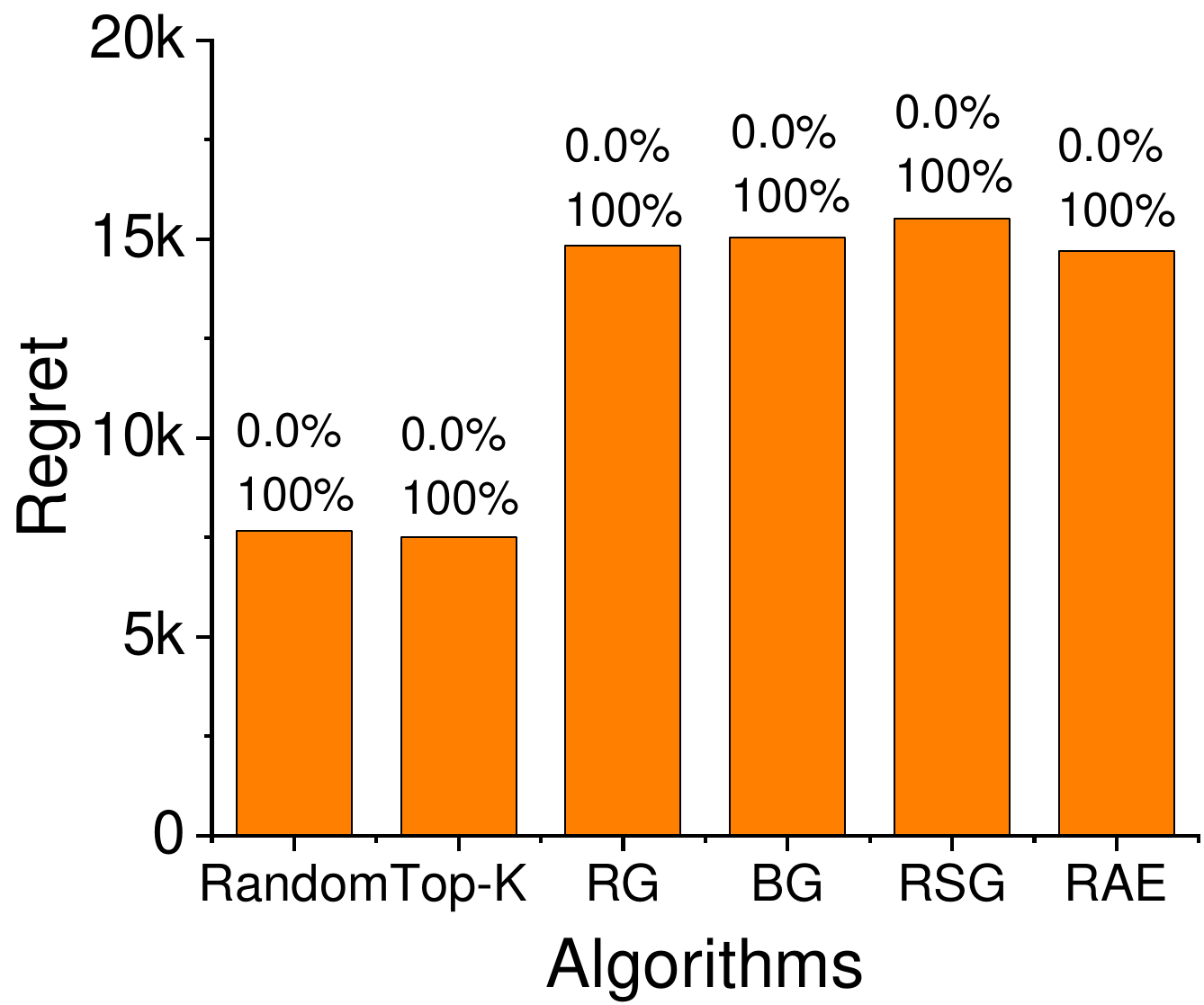} &
        \includegraphics[width=0.17\linewidth]{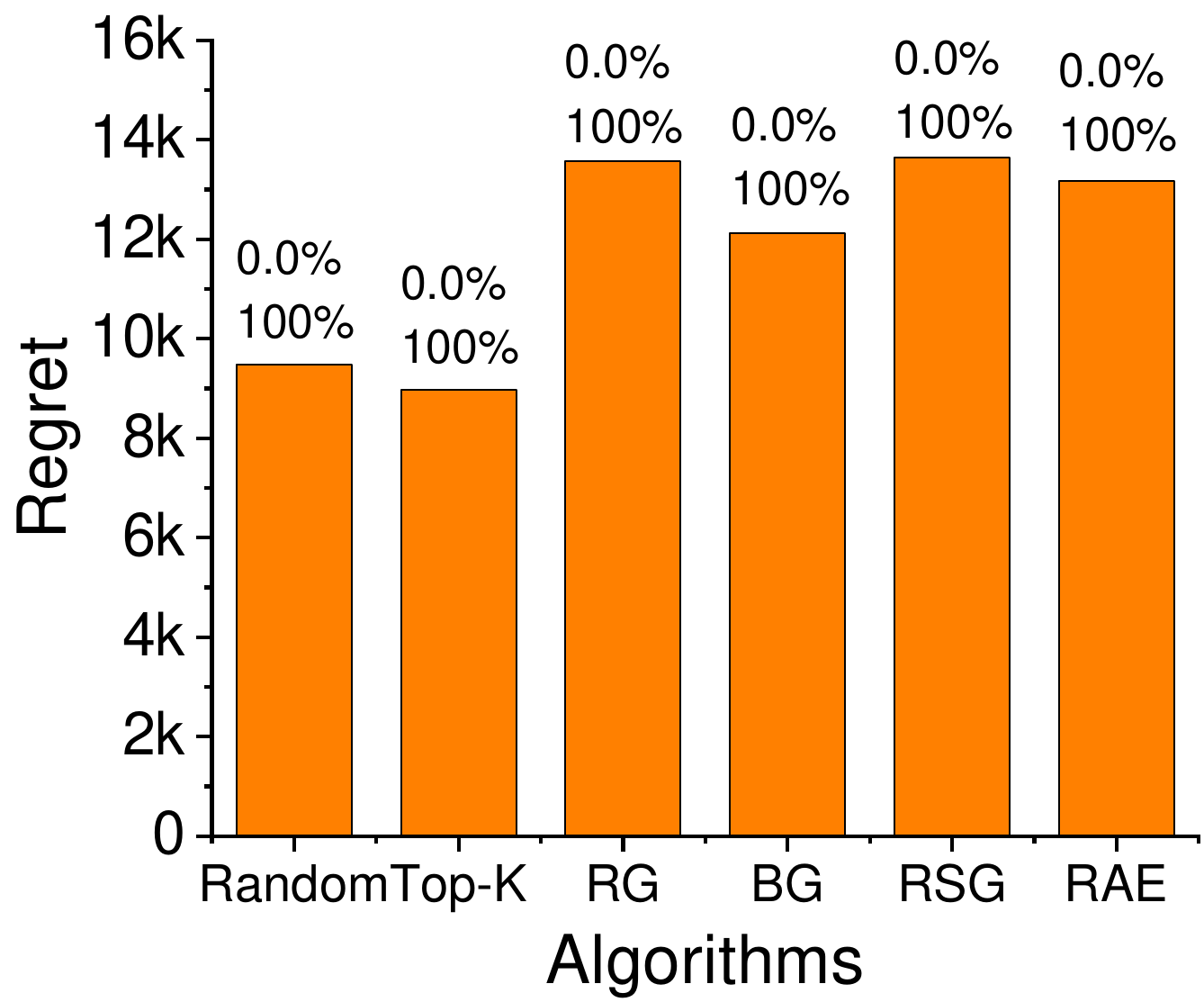} &
        \includegraphics[width=0.17\linewidth]{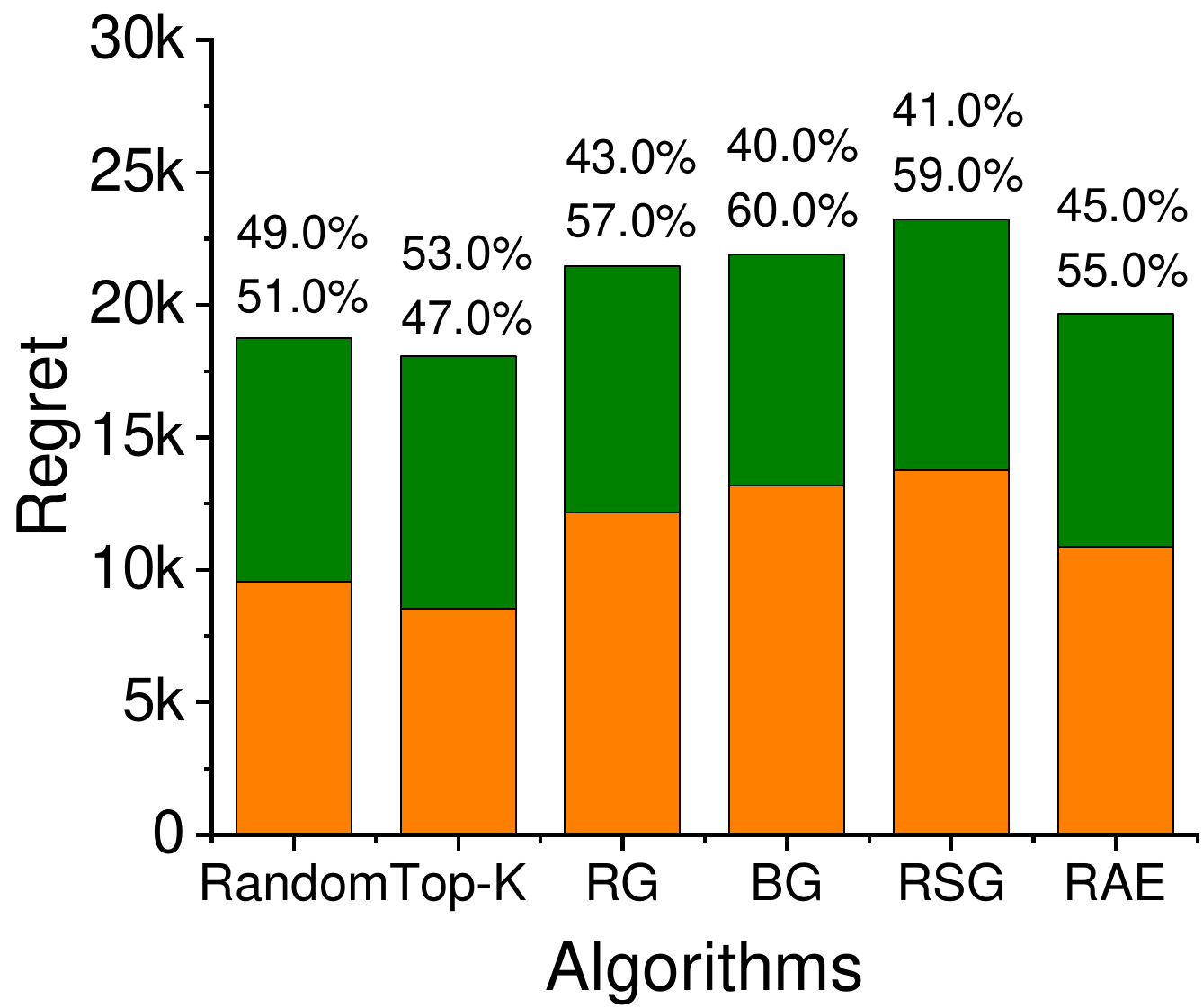} &
        \includegraphics[width=0.17\linewidth]{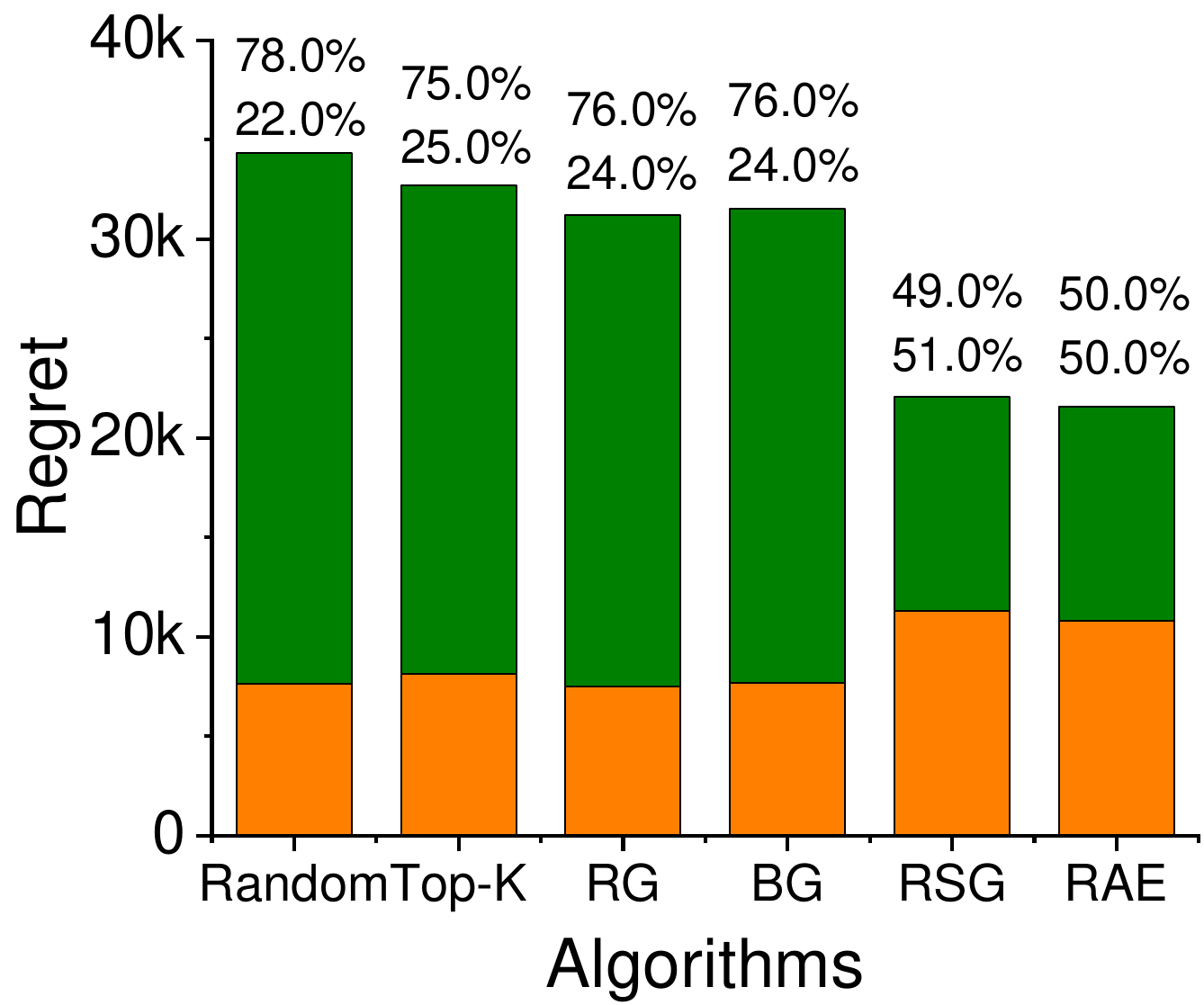} \\
        {\tiny (k) $\delta = 40 \%$} &
        {\tiny (l) $\delta= 60 \%$} &
        {\tiny (m) $\delta = 80 \%$} &
        {\tiny (n) $\delta = 100 \%$} &
        {\tiny (o) $\delta= 120 \%$} \\[5pt]
        
         \includegraphics[width=0.17\linewidth]{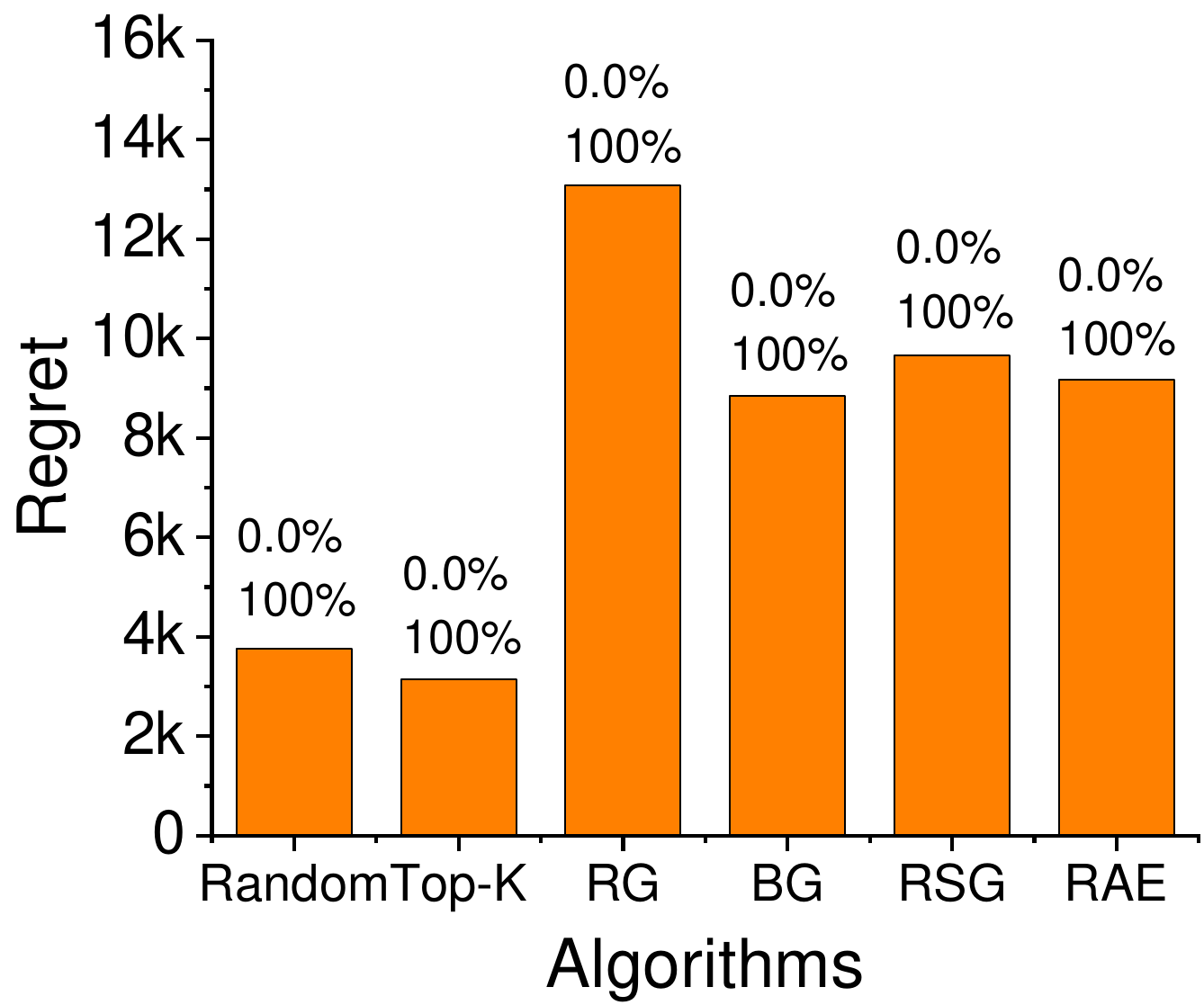} &
        \includegraphics[width=0.17\linewidth]{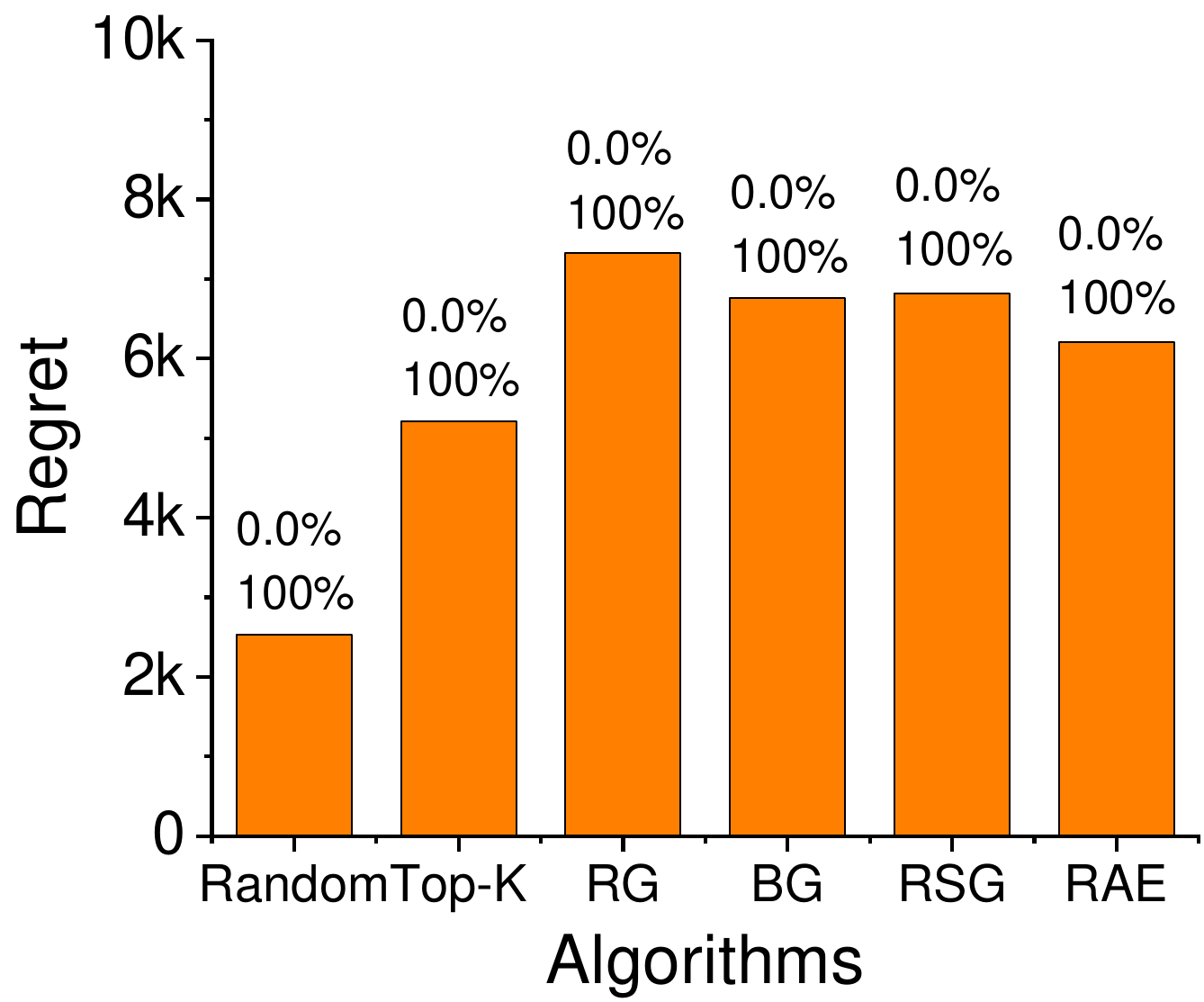} &
        \includegraphics[width=0.17\linewidth]{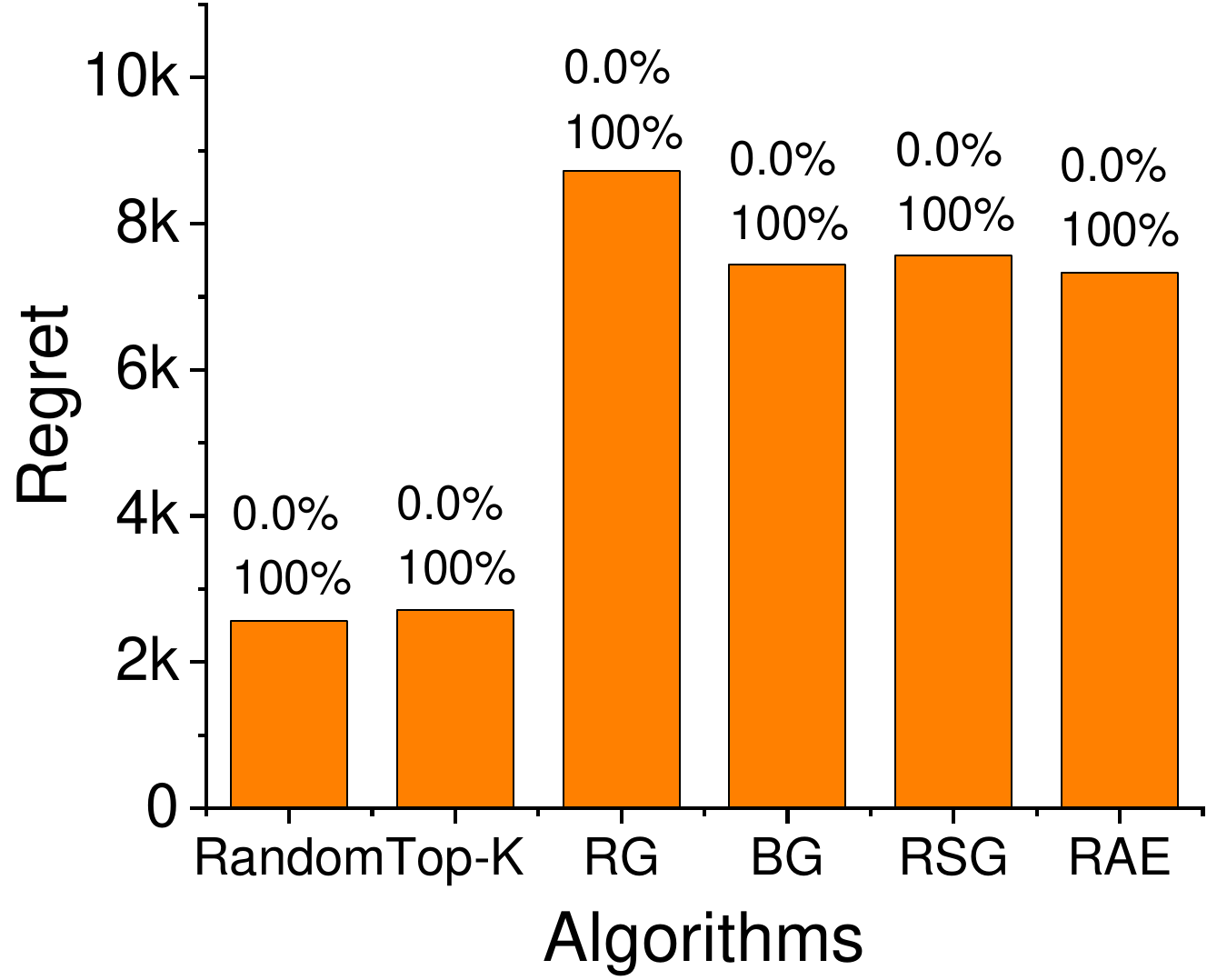} &
        \includegraphics[width=0.17\linewidth]{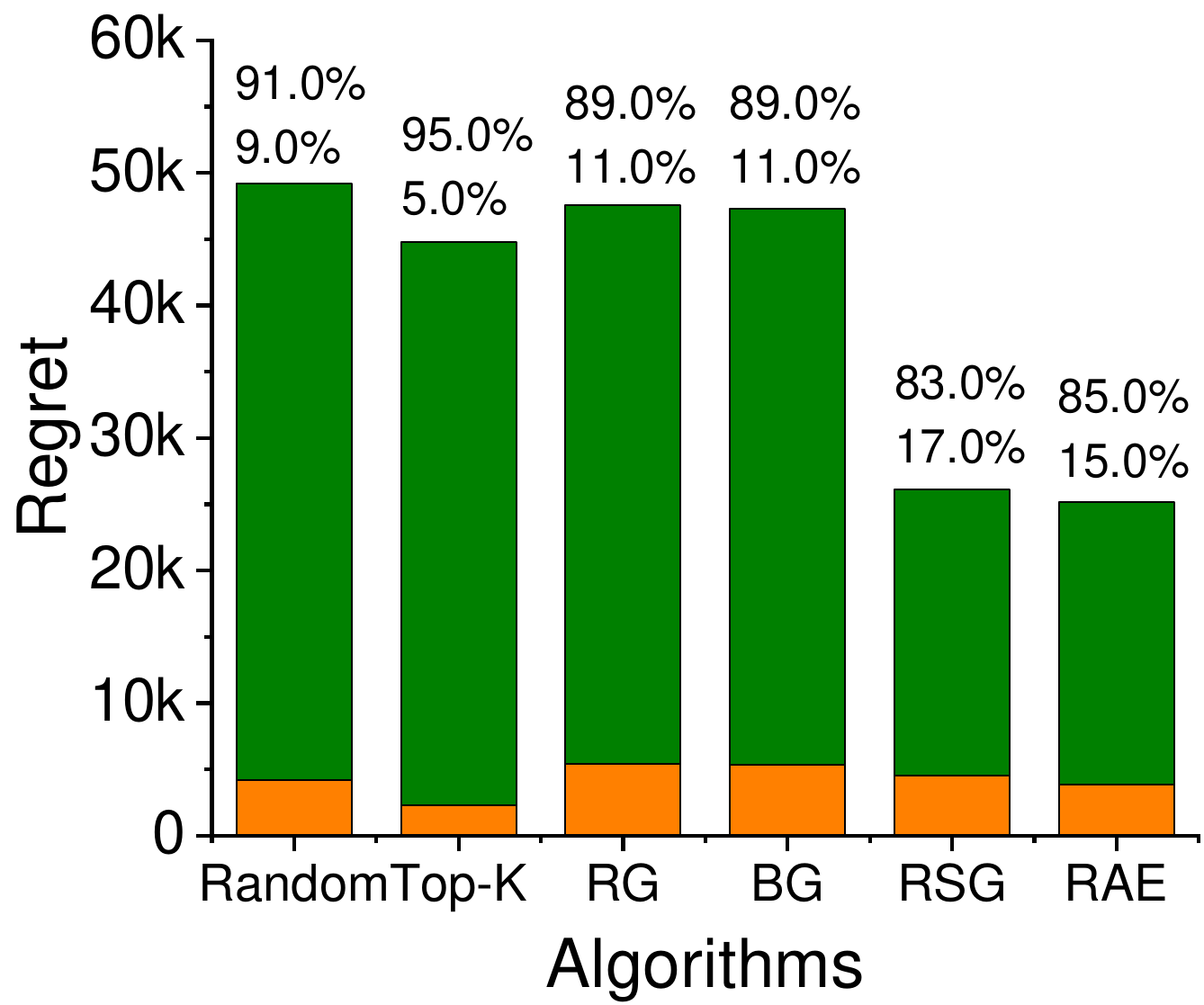} &
        \includegraphics[width=0.17\linewidth]{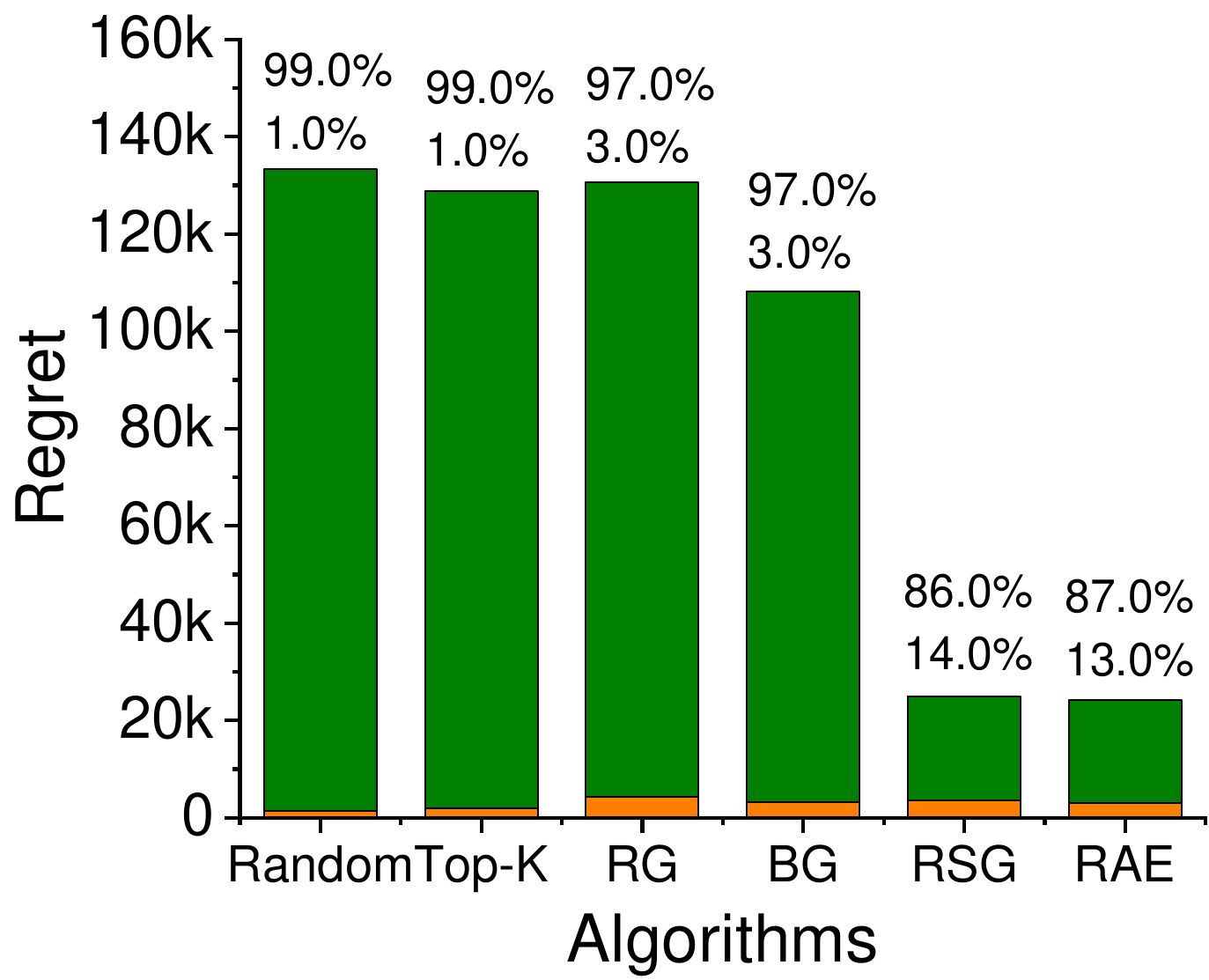} \\
        {\tiny (p) $\delta = 40 \%$} &
        {\tiny (q) $\delta= 60 \%$} &
        {\tiny (r) $\delta = 80 \%$} &
        {\tiny (s) $\delta = 100 \%$} &
        {\tiny (t) $\delta= 120 \%$} \\[5pt]
        
         \includegraphics[width=0.17\linewidth]{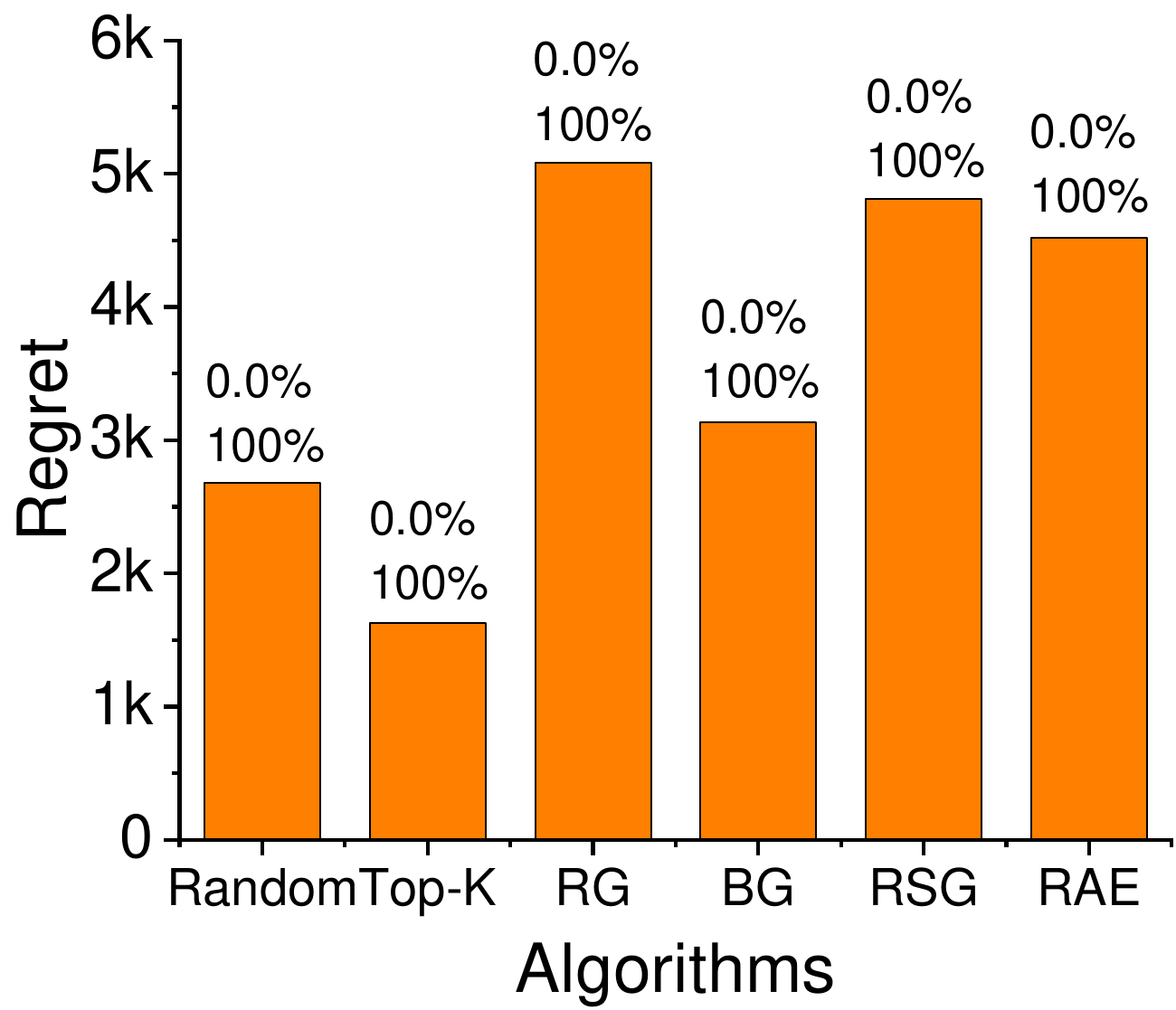} &
        \includegraphics[width=0.17\linewidth]{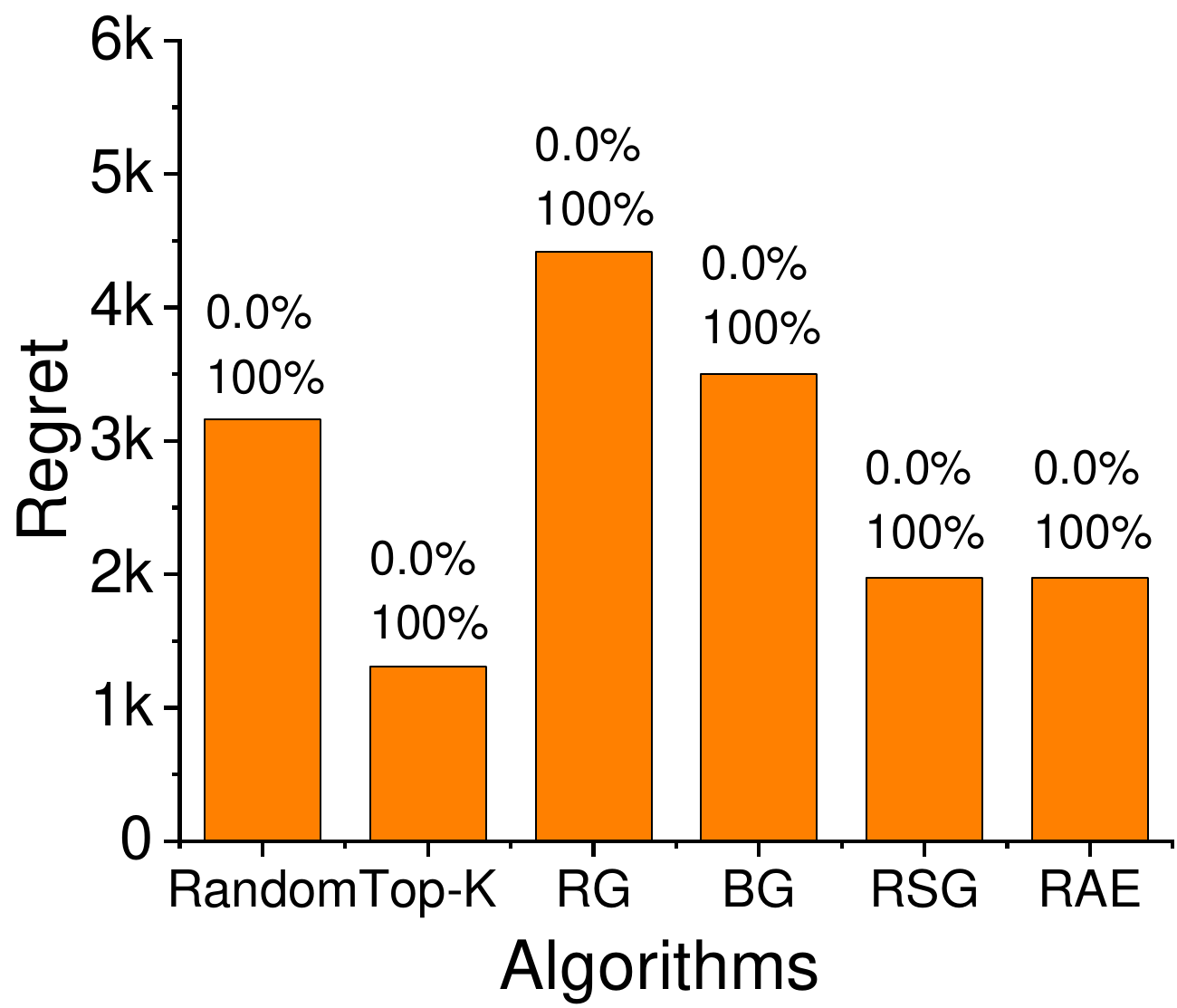} &
        \includegraphics[width=0.17\linewidth]{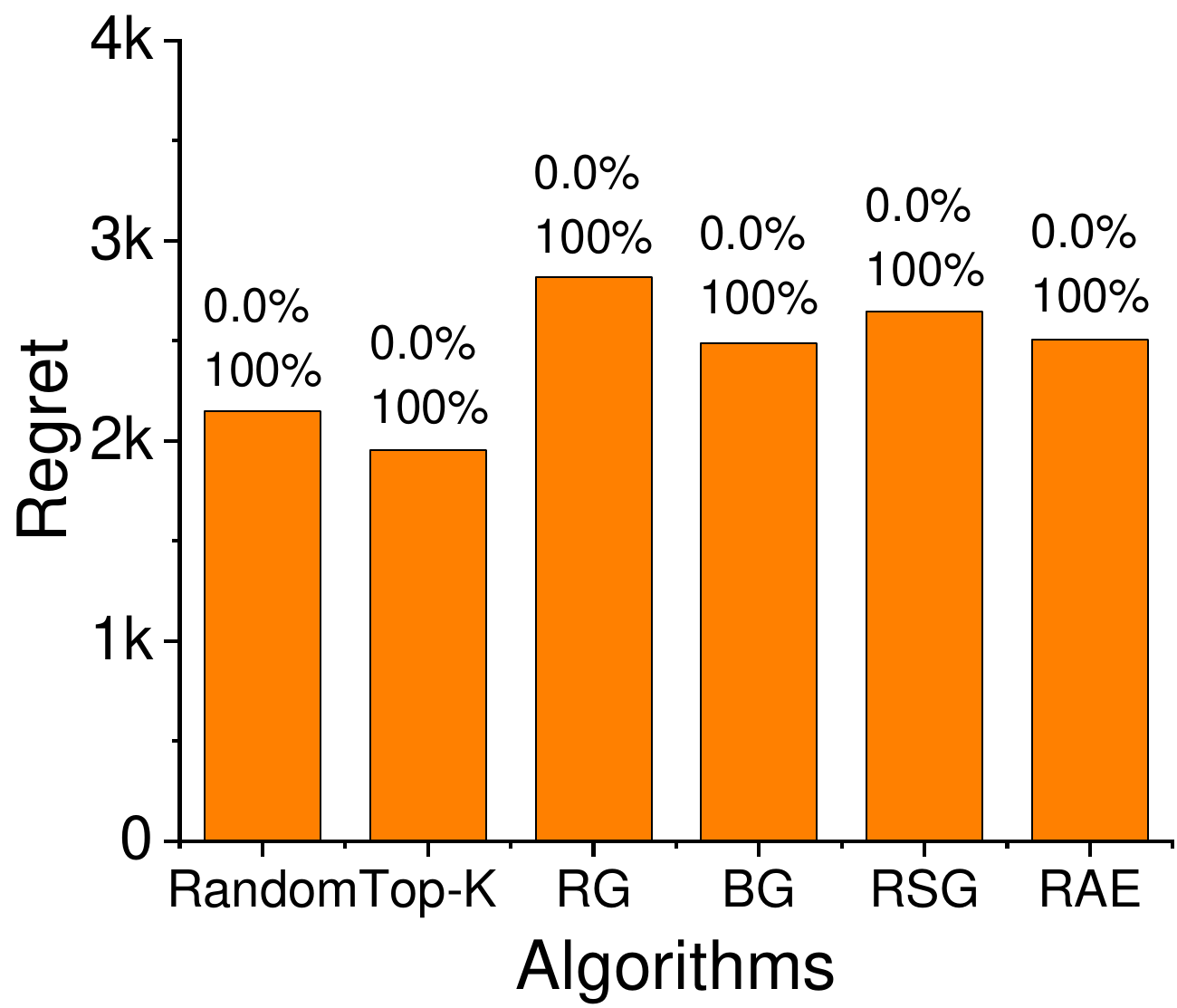} &
        \includegraphics[width=0.17\linewidth]{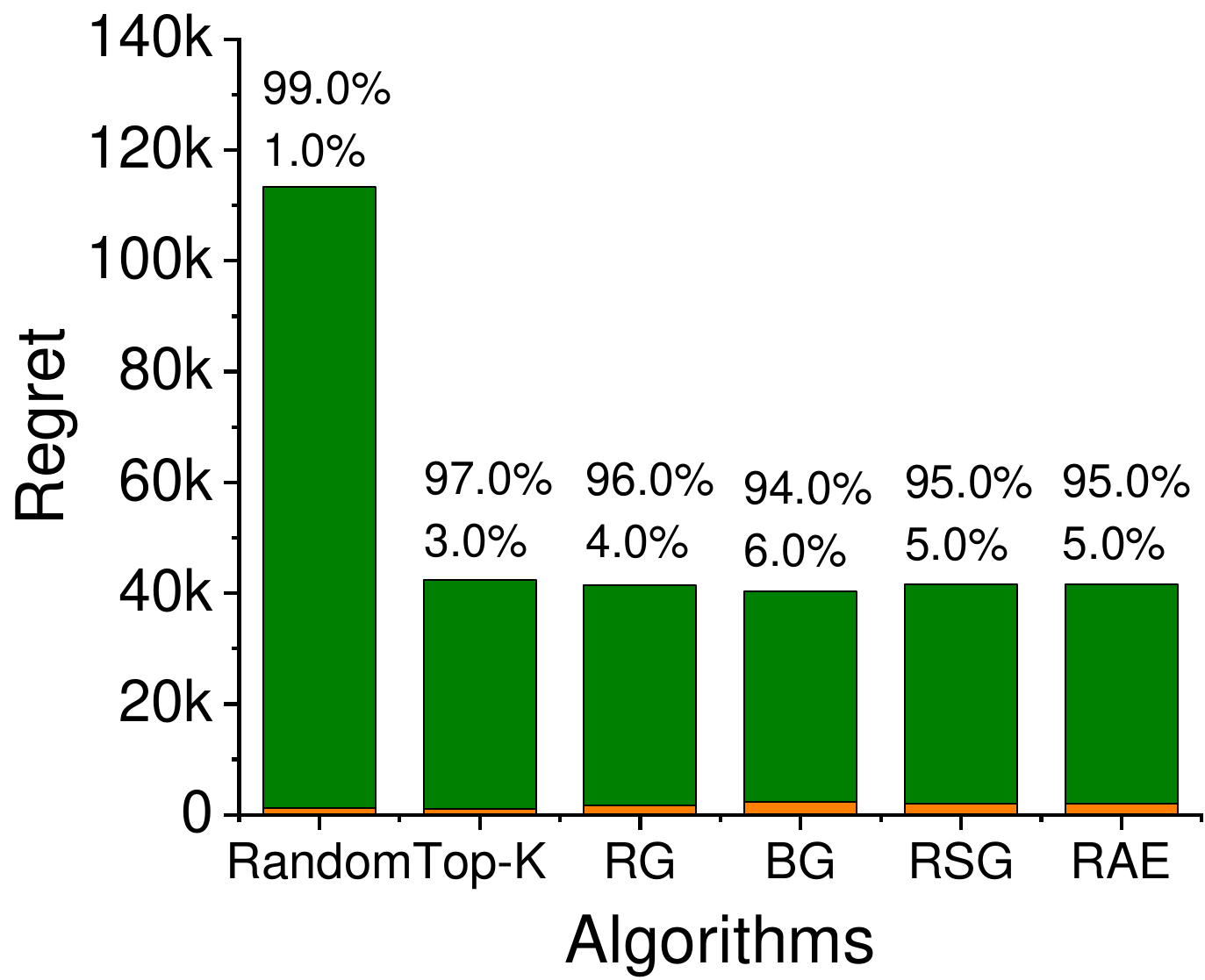} &
        \includegraphics[width=0.17\linewidth]{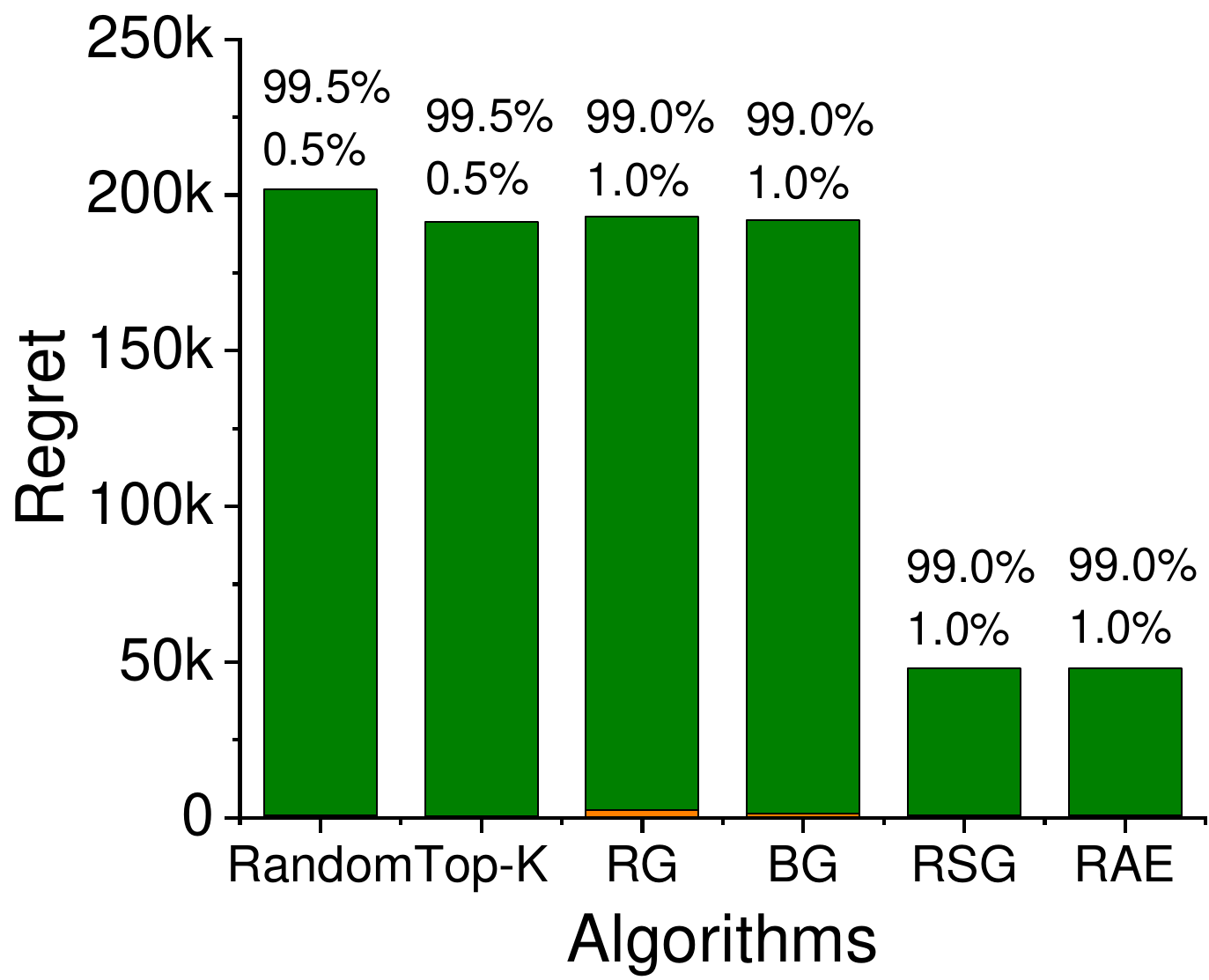} \\
        {\tiny (u) $\delta = 40 \%$} &
        {\tiny (v) $\delta= 60 \%$} &
        {\tiny (w) $\delta = 80 \%$} &
        {\tiny (x) $\delta = 100 \%$} &
        {\tiny (y) $\delta= 120 \%$} \\[5pt]
    \end{tabular}
   \caption{Regret varying $\delta$, when $\lambda = 1\%, \mathcal{|A|} = 100$ $(a, b, c, d, e)$, when $\lambda = 2\%, \mathcal{|A|} = 50$ $(f, g, h, i, j),$ when $\lambda = 5\%, \mathcal{|A|} = 20$ $(k,l,m, n,o)$, when $\lambda = 10\%, \mathcal{|A|} = 10$ $(p, q, r, s, t)$, when $\lambda = 20\%, \mathcal{|A|} = 5$ $(u, v, w, x, y),$ for LA dataset }
    \label{Fig:LA_Result}
\end{figure}

\textbf{ Case 1: $\delta \leq 80\%,$ $\lambda \leq 5\%$ (parts $(a,b,c,f,g,h,k,\ell,m)$ of Figure \ref{Fig:LA_Result}.}
Corresponding to case $1$, we have low $\lambda$, low $\delta$, and this arises a situation where both the global and individual influence demand is low. The influence provider has a large number of advertisers with low individual influence demand. we have three main observations. First, no unsatisfied regret exists because every zone has sufficient billboard slots to allocate. When $\delta$ increases, the excessive regret for all the algorithms drops.  Second, among baseline methods, `Top-$k$' controls excessive regret better. This is because `Top-$k$' allocates the most influential billboard slots individually to the advertisers. So, it is easy to fulfill the influence demand of all the advertisers with fewer billboard slots. Now, among the proposed approach the `RAE' performs better than the `RG', and `RSG' methods. Third, all the advertisers are satisfied in a scenario in which $\delta = 40\%$ and $\lambda = 1\%$. The `Top-$k$' and `Random' outperform the proposed approaches because they control excessive regret better.

\textbf{ Case 2: $\delta \leq 80\%,$, $\lambda \geq 10\%$ (parts $(p,q,r,u,v,w)$ of Figure \ref{Fig:LA_Result}.} Corresponding to case $2$, we have low $\delta$ and high $\lambda$, representing a situation where influence provider have fewer advertisers with high individual influence demand. We have two observations. First, although higher individual influence demand, all the advertisers are satisfied. This is because when $\lambda$ increases, the number of advertisers becomes less, and influence provider have to satisfy fewer advertisers. Second, total regret decreases when the $\delta$ value increases from $40\%$ to $80\%$ for all the proposed approaches. As advertisers have higher individual influence demand, the influence provider can allocate more billboard slots to each. So, the `RAE' has a more extensive search space to exchange billboard slots between each advertiser. Consequently, `RAE' outperforms all the proposed approaches, although `Random' and `Top-$k$' still perform well.

\par

\textbf{ Case 3: $\delta \geq 100\%,$, $\lambda \leq 5\%$ (parts $(d,e,i,j,n,o)$ of Figure \ref{Fig:LA_Result}.} This represents a situation where global influence demand is high and individual influence demand is low. The influence provider has a larger advertiser base in which advertisers have a small individual influence demand. We have two main observations. First, with a high $\delta$ value $100\%$ to $120\%$, the advertiser's demand equals or exceeds the influence provider supply. So, the proposed approaches can only satisfy some advertisers as the excessive regret cannot be diminished. When $\delta = 120\%$, none of the algorithms can satisfy all the advertisers as their influence demand is much less than the influence supply by the influence provider. Second, the `RSG' and `RAE' impact is much better than other proposed and baseline methods in reducing total regret. It is important to highlight that the deployment of slots to the advertisers when the $\delta$ value is very high. However, there is a risk of losing more advertisers as global influence demand is very high compared to the influence provider supply. The `RSG' and `RAE' handle the risk best. They averagely outperform the `RG' and `BG' by four times and two times, respectively.

\textbf{ Case 4: $\delta \geq 100\%,$, $\lambda \geq 10\%$ (parts $(s,t,x,y)$ of Figure \ref{Fig:LA_Result}.} According to case $4$, we have a situation where both the global and individual influence demand is very high. Hence, the influence provider has fewer advertisers with higher individual influence demand. We have two main observations. First, higher $\delta$ and $\lambda$ will lead to higher unsatisfied regret to the influence provider. Hence, all the algorithms suffer from higher regret, and no one can satisfy all the advertisers. The influence provider sacrifices more advertisers due to a low influence supply. Second, when $\lambda$ increases from $10\%$ to $20\%$, the excessive regret decreases while the total regret rises due to higher unsatisfied regret. Here, one must highlight that the `BG' leads to much higher unsatisfied regret than other proposed approaches. The `RAE' and `RSG' still perform better but sacrifice more advertisers to minimize unsatisfied regret. They outperform the `RG' and `BG' by about four times and three times, respectively.

\paragraph{\textbf{Revisit $RQ1$, $RQ2$, and $RQ3$.}}
After analyzing four NYC and LA datasets cases, we can answer RQ1, RQ2, and RQ3. (1) As previously discussed, excessive regret dominates total regret when the global demand is deficient. So, it is difficult to select advertisers for deploying billboard slots to minimize the total regret. Based on our observations, when the advertiser's demand is equal or at most two times larger than the average influence of billboard slots, we experience higher excessive regret. However, with the increment of advertiser's demand by five times or more, the excessive regret decreases, as shown in Figure \ref{Fig:NYC_Result},\ref{Fig:LA_Result} $(a,b,c)$. (2) Increasing global demand reduces excessive influence, but at the same time, it will lead to unsatisfied penalties (e.g., Figure \ref{Fig:NYC_Result},\ref{Fig:LA_Result} $(d,e,i,j)$). We observe that a large number of small or medium individual-demanded advertisers are ideal for influence providers. These advertisers are more flexible in allocating billboard slots and will return the minimum amount of unsatisfied regret to the influence provider. (3) To address RQ3, we vary $\delta$ and $\lambda$ from $40\%$ to $120\%$ and $1\%$ to $20\%$. Corresponding observations are already discussed in four cases, as shown in Figures \ref{Fig:NYC_Result} and \ref{Fig:LA_Result}.

\subsection{Efficiency Study}\label{ES}
\begin{figure}[h]
    \centering
    \begin{tabular}{cccccccccccc}
       \small{Random} & \includegraphics[width=0.06\linewidth]{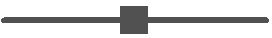} & Top-$k$ & \includegraphics[width=0.06\linewidth]{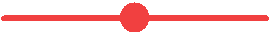} & RG & \includegraphics[width=0.06\linewidth]{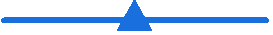} & BG & \includegraphics[width=0.06\linewidth]{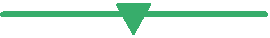} RSG & \includegraphics[width=0.06\linewidth]{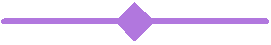} RAE & \includegraphics[width=0.06\linewidth]{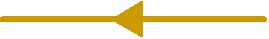}\\
    \end{tabular}
    \begin{tabular}{ccccc}     
        \includegraphics[width=0.17\linewidth]{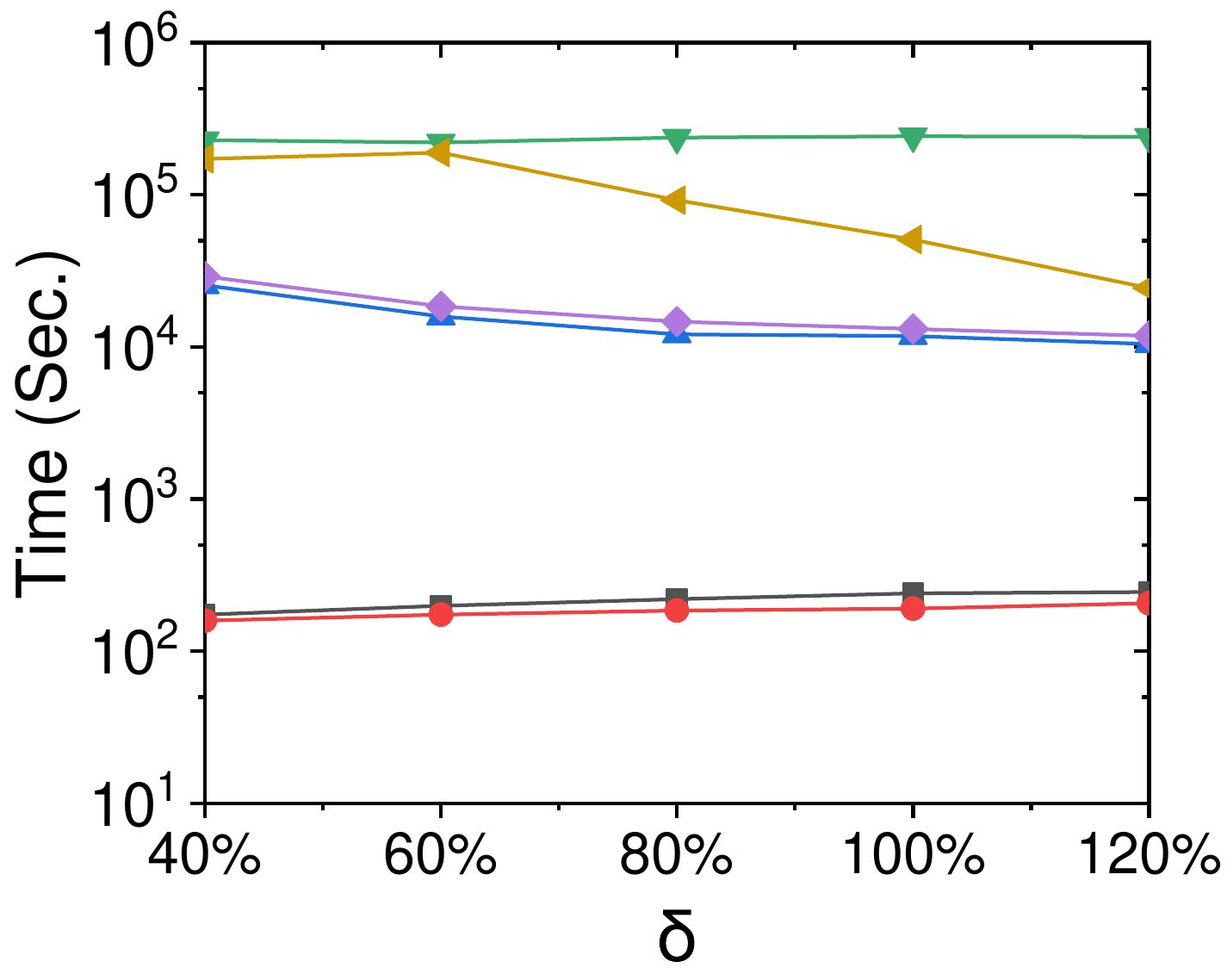} &
        \includegraphics[width=0.17\linewidth]{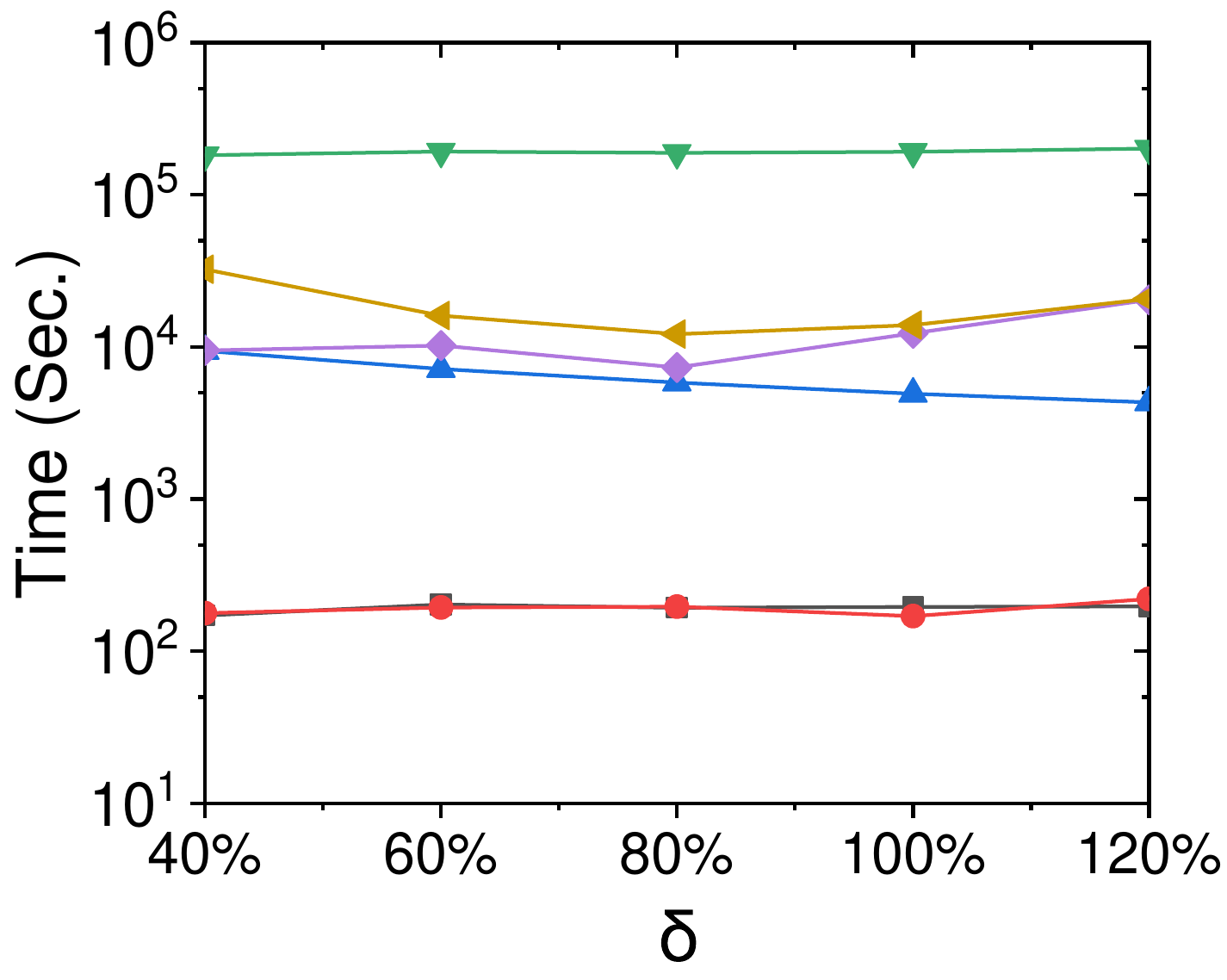} &
        \includegraphics[width=0.17\linewidth]{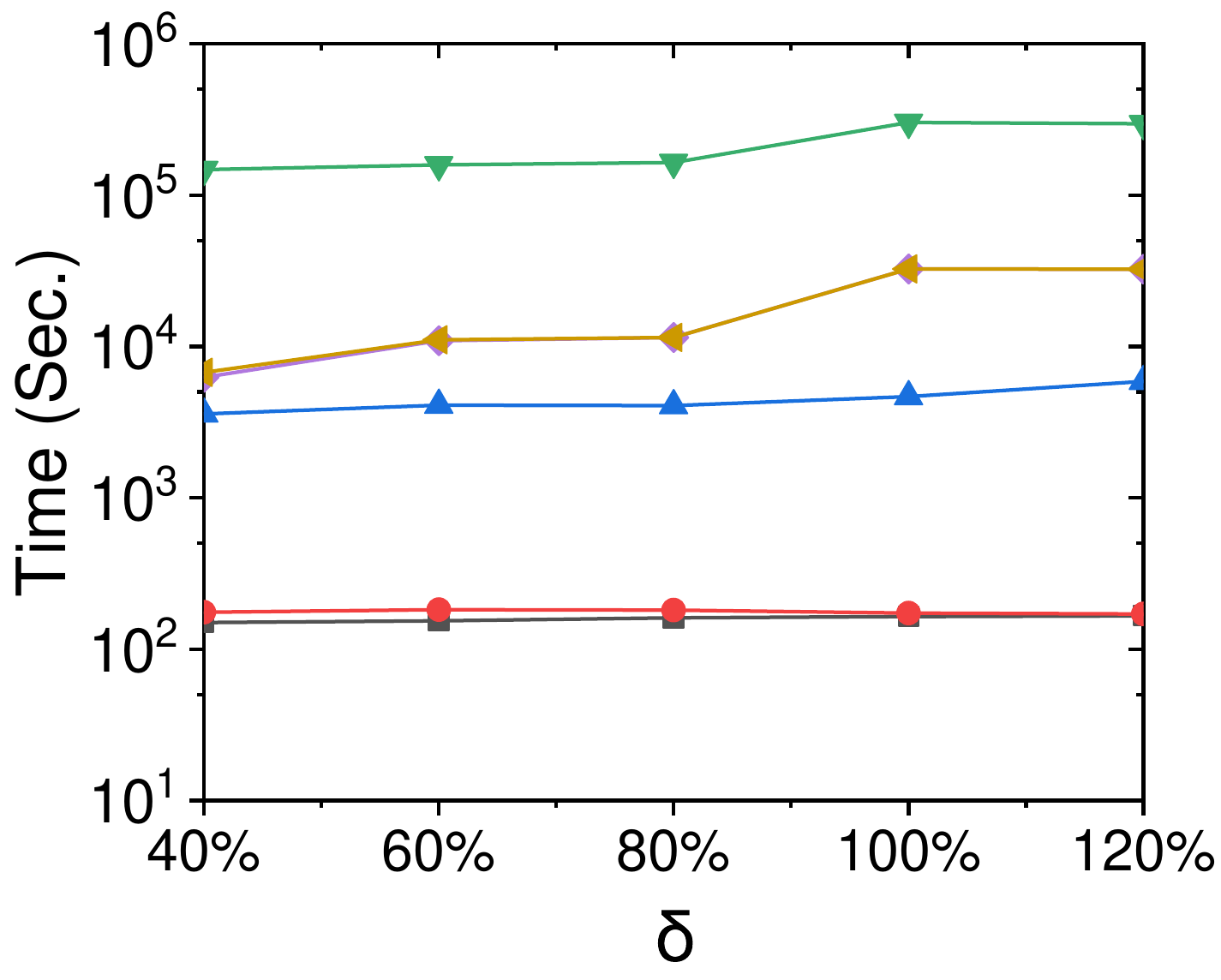} &
        \includegraphics[width=0.17\linewidth]{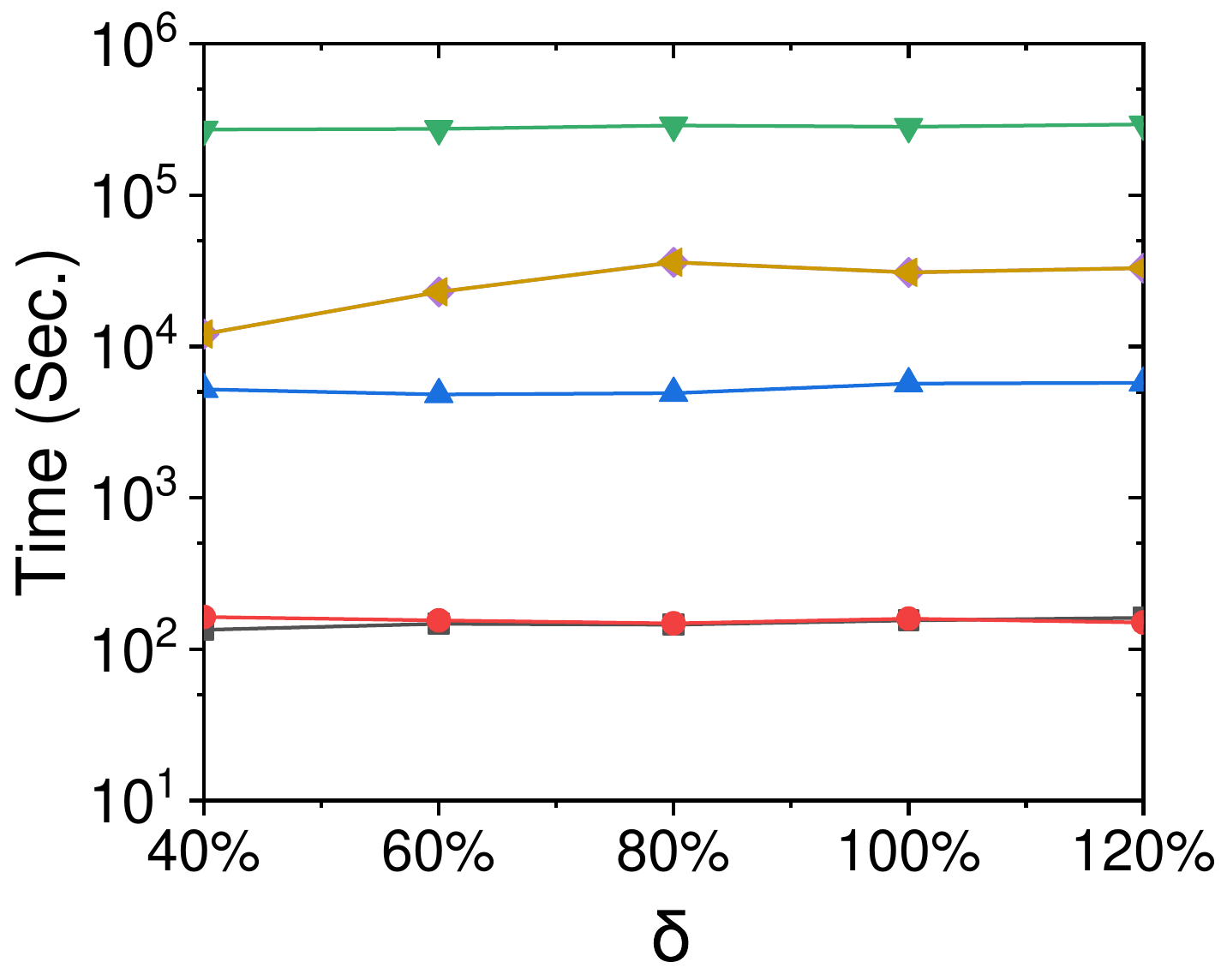} &
        \includegraphics[width=0.17\linewidth]{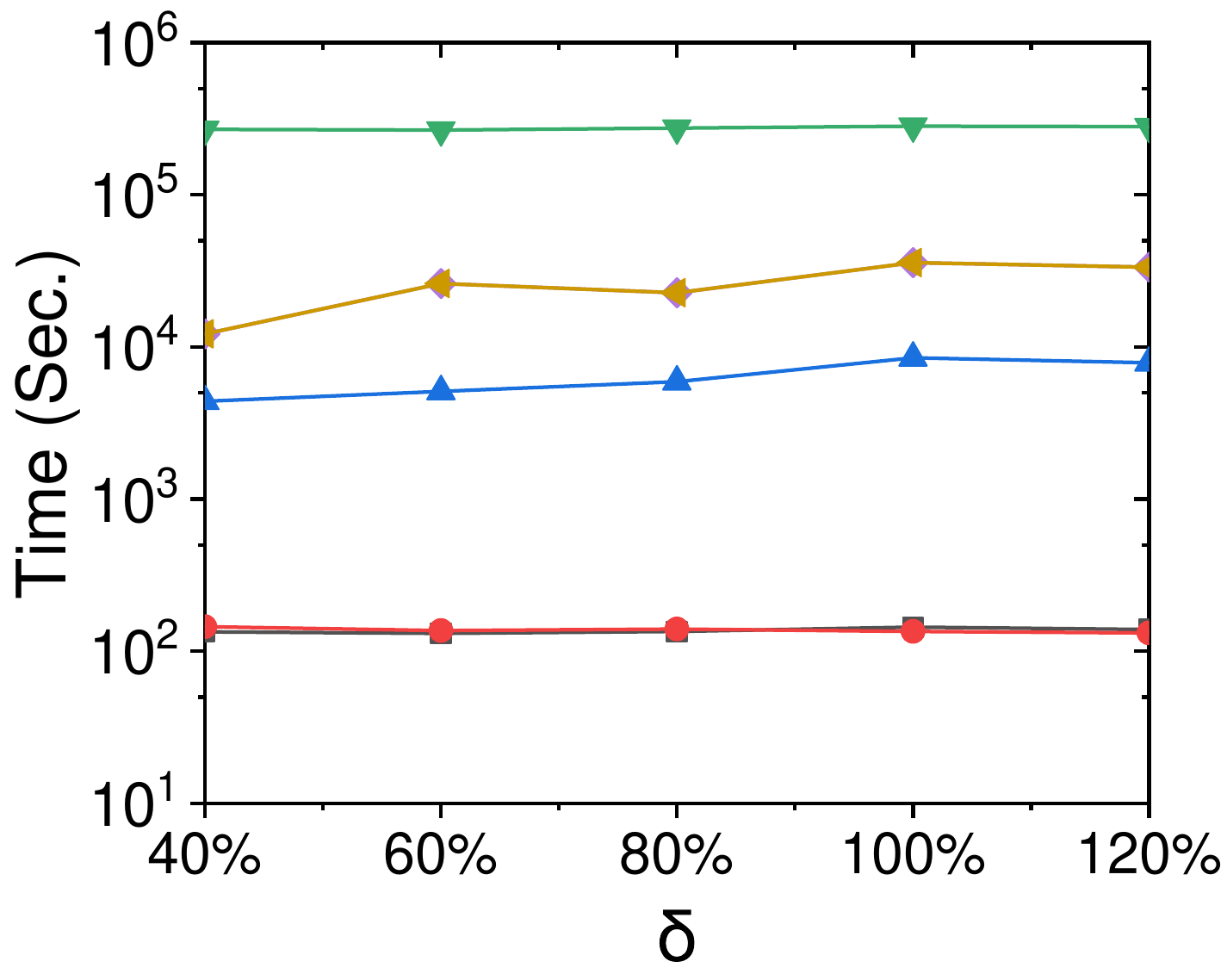} \\
        {\tiny (a) $\lambda = 1 \%$, $|\mathcal{A}| = 100$} &
        {\tiny (b) $\lambda = 2 \%$, $|\mathcal{A}| = 50$} &
        {\tiny (c) $\lambda = 5 \%$, $|\mathcal{A}| = 20$} &
        {\tiny (d) $\lambda = 10 \%$, $|\mathcal{A}| = 10$} &
        {\tiny (e) $\lambda = 20 \%$, $|\mathcal{A}| = 5$} \\[5pt]
    \end{tabular}
    \caption{Efficiency Study on NYC}
    \label{Fig:NYC_TIME}
\end{figure}

\begin{figure}[h]
    \centering
    \begin{tabular}{cccccccccccc}
       \small{Random} & \includegraphics[width=0.06\linewidth]{Symbol/Random.png} & Top-$k$ & \includegraphics[width=0.06\linewidth]{Symbol/Top-k.png} & RG & \includegraphics[width=0.06\linewidth]{Symbol/RG.png} & BG & \includegraphics[width=0.06\linewidth]{Symbol/BG.png} RSG & \includegraphics[width=0.06\linewidth]{Symbol/RSG.png} RAE & \includegraphics[width=0.06\linewidth]{Symbol/RAE.png}\\
    \end{tabular}
    \begin{tabular}{ccccc}     
        \includegraphics[width=0.17\linewidth]{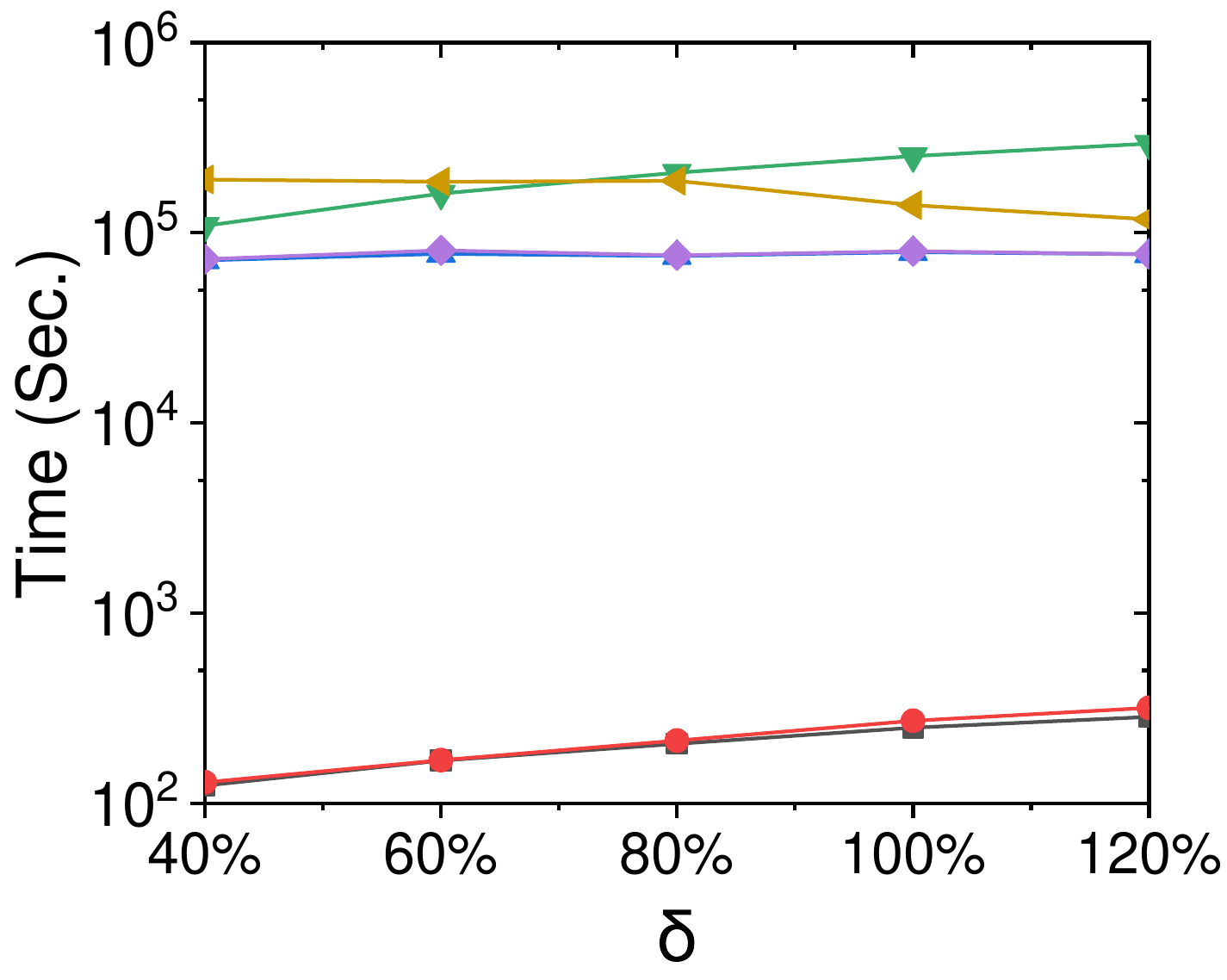} &
        \includegraphics[width=0.17\linewidth]{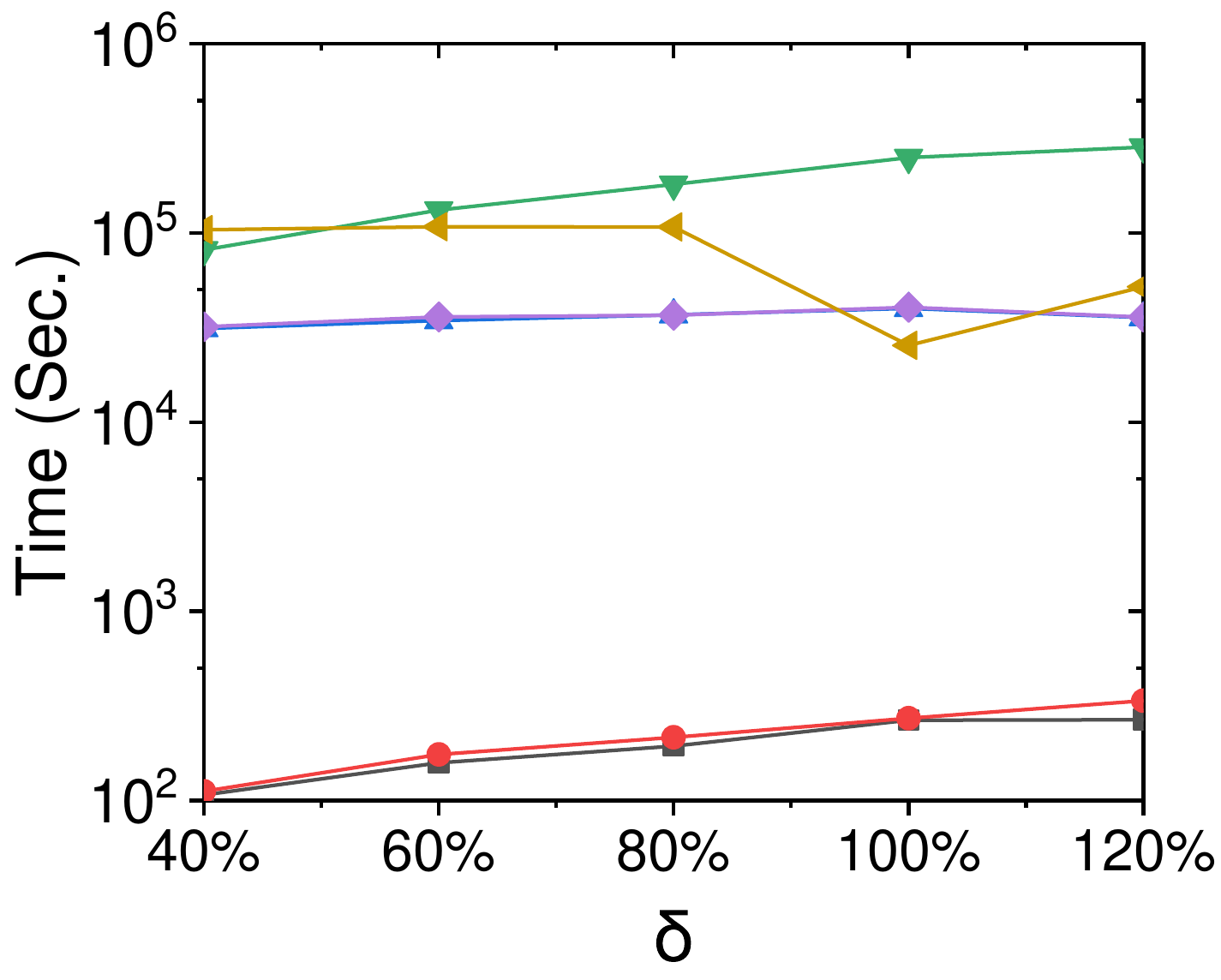} &
        \includegraphics[width=0.17\linewidth]{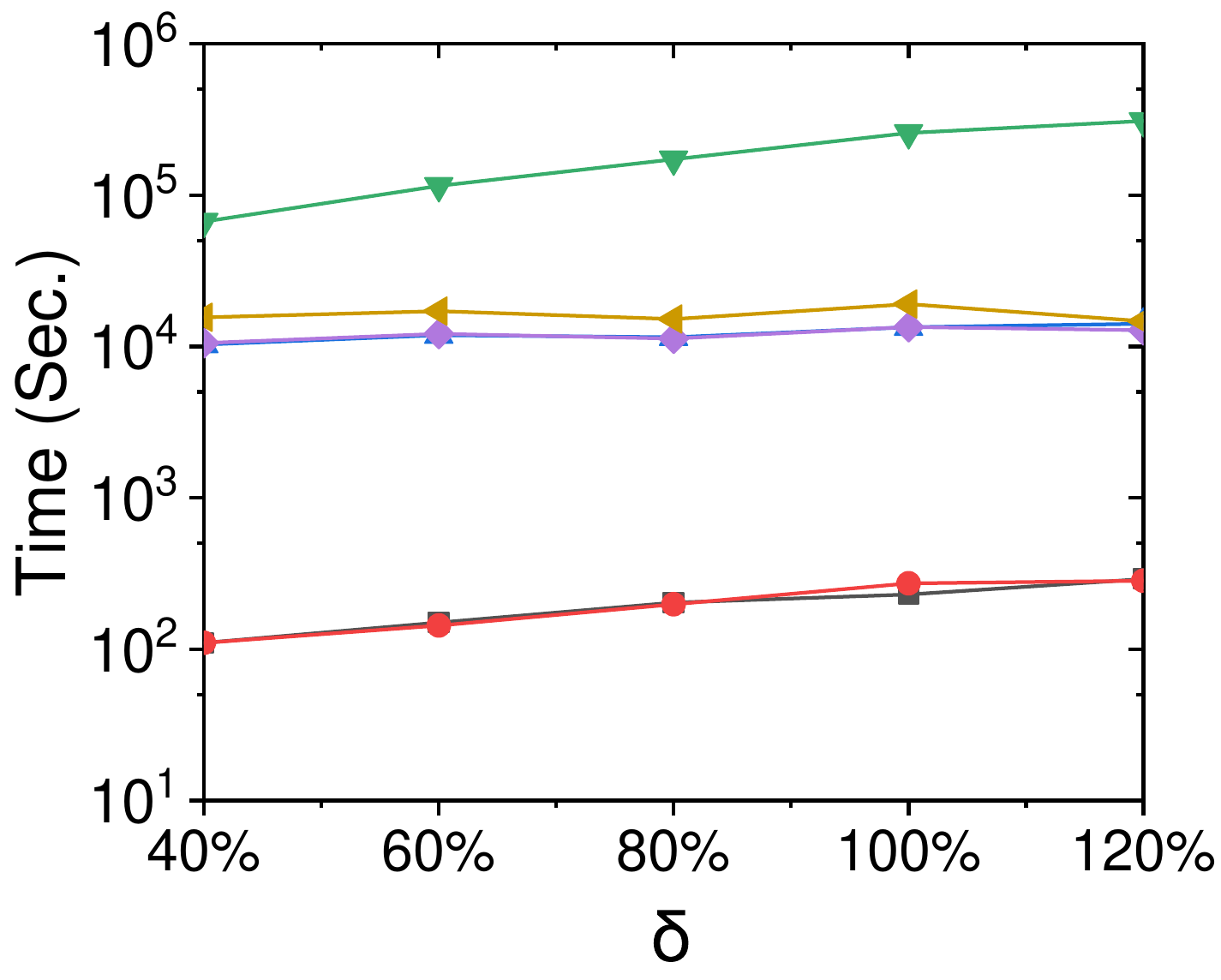} &
        \includegraphics[width=0.17\linewidth]{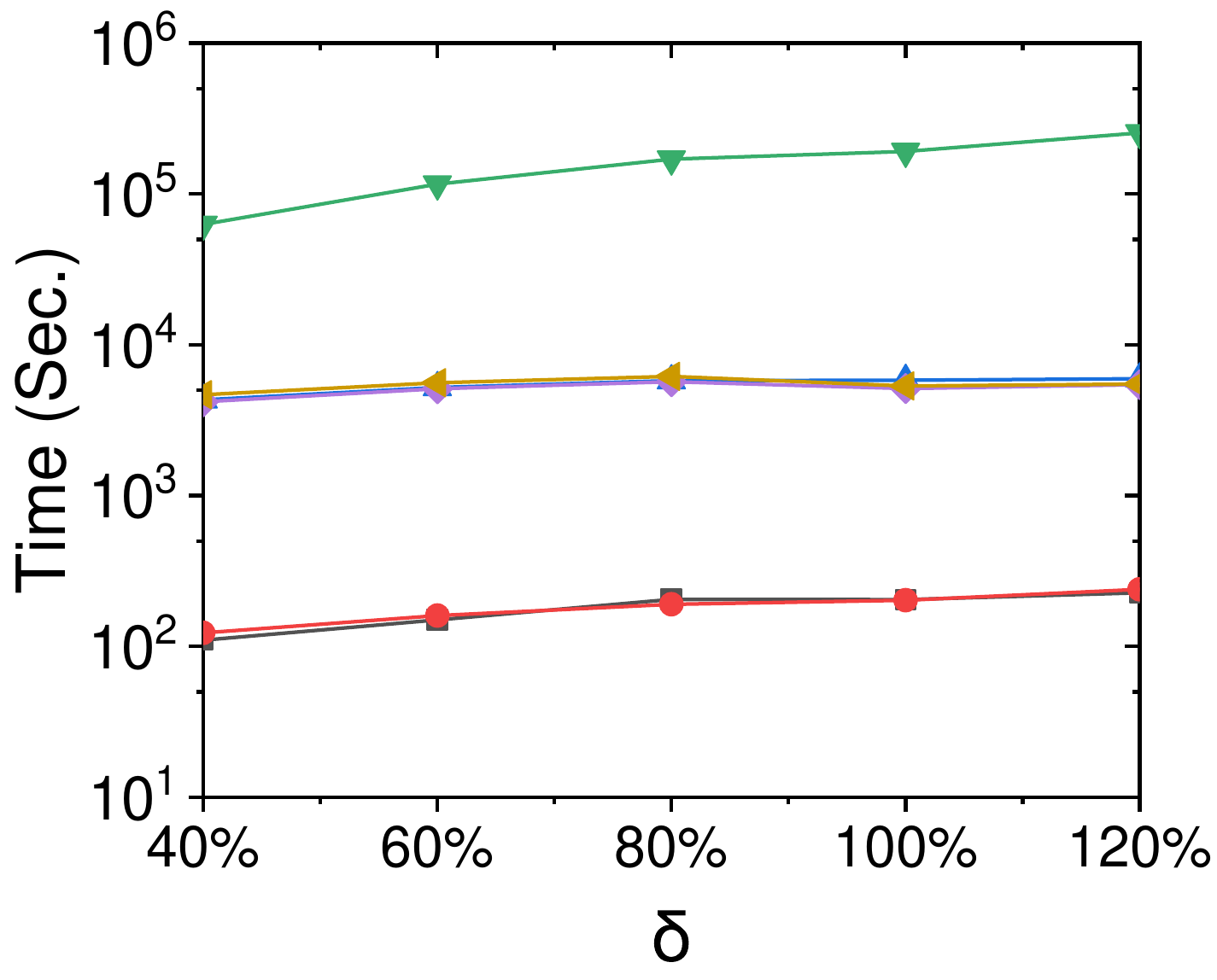} &
        \includegraphics[width=0.17\linewidth]{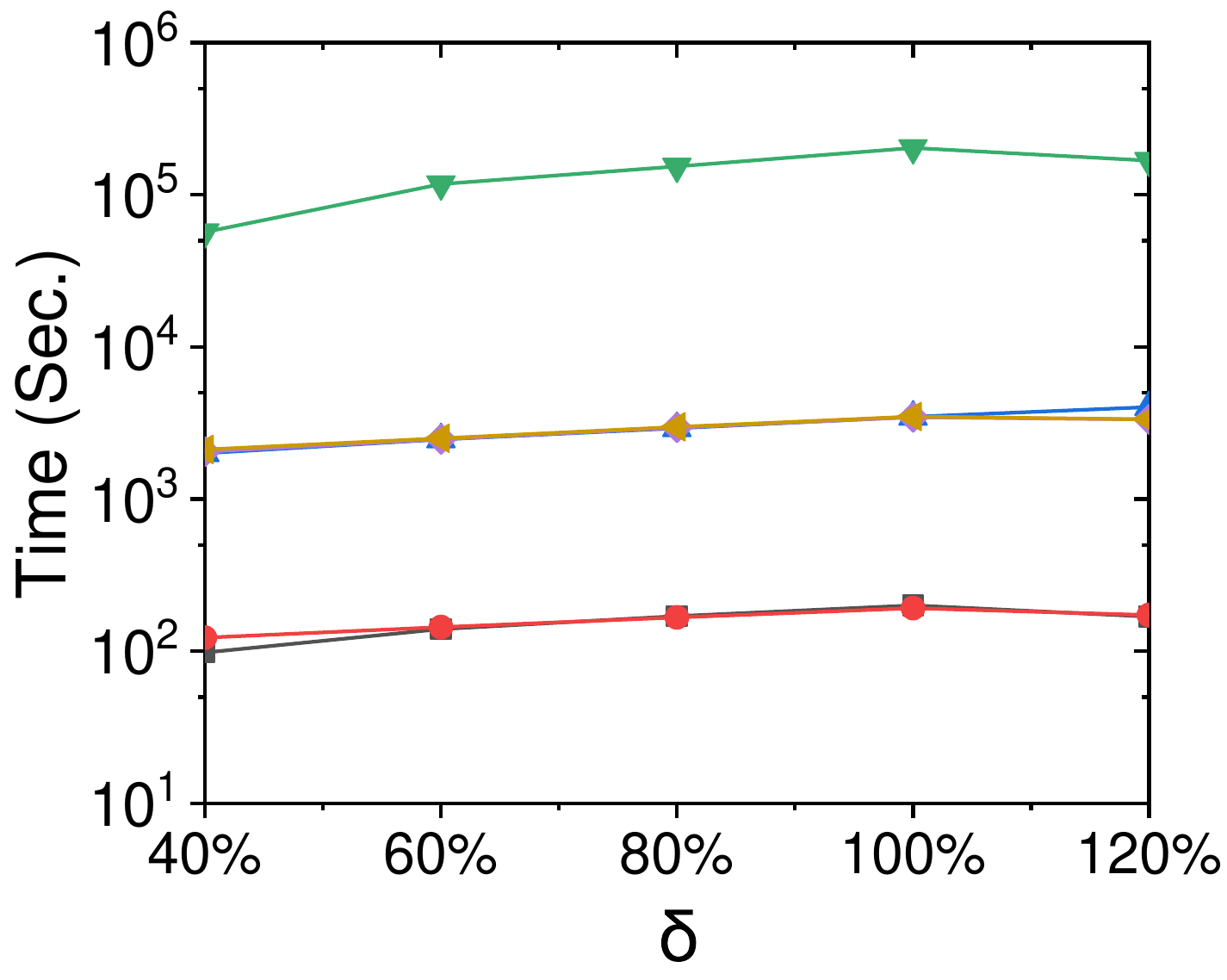} \\
        {\tiny (a) $\lambda = 1 \%$, $|\mathcal{A}| = 100$} &
        {\tiny (b) $\lambda = 2 \%$, $|\mathcal{A}| = 50$} &
        {\tiny (c) $\lambda = 5 \%$, $|\mathcal{A}| = 20$} &
        {\tiny (d) $\lambda = 10 \%$, $|\mathcal{A}| = 10$} &
        {\tiny (e) $\lambda = 20 \%$, $|\mathcal{A}| = 5$} \\[5pt]
    \end{tabular}
    \caption{Efficiency Study on LA}
    \label{Fig:LA_TIME}
\end{figure}

Efficiency is important in such cases where more than a thousand advertisers daily come to the influence provider with their zone-specific influence requirements, and the influence provider needs to handle more than a thousand billboards in a city like New York and Los Angeles. We have conducted experiments over different $\delta$ and $\lambda$ values for both NYC and LA datasets, as reported in Figures \ref{Fig:NYC_TIME} and \ref{Fig:LA_TIME}. We have a couple of observations.

First, the `RG' takes less time than other proposed approaches to deploy billboard slots to advertisers. The `RSG' and `RAE' approaches take more time than the `RG' because they use the allocation provided by the `RG' as their initial allocation and further try to improve the effectiveness by releasing advertisers or exchanging billboard slots among advertisers. The `BG' takes more time than the `RG' because, in each iteration, it computes the marginal gain for all the billboard slots, while in the `RG', each iteration only computes the marginal gain for a randomly chosen subset of slots.

Second, when the $\delta$ value increases from $40\%$ to $120\%$, all algorithms' computational cost increases except the `RAE'. This is because when $\delta$ rises, the individual influence demand of each advertiser also increases. Then, the influence provider must allocate more billboard slots to the advertisers, producing more extensive search space and taking more time. Additionally, the number of unsatisfied advertisers and unsatisfied regret increases with the rising $\delta$ value. Hence, the `RSG' takes more time to release unsatisfied advertisers and allocate released slots to other unsatisfied ones.

Third, with an increase of $\lambda$ value from $1\%$ to $20\%$, all algorithms' computational cost decreases except the `BG'. When $\lambda$ increases, the number of advertisers decreases, and the influence provider needs to allocate billboard slots to a few advertisers. However, the time requirement for the `BG' is consistent with the increasing $\lambda$ because of the large number of marginal gain computations in each iteration. 

Fourth, the `RAE' approach increases computational time with the $\delta$ increase. However, in some cases, when exchanging billboard slots does not minimize the total regret, or no further improvement occurs during allocation, the algorithm exits, and its computational time is not much increased as represented in Figure \ref{Fig:NYC_TIME} and \ref{Fig:LA_TIME}.

Fifth, One thing that needs to be highlighted is that when the $\delta$ value increases, the computational time difference between different $\delta$ values is huge in the case of the `RAE'. However, this difference is much less in the case of all other baseline and proposed algorithms.

\subsection{Additional Parameters Study}\label{APS}
\begin{figure}[h]
    \centering
    \begin{tabular}{lclc}
       Unsatisfied Regret & \includegraphics[width=0.11\linewidth]{Result/Unsatisfied.png} & Excessive Regret & \includegraphics[width=0.11\linewidth]{Result/Excessive.png} \\
    \end{tabular}
    \begin{tabular}{ccccc}     
        \includegraphics[width=0.17\linewidth]{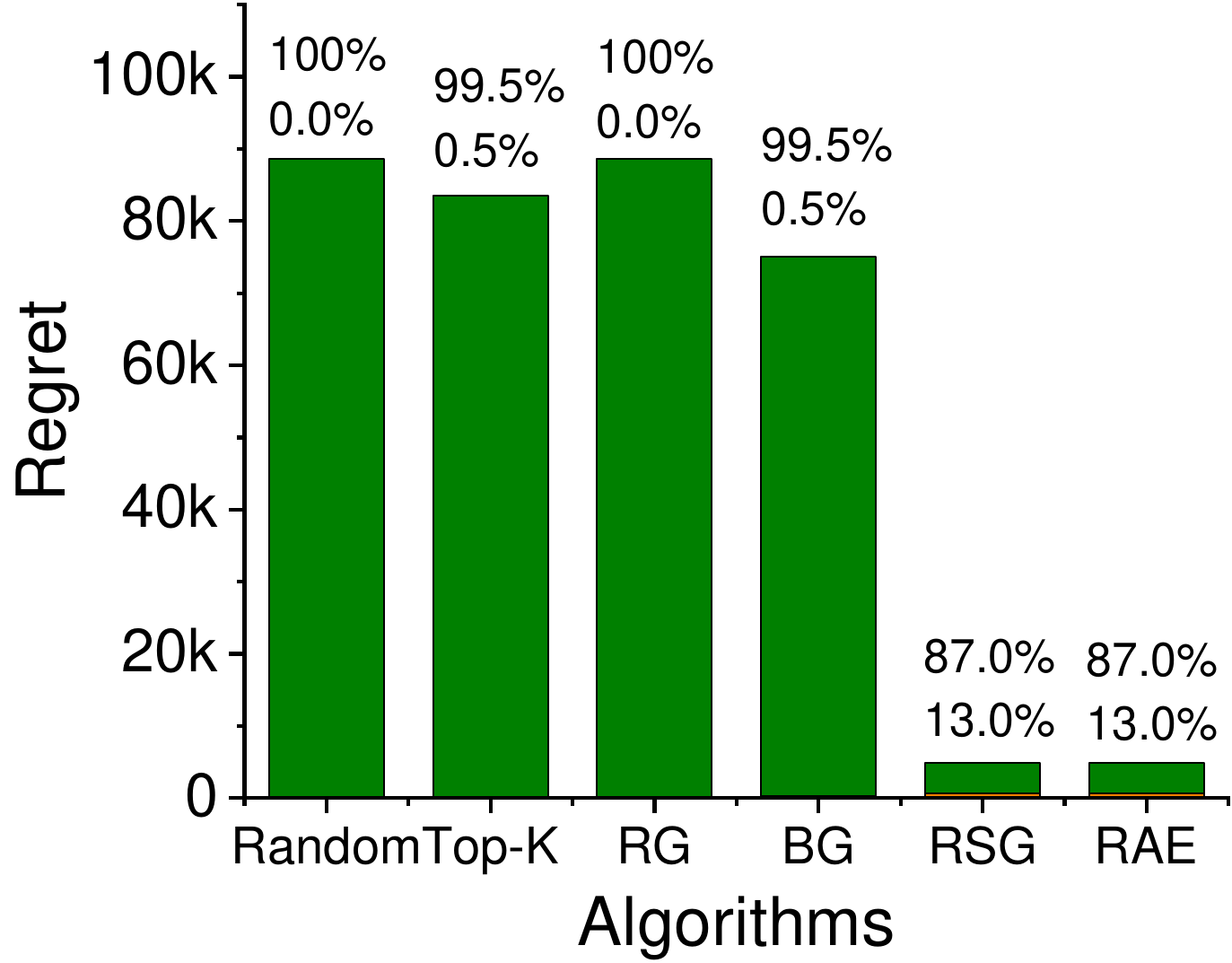} &
        \includegraphics[width=0.17\linewidth]{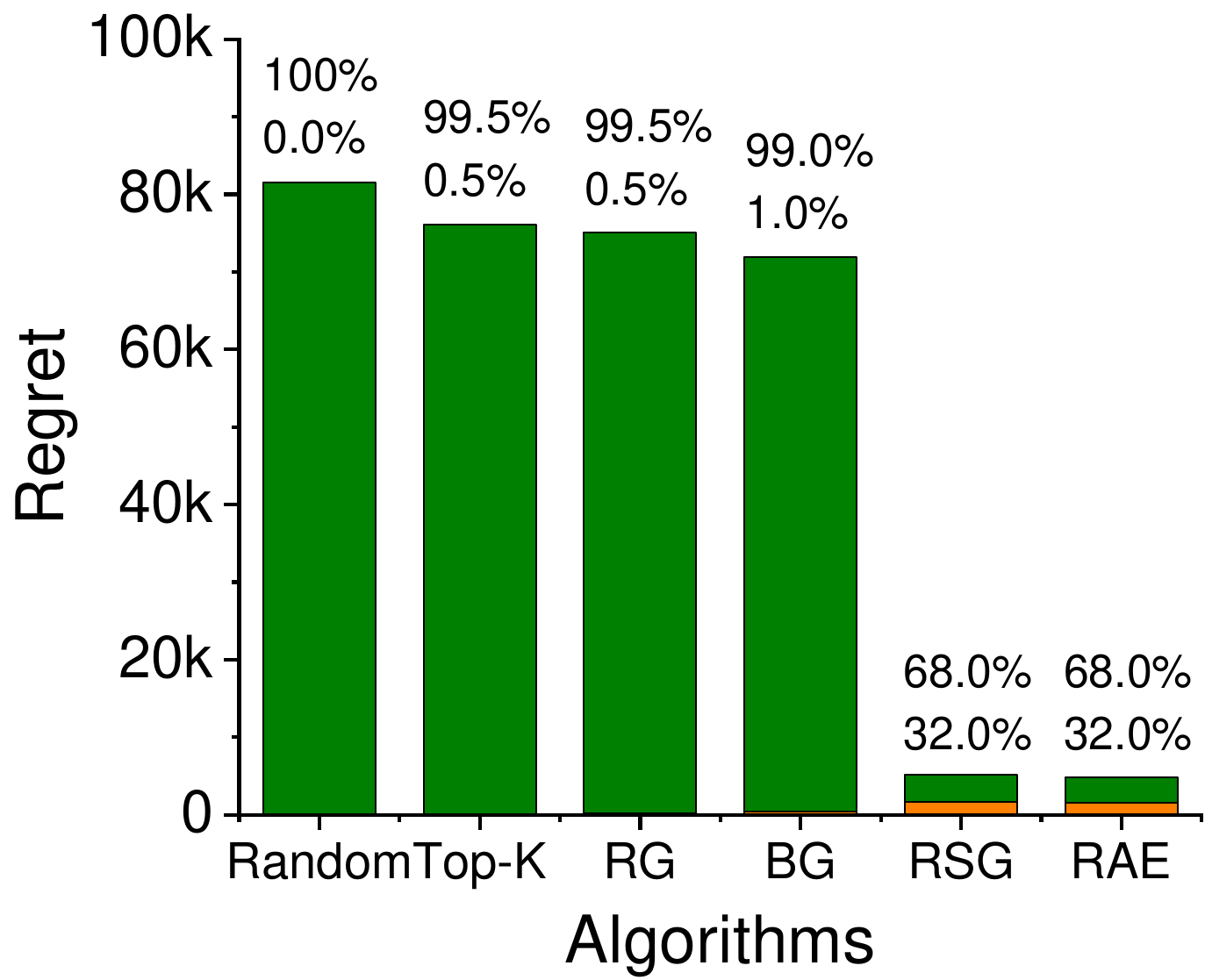} &
        \includegraphics[width=0.17\linewidth]{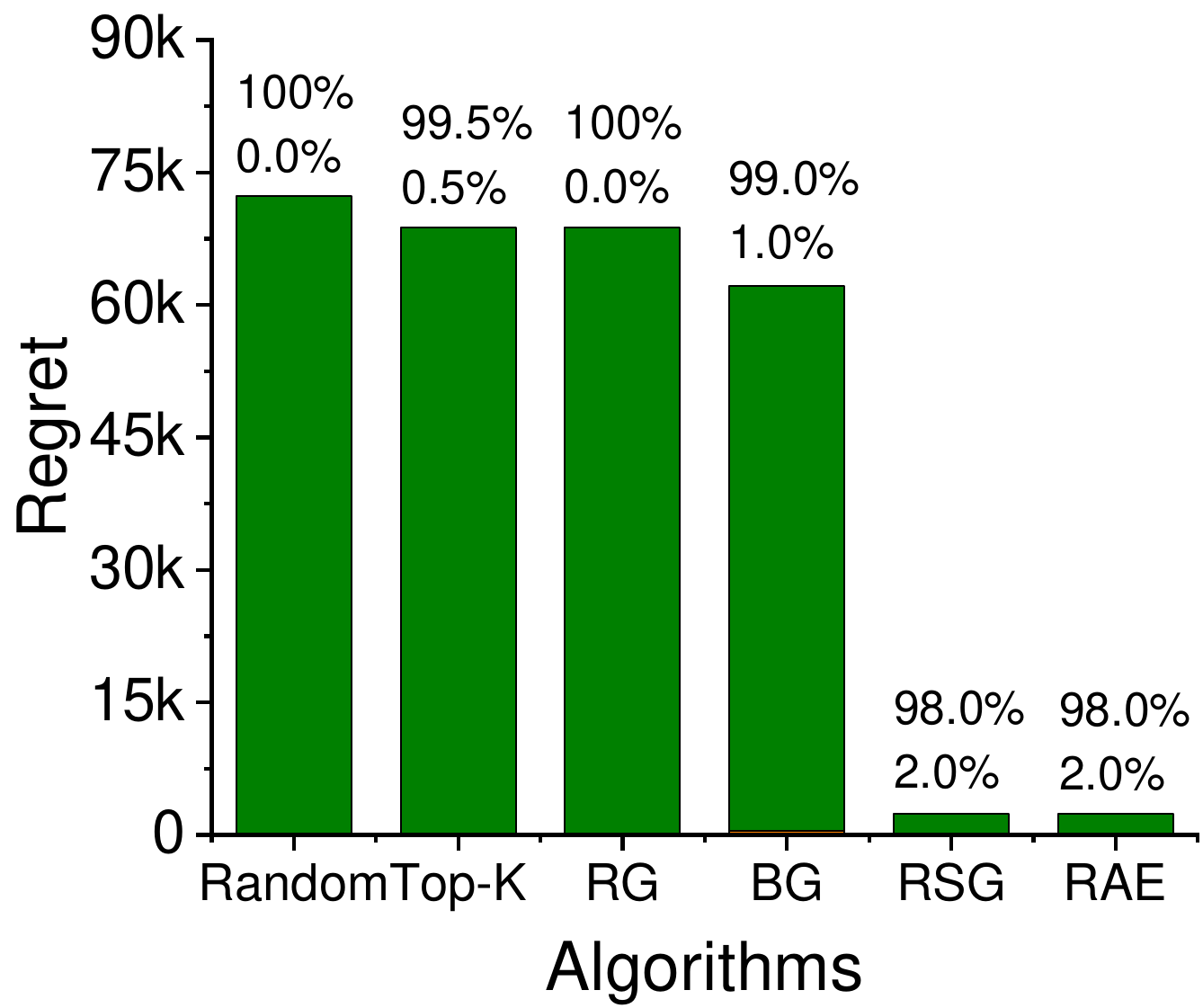} &
        \includegraphics[width=0.17\linewidth]{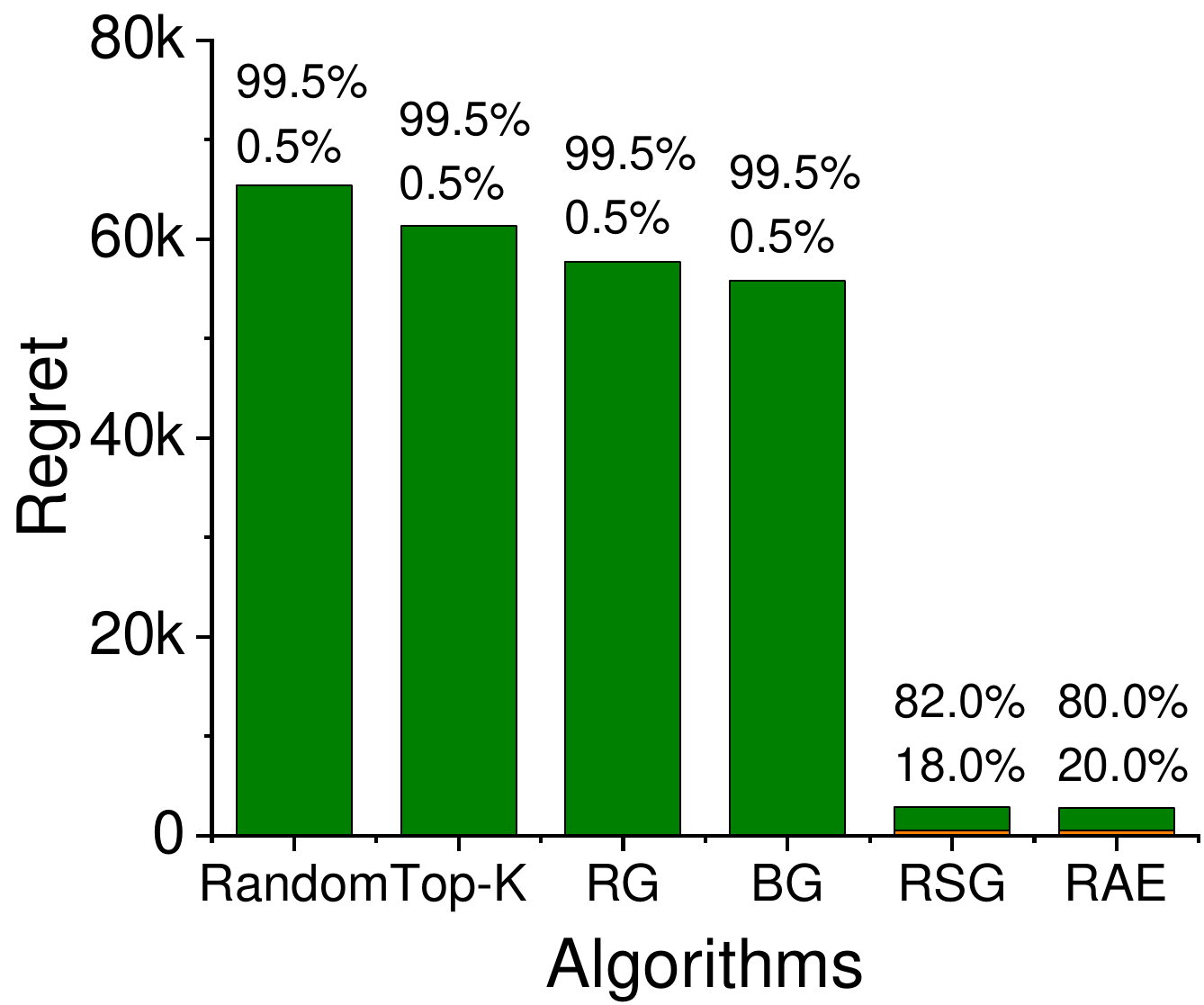} &
        \includegraphics[width=0.17\linewidth]{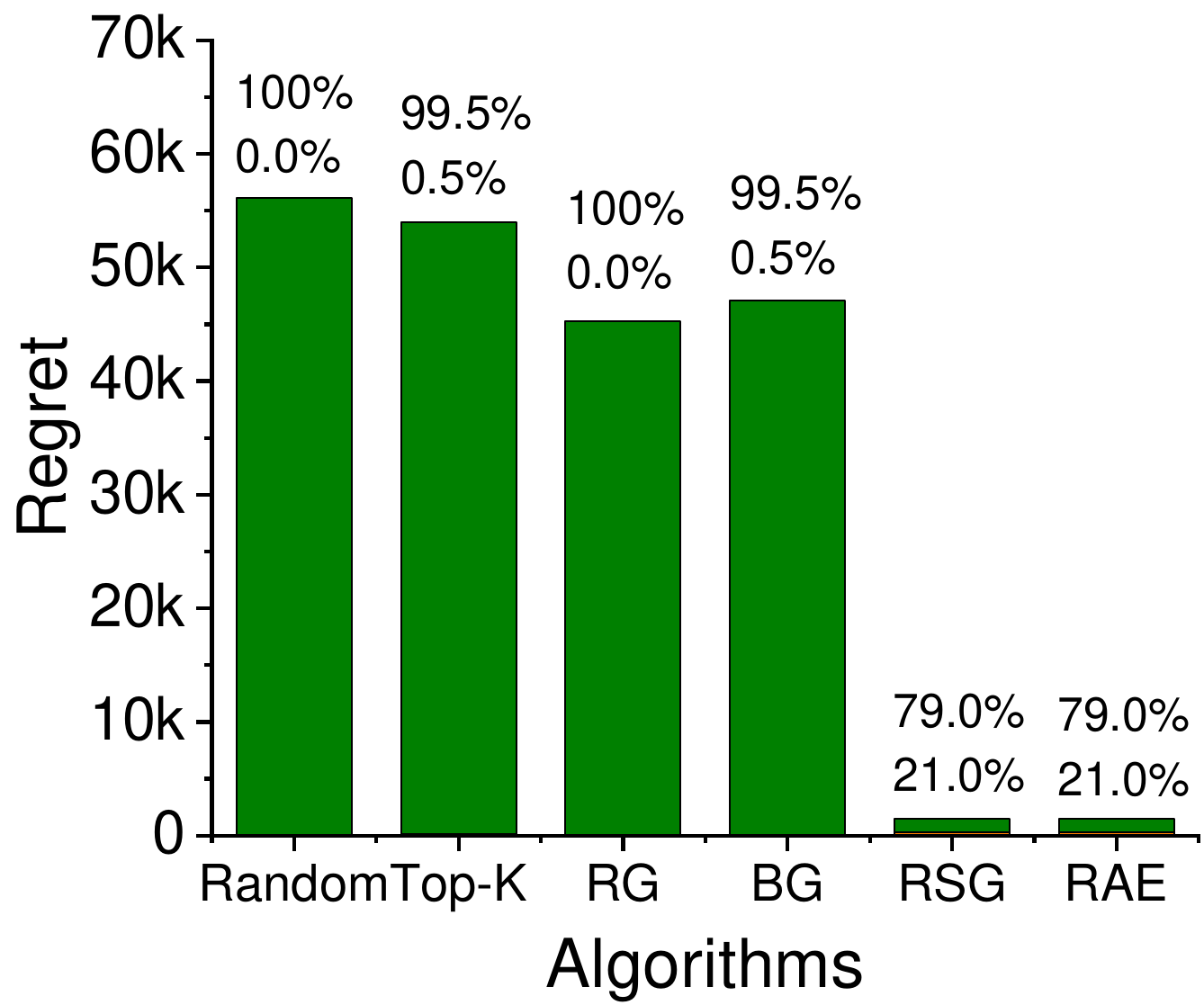} \\
        {\tiny (a) $\gamma = 0 $} &
        {\tiny (b) $\gamma = 0.25 $} &
        {\tiny (c) $\gamma = 0.5$} &
        {\tiny (d) $\gamma = 0.75$} &
        {\tiny (e) $\gamma= 1$} \\[5pt]
    \end{tabular}
   \caption{Regret on varying $\gamma$, when $\lambda = 5\%,~ \mathcal{|A|} = 20$, $\epsilon =0.01$ for NYC dataset }
    \label{Fig:NYC_Gamma}
\end{figure}

\begin{figure}[h]
    \centering
    \begin{tabular}{lclc}
       Unsatisfied Regret & \includegraphics[width=0.11\linewidth]{Result/Unsatisfied.png} & Excessive Regret & \includegraphics[width=0.11\linewidth]{Result/Excessive.png} \\
    \end{tabular}
    \begin{tabular}{ccccc}     
        \includegraphics[width=0.17\linewidth]{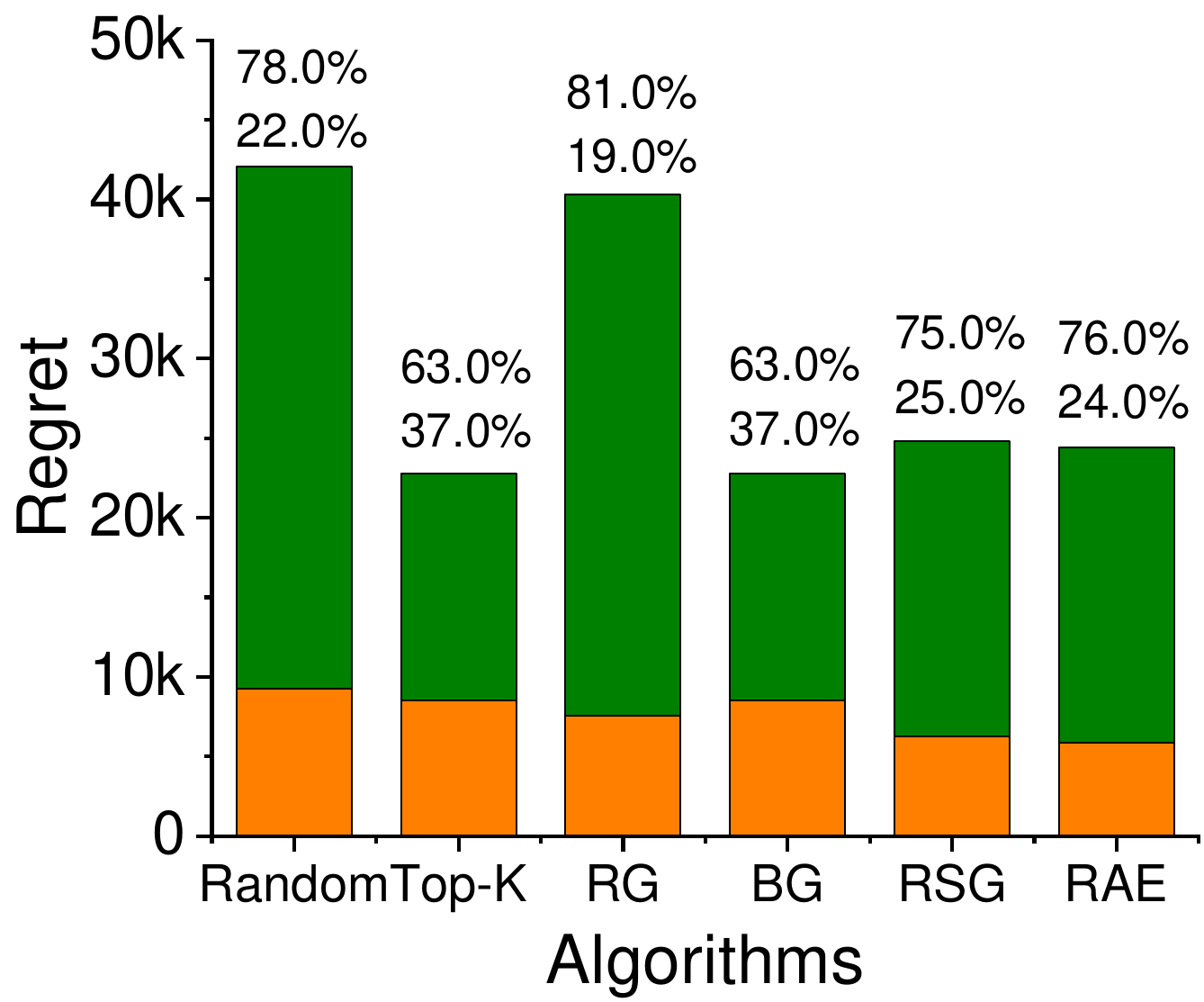} &
        \includegraphics[width=0.17\linewidth]{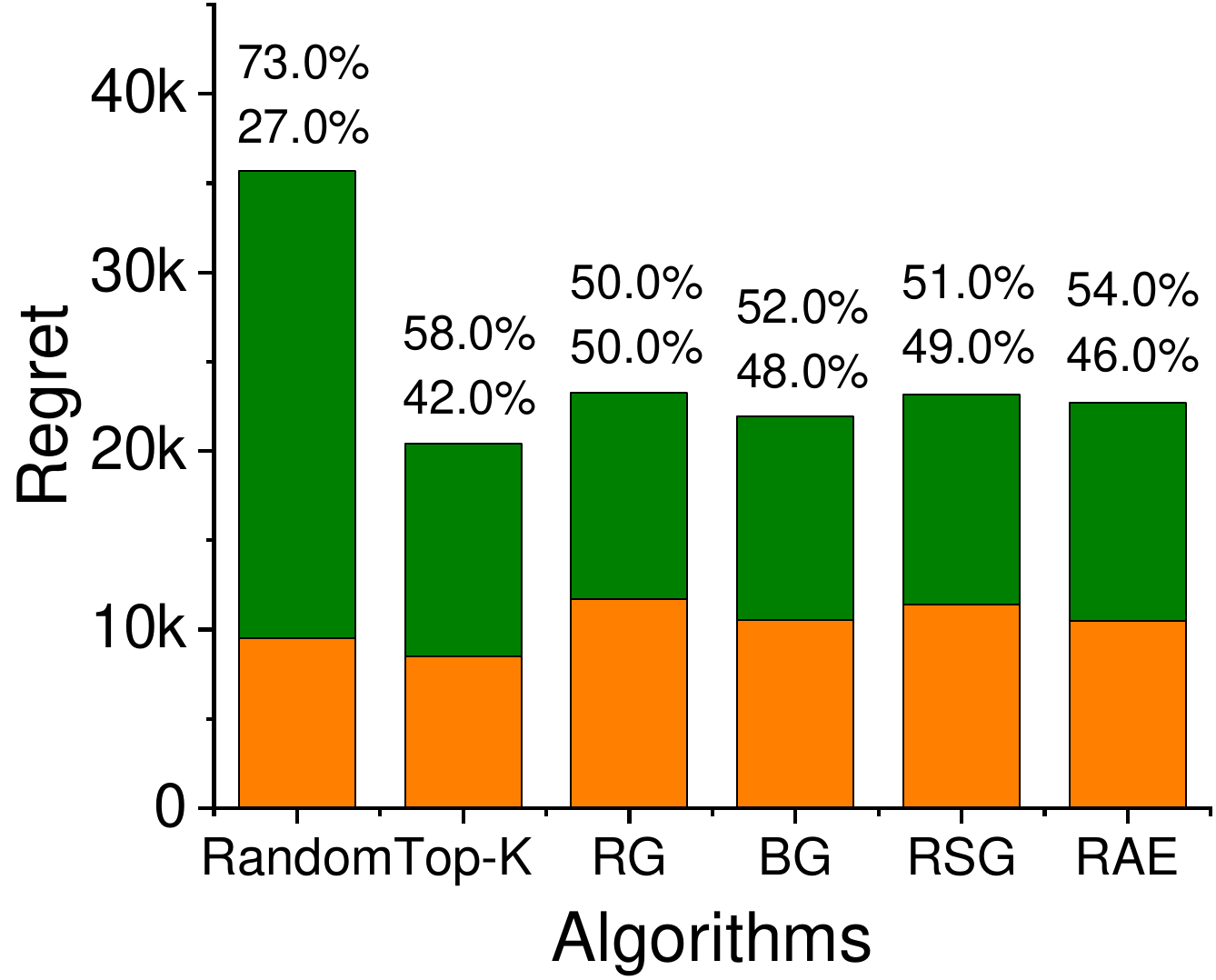} &
        \includegraphics[width=0.17\linewidth]{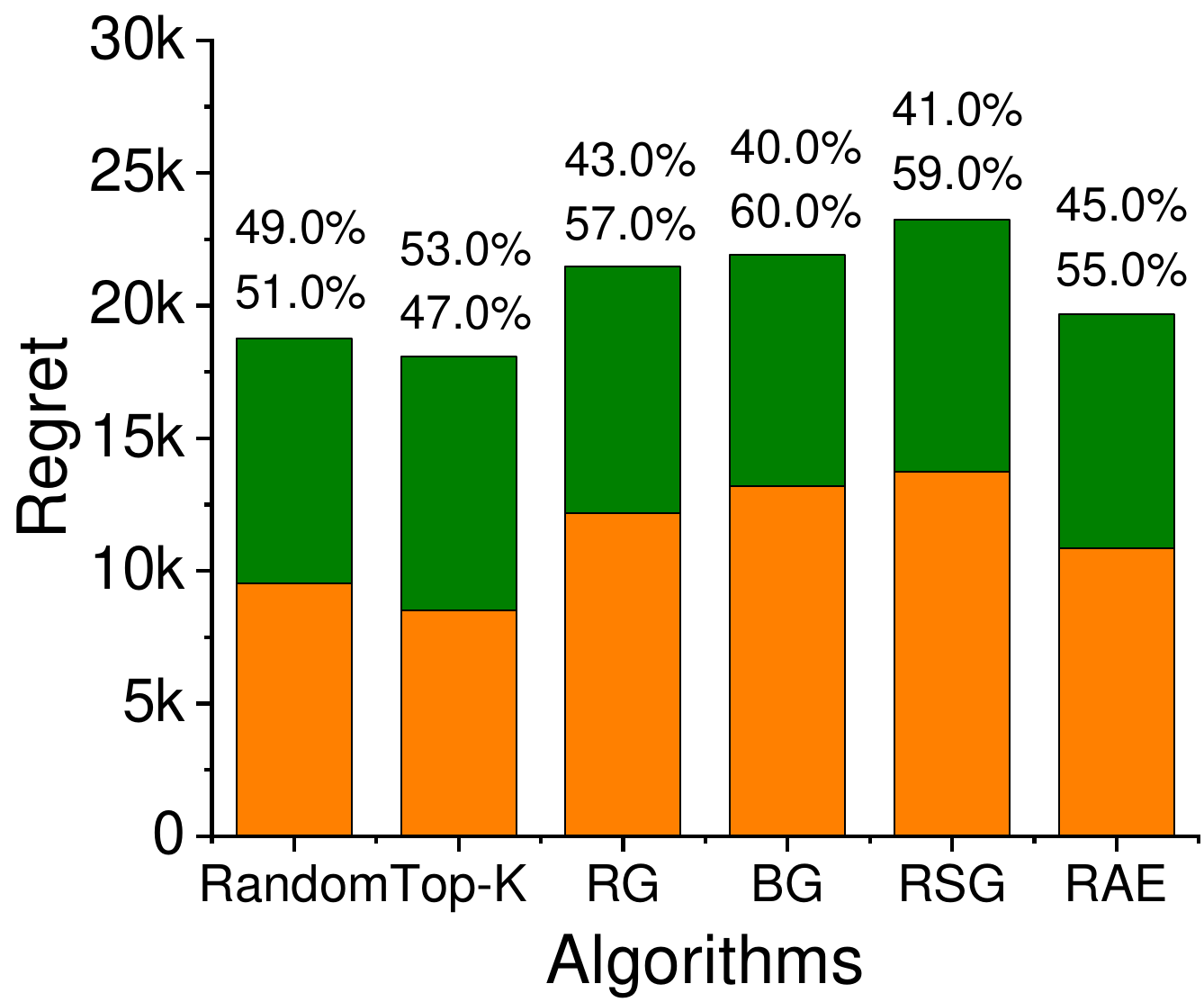} &
        \includegraphics[width=0.17\linewidth]{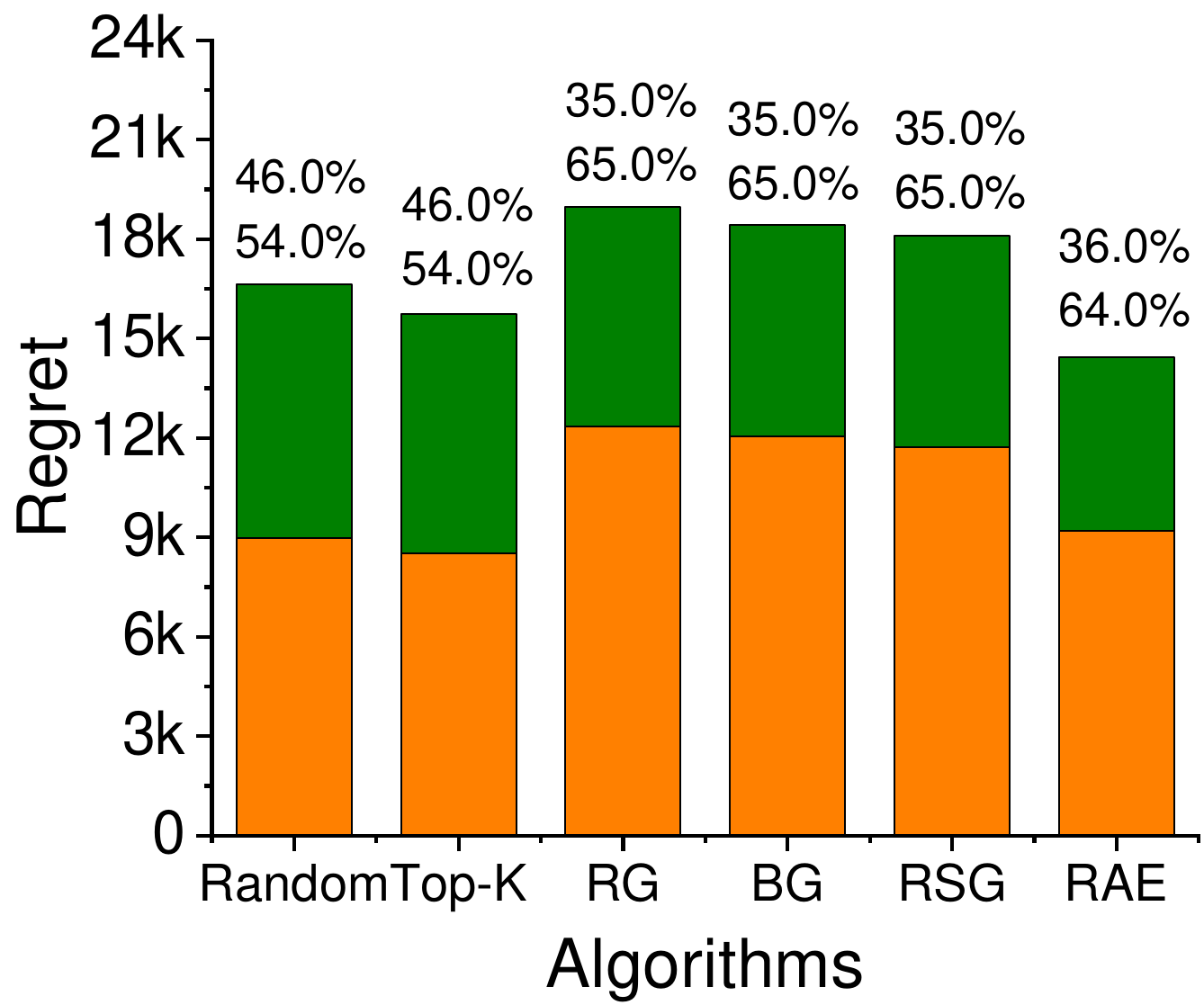} &
        \includegraphics[width=0.17\linewidth]{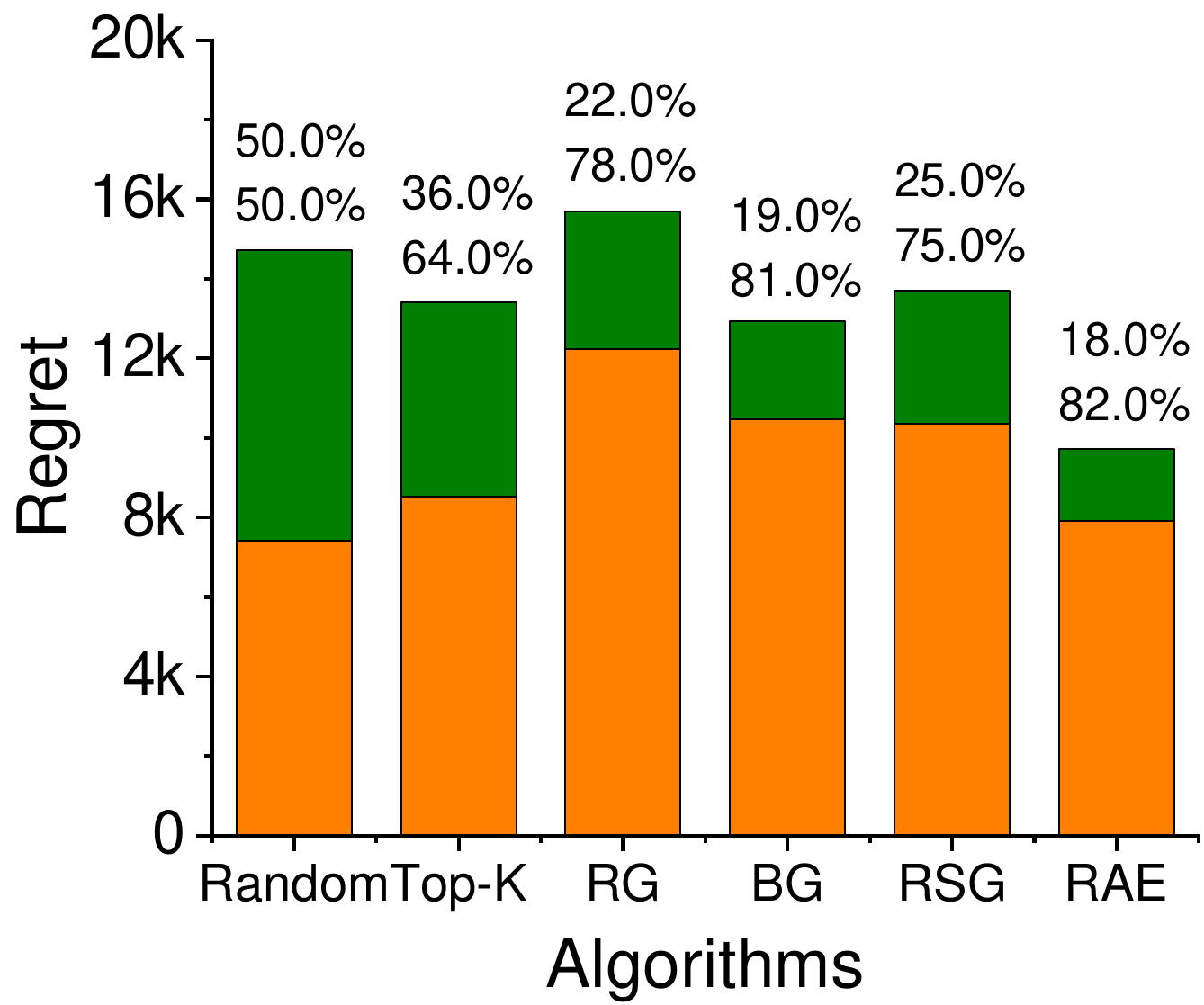} \\
        {\tiny (a) $\gamma = 0 $} &
        {\tiny (b) $\gamma = 0.25 $} &
        {\tiny (c) $\gamma = 0.5$} &
        {\tiny (d) $\gamma = 0.75$} &
        {\tiny (e) $\gamma= 1$} \\[5pt]
    \end{tabular}
   \caption{Regret on varying $\gamma$, when $\lambda = 5\%,~ \mathcal{|A|} = 20$, $\epsilon =0.01$ for LA dataset }
    \label{Fig:LA_Gamma}
\end{figure}

\begin{figure}[h]
    \centering
    \begin{tabular}{lclc}
       Unsatisfied Regret & \includegraphics[width=0.11\linewidth]{Result/Unsatisfied.png} & Excessive Regret & \includegraphics[width=0.11\linewidth]{Result/Excessive.png} \\
    \end{tabular}
    \begin{tabular}{ccccc}     
        \includegraphics[width=0.17\linewidth]{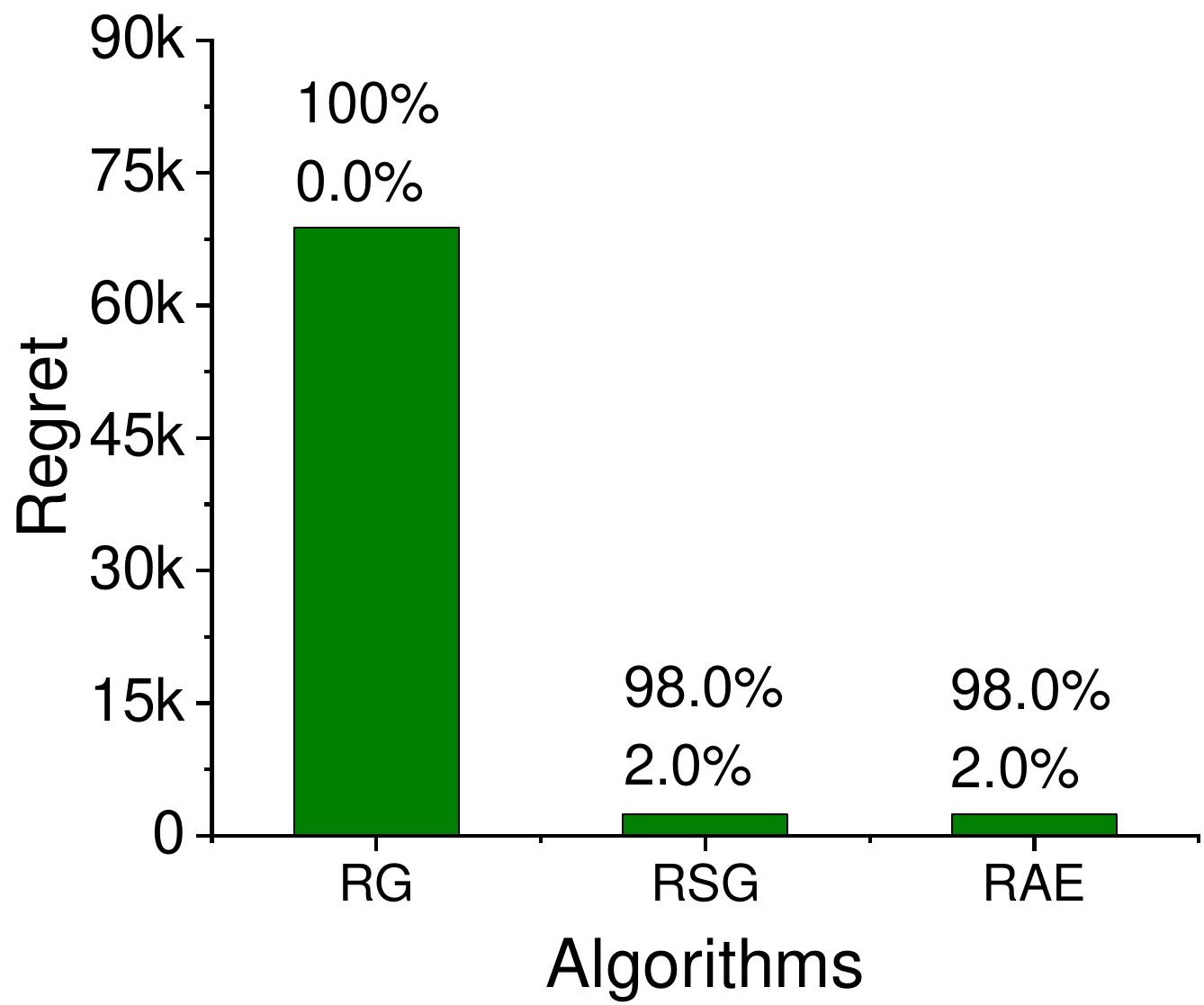} &
        \includegraphics[width=0.17\linewidth]{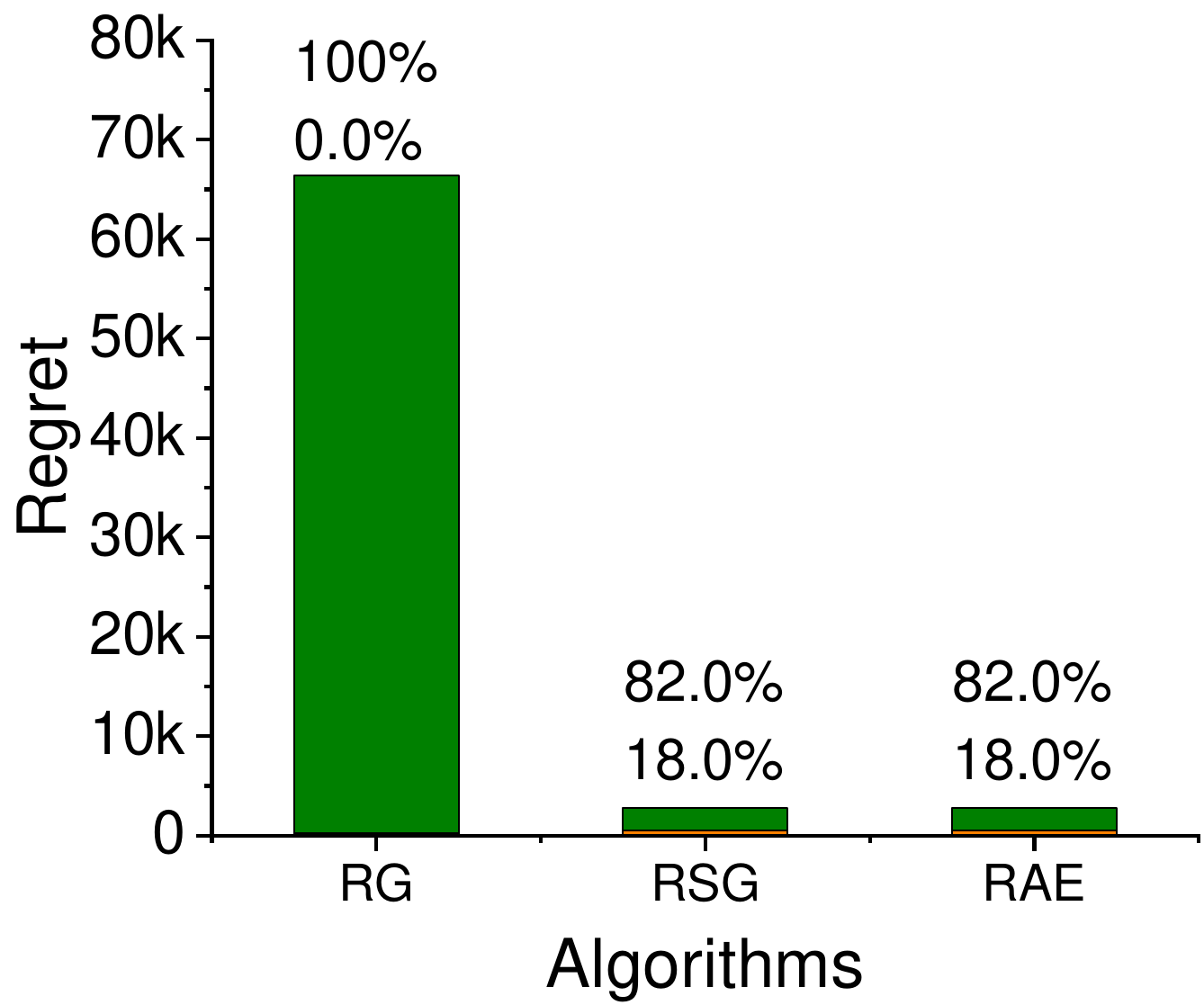} &
        \includegraphics[width=0.17\linewidth]{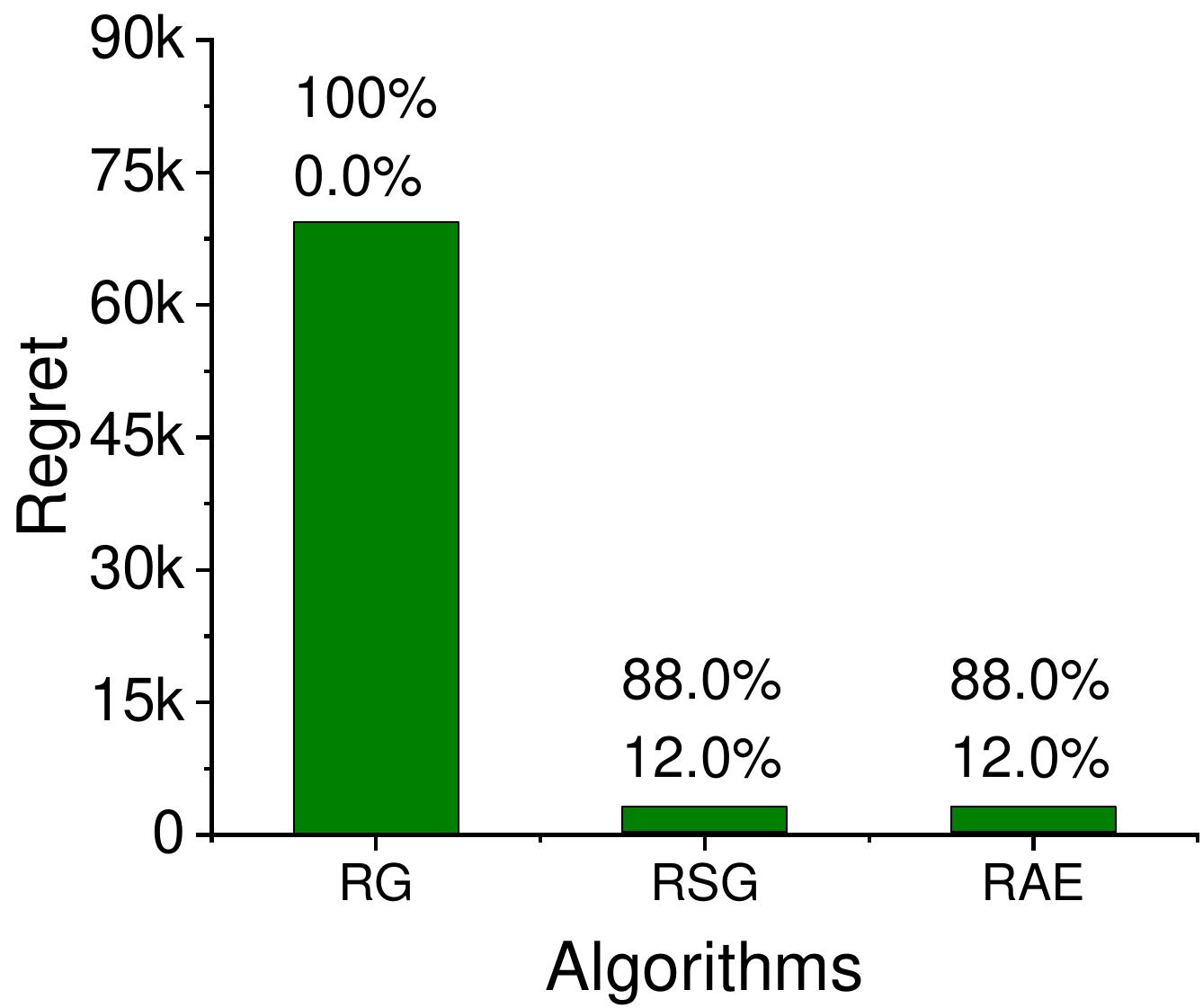} &
        \includegraphics[width=0.17\linewidth]{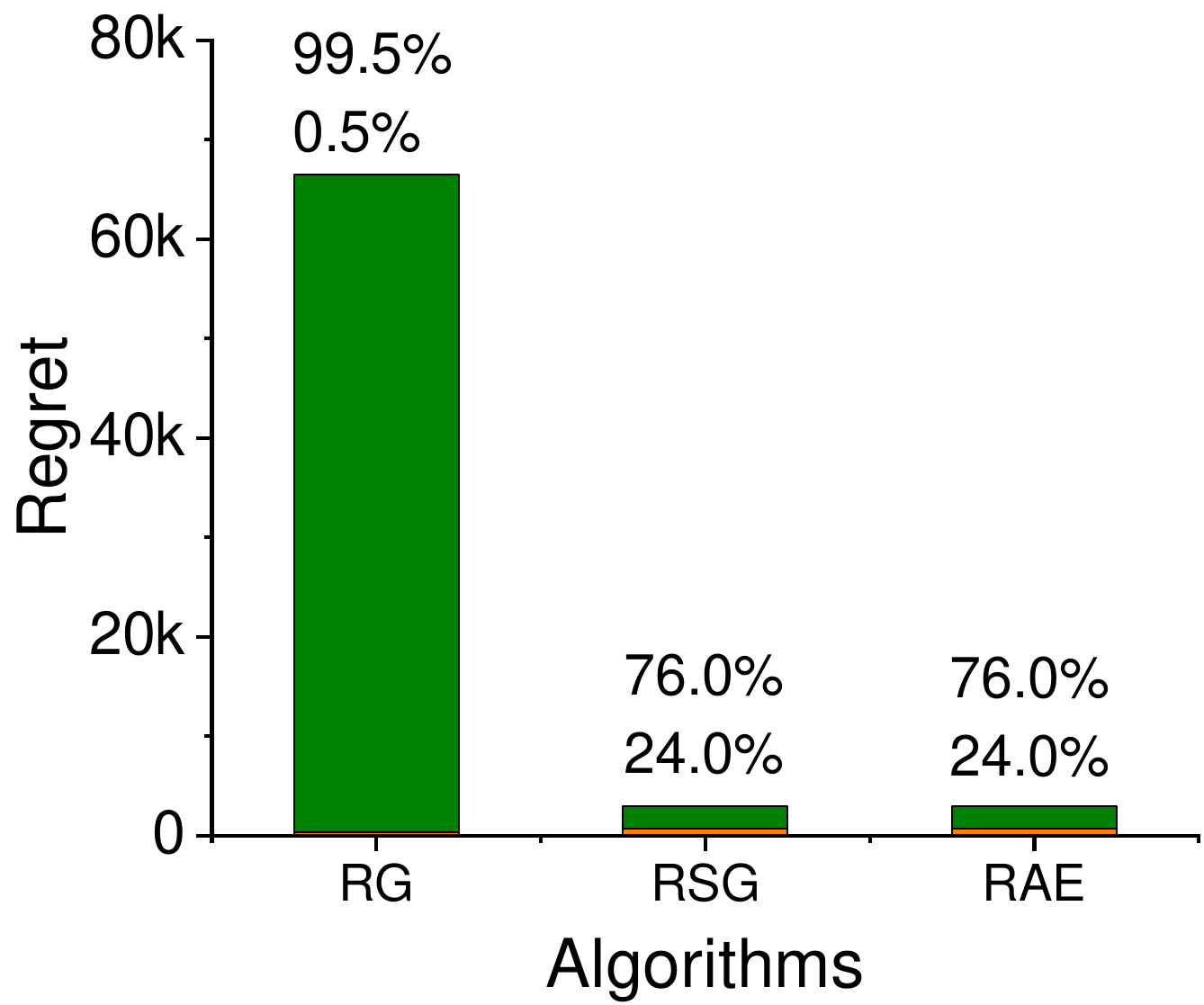} &
        \includegraphics[width=0.17\linewidth]{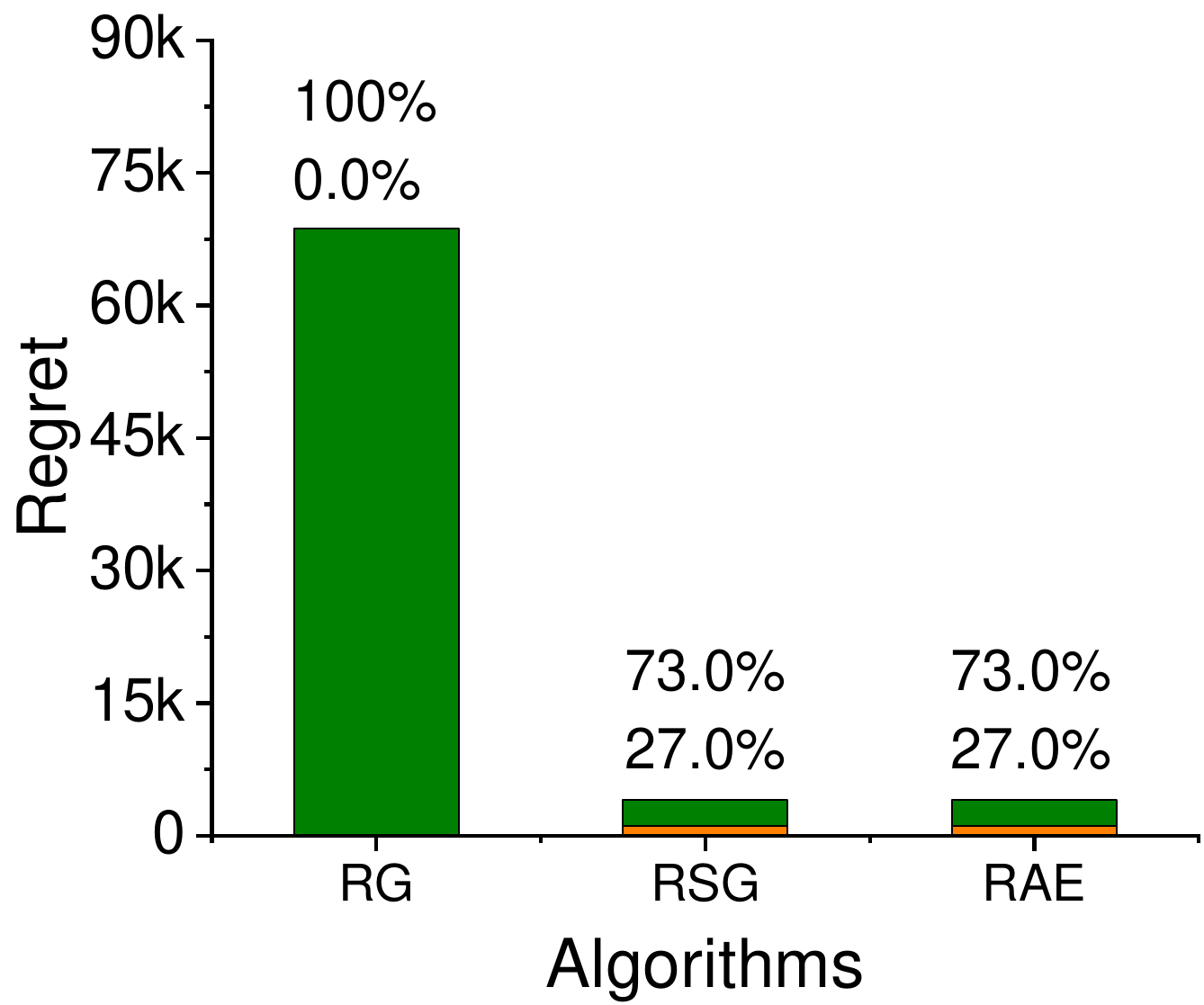} \\
        {\tiny (a) $\epsilon = 0.01 $} &
        {\tiny (b) $\epsilon = 0.05 $} &
        {\tiny (c) $\epsilon = 0.1$} &
        {\tiny (d) $\epsilon = 0.15$} &
        {\tiny (e) $\epsilon = 2$} \\[5pt]
    \end{tabular}
   \caption{Regret on varying $\epsilon$, when $\lambda = 5\%,~ \mathcal{|A|} = 20$, $\gamma =0.5$ for NYC dataset }
    \label{Fig:NYC_Epsilon}
\end{figure}

\begin{figure}[h]
    \centering
    \begin{tabular}{lclc}
       Unsatisfied Regret & \includegraphics[width=0.11\linewidth]{Result/Unsatisfied.png} & Excessive Regret & \includegraphics[width=0.11\linewidth]{Result/Excessive.png} \\
    \end{tabular}
    \begin{tabular}{ccccc}     
        \includegraphics[width=0.17\linewidth]{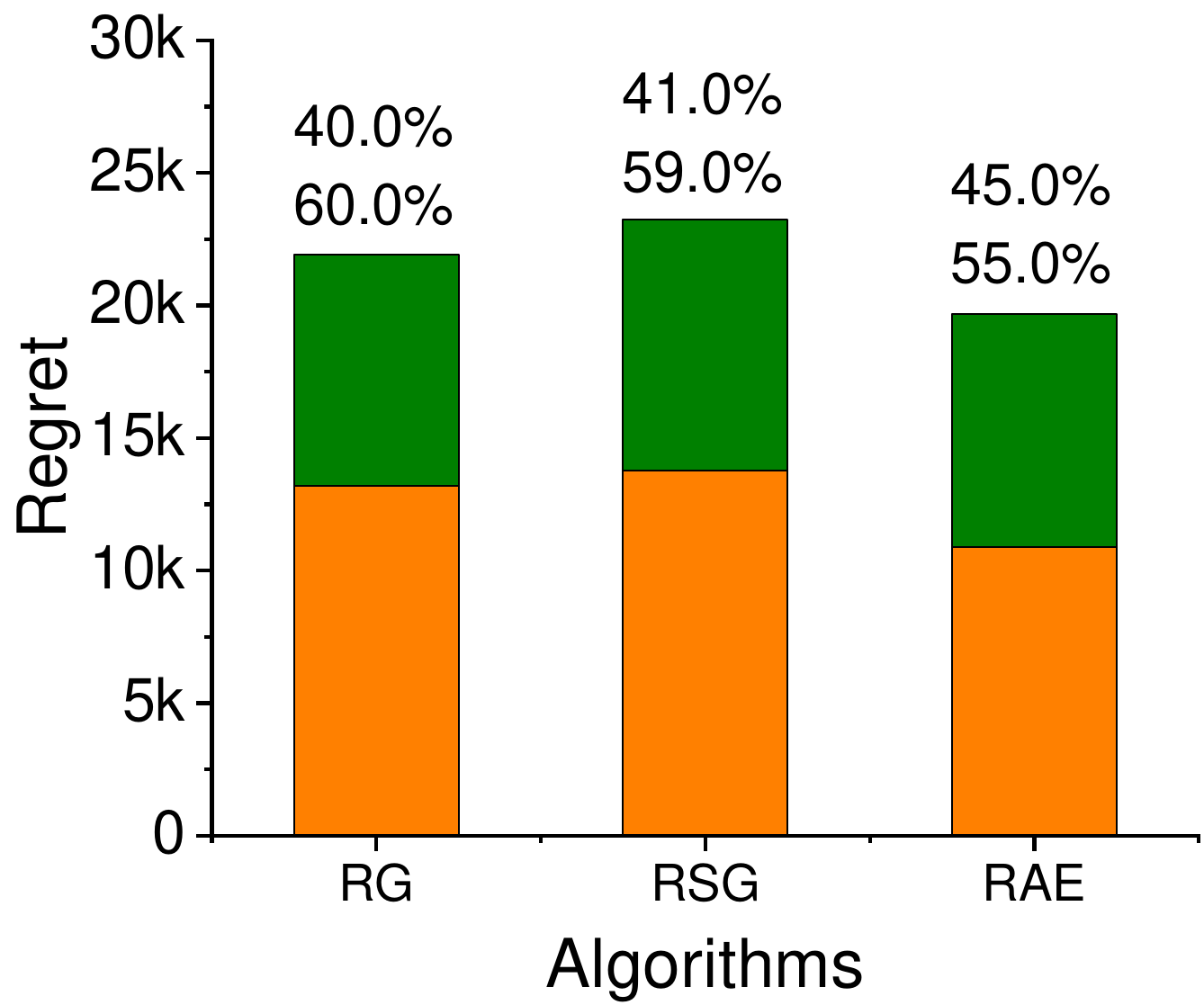} &
        \includegraphics[width=0.17\linewidth]{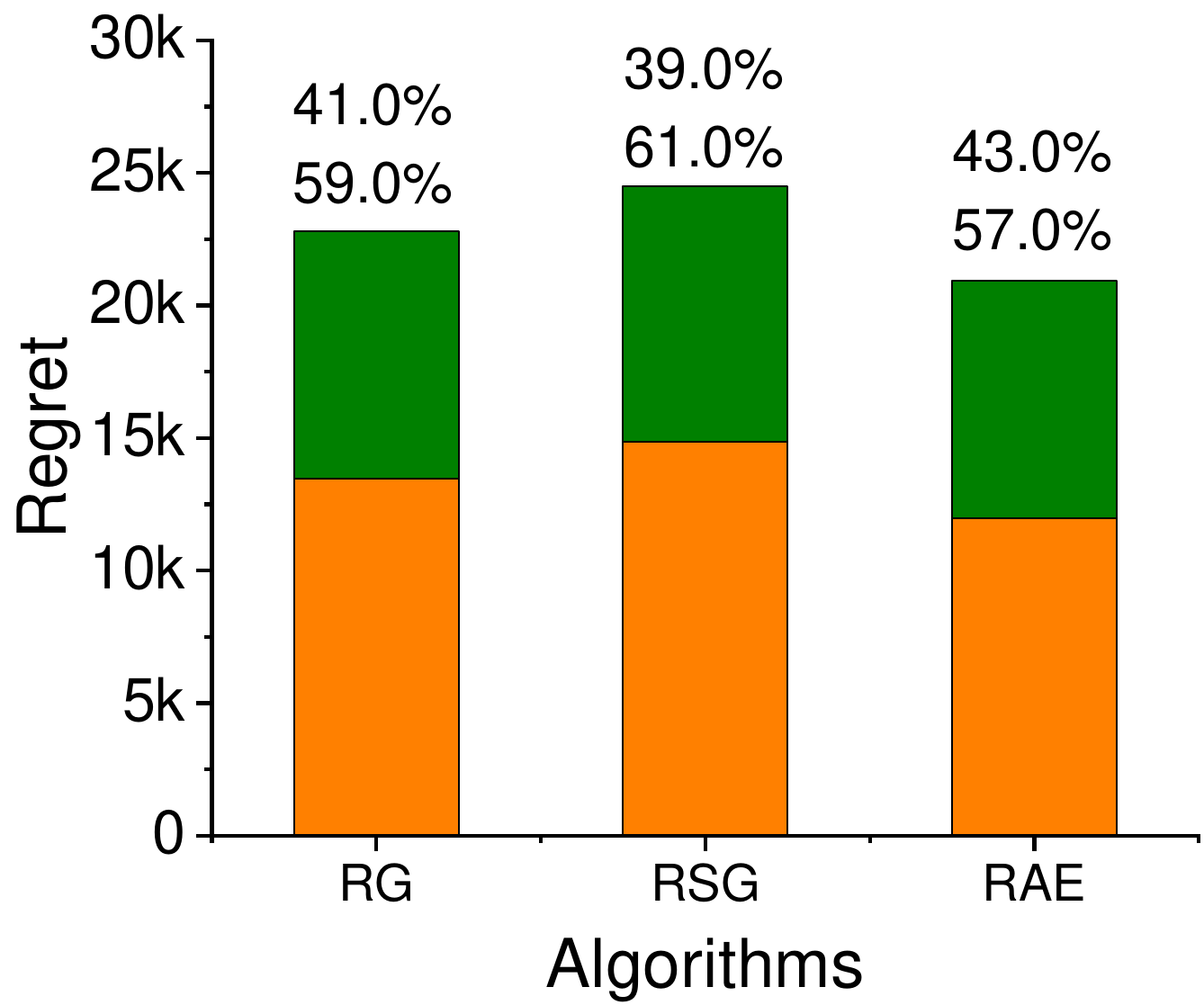} &
        \includegraphics[width=0.17\linewidth]{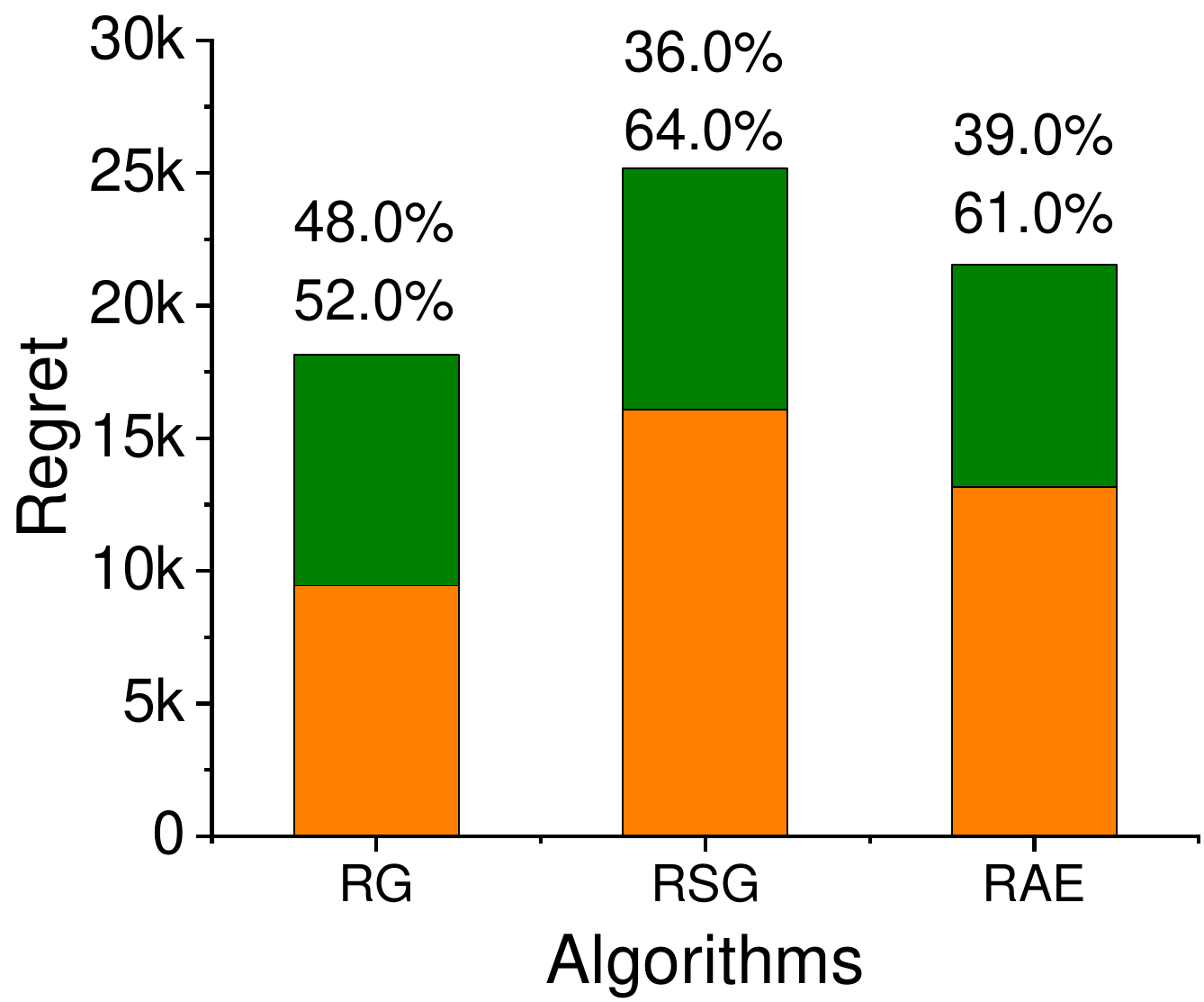} &
        \includegraphics[width=0.17\linewidth]{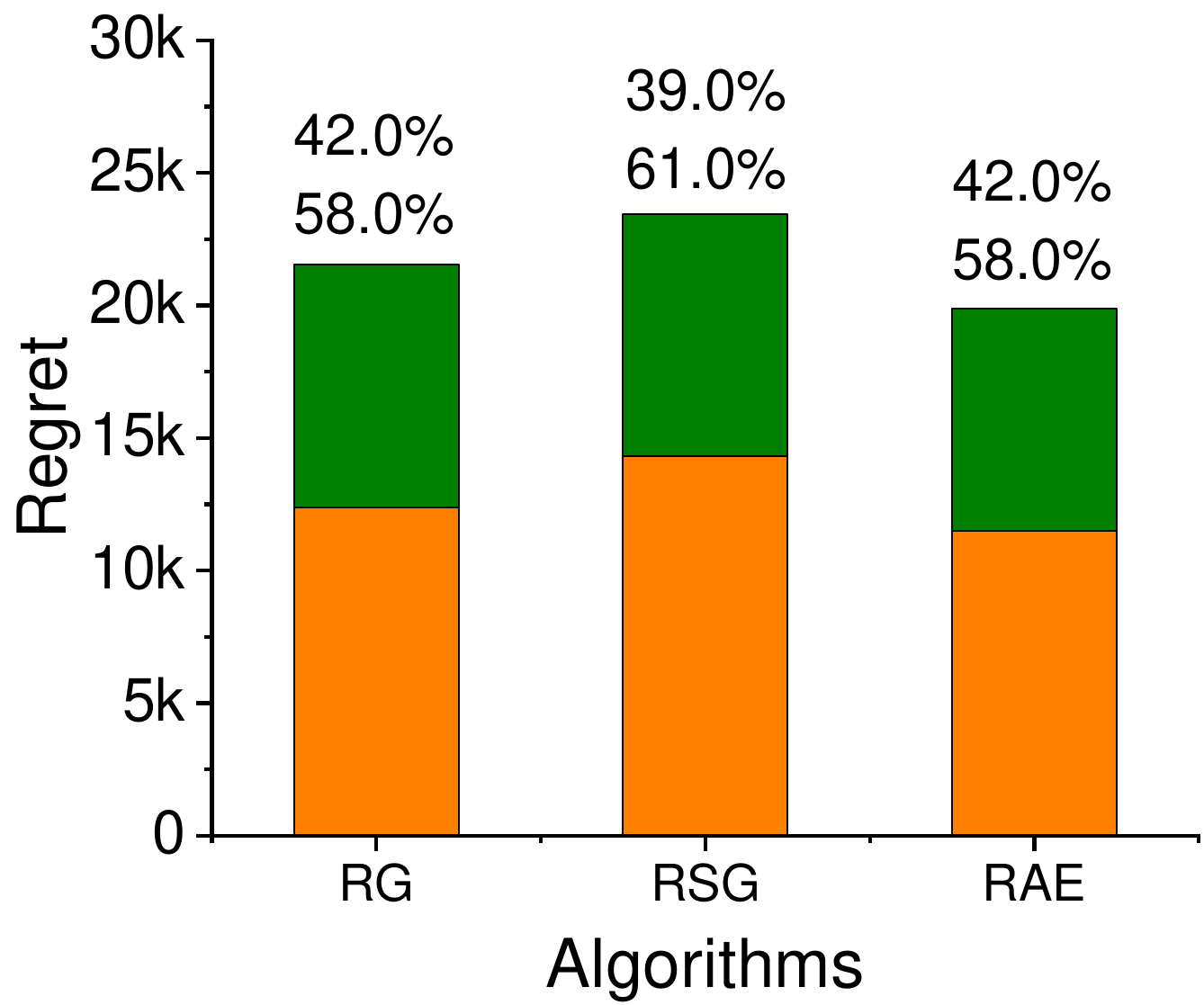} &
        \includegraphics[width=0.17\linewidth]{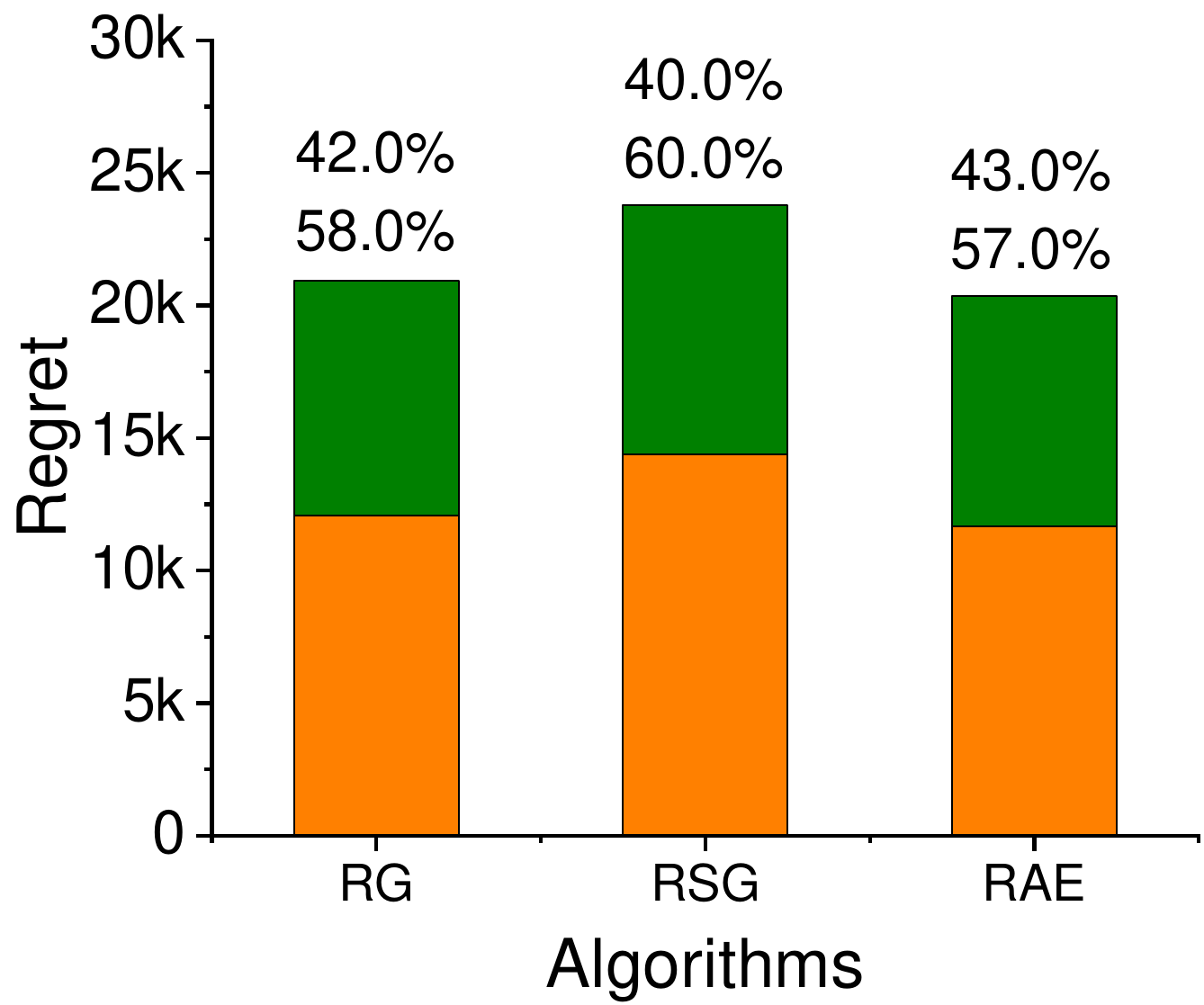} \\
        {\tiny (a) $\epsilon = 0.01 $} &
        {\tiny (b) $\epsilon = 0.05 $} &
        {\tiny (c) $\epsilon = 0.1$} &
        {\tiny (d) $\epsilon = 0.15$} &
        {\tiny (e) $\epsilon = 2$} \\[5pt]
    \end{tabular}
   \caption{Regret on varying $\epsilon$, when $\lambda = 5\%,~ \mathcal{|A|} = 20$, $\gamma =0.5$ for LA dataset }
    \label{Fig:LA_Epsilon}
\end{figure}

This section discusses the additional parameters used in our experiments, e.g., $\eta$, $\gamma$, and $\epsilon$. First, we observe that the total regret for all the proposed and baseline methods increases with the increase of distance $(\eta)$. As one billboard slot can influence more trajectories, the influence provider's supply, $\sigma^{h}$, increases. While increasing $\sigma^{h}$ but fixing the value of $\delta$ and $\lambda$, the advertiser's demand increases, and consequently, regret rises proportionally. Second, we observed that $\gamma$ plays a key role in the increase of unsatisfied regret. As we previously discussed in Definition \ref{Def:Reg_Model}, $\gamma$ controls the unsatisfied regret for the unsatisfied advertisers. When the $\gamma$ value is very small, influence providers suffer from higher regret. However, when the $\gamma$ value increases, the unsatisfied regret decreases, as represented in Figure \ref{Fig:NYC_Gamma} and \ref{Fig:LA_Gamma}. Third, the accuracy speed-up parameter, $\epsilon$, plays a crucial role in the `RG', `RSG', and `RAE' algorithms, and from Figure \ref{Fig:NYC_Epsilon} and \ref{Fig:LA_Epsilon}, it is clear that when $\epsilon$ value increases, total regret also increases, but computational time decreases.

\par

\paragraph{\textbf{Revisit $RQ4$, and $RQ5$.}}
Based on our observations in section \ref{ES} and \ref{APS}, we can answer $RQ4$ and $RQ5$. (1) when we vary the $\delta$ value from $40\%$ to $120\%$, the computational cost for all the algorithms also rises for both the NYC and LA datasets. With the increasing value of $\delta$, each advertiser's individual influence demand and unsatisfied regret also increases. (2) when $\gamma$ value increases from $0$ to $1$, the total regret of an advertiser decreases. $\gamma = 0$ provides higher regret while $\gamma = 1$ provides lower regret as only the fraction of influence provided to the advertiser is returning as regret. Next, with the increase $\epsilon$, an advertiser's total regret increases because the randomly selected sample size decreases and degrades the solution quality.
\section{Conclusion and Future Directions} \label{Sec:CFD}
In this paper, we have studied the problem of regret minimization in billboard advertisements under zonal influence constraint in a multi-advertiser setting. We have proposed a Budget-Effective Greedy algorithm to allocate billboard slots to the advertisers. To improve its performance further, we proposed the `RG', `RSG', and `RAE' algorithms. All the methods have been analyzed to understand their time and space requirements. They have been implemented with real-world datasets and compared with existing methods to show their effectiveness and efficiency. Our future study of this problem will focus on developing more efficient local search techniques to improve the performance further.

\bibliography{sample-bibliography}

\end{document}